\documentclass[%
 reprint,
nofootinbib,
 amsmath,amssymb,
 aps,
prd,
]{revtex4-2}

\usepackage{newtxtext,newtxmath}
\usepackage[T1]{fontenc}
\usepackage{ae,aecompl}
\usepackage{float}

\usepackage{color}

\usepackage{graphicx}% Include figure files
\usepackage{dcolumn}% Align table columns on decimal point
\usepackage{bm}% bold math
\usepackage{subcaption}
\usepackage{caption}
\usepackage{color}
\usepackage{hyperref}
\usepackage{makecell}
\usepackage{tabularx}
\usepackage{aas_macros}
\usepackage{multirow}

\newcommand{\aplus}{$A_{+}\,$}
\newcommand{\aetwo}{$A^{\mathrm{E2}}_{20}$}
\newcommand{\ud}{\mathrm{d}}
\newcommand{\msun}{\,\mathrm{M}_\odot}

\begin{document}

\preprint{}

\title[GWs from rotating CCSNe]{Trends in gravitational wave emission in axisymmetric simulations of rotating core-collapse supernovae}

\author{Bailey Sykes}
 \email{bailey.sykes@monash.edu}
\author{Bernhard M\"{u}ller}%
 \email{bernhard.mueller@monash.edu}
\affiliation{%
 School of Physics and Astronomy, Monash University, Clayton, VIC, 3010, Australia
}%

\date{\today}

\begin{abstract}
The quantitative impact of strong rotation on the amplitudes and frequencies of the post-bounce gravitational wave (GW) signal from core-collapse supernovae (CCSNe) is still not fully understood. To study trends in frequencies and amplitudes, and possibly spectacular phenomena like resonant amplification, we perform a series of axisymmetric long-duration magnetohydrodynamic CCSN simulations of a $17 \msun$ progenitor using a finely spaced grid in initial rotation rate from $0.29\,\mathrm{rad}\,\mathrm{s}^{-1}$ to  $3.48\,\mathrm{rad}\,\mathrm{s}^{-1}$. We find that these rotating models produce GWs at frequencies of up to $3\,\mathrm{kHz}$, higher than in typical non-rotating models in the literature. The high frequencies arise due to small polar radii of rapidly rotating proto-neutron stars and stabilization by angular momentum gradients at lower latitude. GW frequencies and amplitudes tend to decrease with faster rotation. Different from two complementary simulations without magnetic fields, the magnetohydrodynamic models are characterized by an absence of $p$-modes above the dominant high-frequency emission band. We find no indication of resonant mode amplification for any rotation rate, although a temporo-spatial and space-frequency analysis reveals some interesting couplings of quadrupolar motions across the proto-neutron star and the gain region. We find that linear mode analysis based on the spherically averaged structure becomes unsuitable in this regime of rapid rotation. Generalized, multi-dimensional perturbative techniques need to be developed to study the mode structure and mode interaction in the collapse of rapidly rotating massive stars.

\end{abstract}

\maketitle

\section{Introduction}

For a sufficiently massive star with zero-age main sequence mass $M_\mathrm{ZAMS} \gtrsim 8 \, \mathrm{M_{\odot}}$, the final phase of its evolution is a violent death in the form of a core-collapse supernova (CCSN). Left behind in the resulting cloud of ashes is a compact remnant: either a neutron star (NS) or a black hole (BH). While the electromagnetically luminous explosion itself is routinely observed, the multi-messenger signals more closely associated with the formation of the compact remnant -- i.e. neutrinos and gravitational waves (GWs) -- remain elusive, with the sole exception of the roughly two-dozen neutrinos from SN1987A. Advances are being made on both fronts, with the ongoing development of more sensitive detectors from both neutrinos \citep{HyperK:2018} and GWs \citep{aLIGO:2015} potentially expanding the scope of multi-messenger astronomy to new types of transients.

In the past decade, the detection of GWs from compact binary mergers has become commonplace, with many hundreds of events recorded to date. With this success, the search for GWs is broadening to include other transients \citep{Powell_Lasky:2025}, with CCSNe being a prominent example. This introduces new challenges however, as poor constraints on the signal morphology make template-based search techniques \citep{gwtc3:2023} less effective. To resolve this issue in the context of GWs from CCSNe, simulations of core-collapse have been performed in both 2D \citep[e.g.,][]{Murphy_Ott_Burrows:2009, Mueller_Janka_Marek:2013, Jardine_Powell_Mueller:2022, Jakobus_et_al:2023, Yakunin_et_al:2015, Richers_et_al:2017, EggenbergerAndersen_et_al:2021} and 3D \citep[e.g.,][]{Ott_et_al:2013, Kuroda_Kotake_Takiwaki:2016,Andresen_Mueller_Mueller_Janka:2017, Andresen_et_al:2019,oConnor_Couch:2018,Powell_Mueller:2020,Powell_Mueller:2024, Radice_et_al:2019,Vartanyan_et_al:2023,  Andresen_Glas_Janka:2021}, although 2D simulations only produce the plus-polarization of the GW and may overestimate the GW amplitude \citep{Andresen_Mueller_Mueller_Janka:2017, Mezzacappa_et_al:2020, Powell_Mueller:2025}. With this large body of simulations available, the GW morphology of standard neutrino-driven explosions is better understood, and parameterized templates can be used in search algorithms \citep{Powell_Mueller:2022,Bruel_et_al:2023, Andresen_Finkel:2024, Powell_Mueller:2025}. What remains however, is a relative paucity of models which include the effects of rotation and magnetic fields. While there is a growing library of such simulations, the parameter space is still poorly sampled, and the impact of rotation on GW emission remains difficult to predict. 

While the exact GW predictions of different simulation codes vary, a few general features of the computed signals appear to be consistent, such as the emergence of a prominent $f$-mode \citep{Torres-Forne_et_al:2018}. This component of the signal appears soon after bounce and gradually increases in frequency as the proto-neutron star (PNS) structure evolves due to a combination of neutrino cooling and accretion. The frequency of the mode varies on the order of a few kHz although this depends on the specific dynamics of the collapse, as well as on the code being used. Analytic `universal relations', such as those by \citet{Torres-Forne_et_al:2018}, \citet{Sotani_Takiwaki_Togashi:2021} and \citet{Mueller_Janka_Marek:2013}, have been used to correlate GW emission with PNS properties. These appear reasonably robust, even for rotating stars, although some deviations have been found in specific cases \citep{Jardine_Powell_Mueller:2022}.

In rotational models, the GW signal also typically includes a strong but transient bounce signal \citep{Fuller_Klion_Abdikamalov_Ott:2015, Dimmelmeier_Ott_Marek_Janka:2008}. This has been noted in several recent simulations \citep[e.g.][]{Powell_Mueller:2020}, but its presence may be sensitive to numerical effects in the core region \citep{Jardine_Powell_Mueller:2022}. Furthermore, the standing accretion shock instability (SASI) is also thought to contribute a low-frequency component to the GW signal of CCSNe \citep{Andresen_et_al:2019, Powell_Mueller:2024, Powell_Mueller:2025, Kuroda_Kotake_Takiwaki:2016}, in the range of a few hundred Hz, for CCSNe both with and without rotation. 

While 3D simulations remain the gold standard for computing GW waveforms, 2D simulations still provide a convenient and computationally feasible means of exploring trends within the progenitor parameter space \citep[e.g.,][]{Richers_et_al:2017}. To this end, in this study we examine GWs from a homogeneous set of 2D simulations that finely scan the dependence on the initial rotation rate. We make particular note of the recent work by \citet{Cusinato_et_al:2025, Cusinato_Obergaulinger_Aloy:2025}, where strong transient GW amplification is found after core bounce of rotating stars due to a resonance between the PNS $f$-mode and rotationally-induced oscillations (i.e. a crossing of the $f$-mode and epicyclic frequencies). Such an amplification would be a clear imprint of CCSNe core dynamics if detected by GW observatories, and may even be evident in the corresponding neutrino signal. By using the same progenitor as \citet{Cusinato_et_al:2025}, performing comparable magnetohydrodynamics (MHD) simulations which employ similar microphysics, we are able to investigate the robustness of this resonance effect.  

This paper will commence with a description of our simulation setup and an overview of all models in Section \ref{sec:setup}. We briefly discuss the collapse/explosion dynamics in Section \ref{sec:exp_dyn}, although we note this discussion is heavily tuned towards results relevant for the subsequent discussion of GWs of Section \ref{sec:gws}. We present our investigation of GWs in three parts: first, in Section \ref{sec:gw_main_set}, we present results for our main set of $14$ simulations with varying rotation rates. This is followed in Section \ref{sec:gw_spatial} by a detailed examination of a spatial decomposition of the GW signal generated in an additional auxiliary simulation. The investigation is rounded out in Section \ref{sec:linear_mode} with an evaluation of linear mode analysis in the context of rotating stars. We critically evaluate our experiments and offer some concluding remarks in Section \ref{sec:conclusions}.

\section{Simulation setup}
\label{sec:setup}
We perform $15$ simulations of rotational core-collapse with the Newtonian MHD version of the \textsc{CoCoNuT-FMT} code \citep{Mueller_Janka_Dimmelmeier:2010, Mueller_Janka_Marek:2012, Mueller_Janka_Marek:2013, Mueller_Varma:2020}. Another two simulations are performed with the non-MHD version, bringing the total to $17$ simulations. Neutrino transport is dealt with by way of the fast multigroup transport (FMT) scheme of \citet{Mueller_Janka:2015}. The SFHo equation of state (EoS) of \citet{Steiner_Hempel_Fischer:2013} is used in the high-density regime near the PNS. 

All models are initialized from the same progenitor: a low-metallicity ($\mathrm{Z} = 0.02\, \mathrm{Z_{\odot}}$) model by \citet{Aguilera-Dena_Langer_Moriya_Schootemeijer:2018} with a main-sequence mass of $\mathrm{M_{ZAMS}} = 17 \, \mathrm{M_{\odot}}$, evolved with rotation and magnetic fields with the \textsc{MESA} stellar evolution code \citep{Paxton_et_al:2011}. For the MHD simulations, the initial magnetic field has a maximum strength of $B_\mathrm{tor} = B_\mathrm{pol} = 10^{10} \, \mathrm{G}$ and is given a twisted toroidal configuration similar to \citet{Sykes_Mueller:2025}, \citet{Varma_Mueller_Obergaulinger:2021}, and \citet{Obergaulinger_Just_Aloy:2018}. Specifically, the vector potential takes the form,
\begin{equation}
    \big( A^{r}, A^{\theta} , A^{\varphi} \big) = \frac{r_{0}^{3} r \cos{\theta}}{2(r^{3} + r_{0}^{3})} \big( 2B_{\mathrm{tor}}, 0, B_{\mathrm{pol}} \big),
\end{equation}
with radial scale $r_{0} = 10^{3} \, \mathrm{km}$.

Simulations are performed in 2D (i.e., axisymmetry) on a spherical grid with $550$ logarithmically spaced radial zones extending from the core to $10^{10} \, \mathrm{cm}$, and $128$ zones in polar angle, resulting in an angular resolution of $1.4^{\circ}$. During the initial collapse phase, the innermost $7.5 \, \mathrm{km}$ are evolved in spherical symmetry to lessen time-stepping constraints and speed up collapse. When central densities reach $10^{13} \, \mathrm{g \ cm^{-3}}$, indicating imminent core bounce, this spherical region is reduced in size to $1.6 \, \mathrm{km}$ to avoid smoothing out asymmetries. All simulations are evolved for at least $1.5 \, \mathrm{s}$ post-bounce.

Of the $17$ simulations we perform, the primary set of $14$ are identical aside from the initial rotation profile. Following \citet{Cusinato_et_al:2025}, we uniformly boost the angular velocity on the entire grid of each model by a constant factor. A summary of these models and their associated rotation boosts is provided in Table \ref{tab:run_summary}. Three models are directly comparable to those of \citet{Cusinato_et_al:2025}, and we adopt their naming convention in those cases (SR, IR, and FR prefixes); otherwise model names are simply prefixed with ``BF'' to indicate the corresponding boost factor.

\newcolumntype{Y}{>{\centering\arraybackslash}X}
\begin{table*}
    \centering
    \begin{tabularx}{0.8\linewidth}{c|l|YYYYYY}
        \hline
        \hline
        Set & Model & $\Omega_{\mathrm{c}} \, [\mathrm{rad \,s^{-1}}]$ & Boost factor & $\mathrm{t_{bounce} \, [ms]}$ & MHD & $f_{1\mathrm{s}} \, [\mathrm{kHz}]$ & $A_{+,\mathrm{max}} \, [\mathrm{cm}]$\\
        \hline
        \multirow{14}{*}{Main} & SR1 &  0.290 & 1 & 142 & Y & 2.9 & 66\\
         & BF1.5 &  0.435 & 1.5 & 142 & Y & 2.8 & 67\\
         & BF2 &  0.581 & 2 & 142 & Y & 2.6 & 58\\
         & BF2.5 &  0.726 & 2.5 & 142 & Y & 2.5 & 80\\
         & BF3 &  0.871 & 3 & 142 & Y & 2.5 & 92\\
         & BF3.3 &  0.958 & 3.3 & 143 & Y & 2.5 & 92\\
         & IR3.5 &  1.02 & 3.5 & 143 & Y & 2.4 & 104\\
         & BF3.7 &  1.08 & 3.7 & 143 & Y & 2.3 & 116\\
         & BF4 &  1.16 & 4 & 143 & Y & 2.5 & 135\\
         & BF4.6 &  1.34 & 4.6 & 144 & Y & 2.4 & 176\\
         & BF5.6 &  1.63 & 5.6 & 145 & Y & 2.4 & 256\\
         & BF7 &  2.03 & 7 & 147 & Y & 2.3 & 394\\
         & BF9.1 &  2.64 & 9.1 & 150 & Y & 2.0 & 636\\
         & FR12 &  3.48 & 12 & 159 & Y & 2.2 & 960\\
        \hline
         Full output & BF3.55 & 1.03 & 3.55 & 143 & Y & 2.4 & 116\\
        \hline
        \multirow{2}{*}{Non-magnetic} & $\mathrm{BF0\_noB}$ & 0.0 & 0 & 142 & N & 2.2/2.9/3.7 & 112\footnote{Model $\mathrm{BF0\_noB}$ is the only case where $A_{+,\mathrm{max}}$ does not occur at bounce. The maximum amplitude can be read from Figure \ref{fig:aplus_noB}.} \\
         & $\mathrm{IR3.5\_noB}$ & 1.02 & 3.5 & 143 & N & 1.6/2.2 & 100\\
        \hline
         
    \end{tabularx}
    \caption{Summary of simulations presented in this work. Simulations in the main set yield our primary results, but are supplemented by the 'Full output' model, which has very short output intervals, as well as a set of two non-MHD models. Unless explicitly stated, simulation configurations are otherwise identical. The table lists, in order, the central pre-collapse angular velocity, $\Omega_\mathrm{c}$, the multiplier for the rotation rate (i.e. boost factor), the time of core bounce relative to the simulation start time, whether MHD or non-magnetic hydrodynamics is used, the peak GW frequency at $t_\mathrm{pb} = 1.0 \, \mathrm{s}$, and finally the maximum GW amplitude attained. The peak frequencies are intended as a rough guide to the trend in GW frequency and should be considered in tandem with, e.g., Figure \ref{fig:dwt_grid}. In non-magnetic models, multiple prominent GW emission bands appear and these are listed with the $f$-mode first, followed by the relevant $p$-modes.}
    \label{tab:run_summary}
\end{table*}

\begin{figure}
    \centering
    \includegraphics[width=\linewidth]{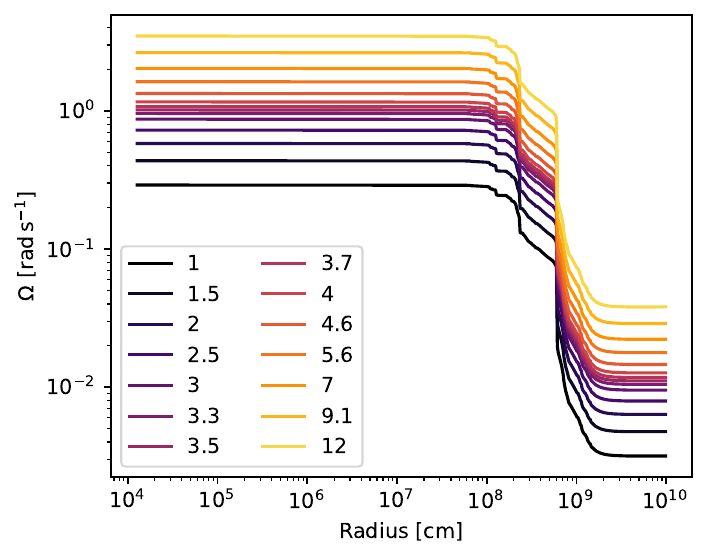}
    \caption{Initial angular velocities of models in the main set, as listed in Table \ref{tab:run_summary}. Models are labeled in the legend by their corresponding angular velocity boost factor over the original progenitor model.}
    \label{fig:init_omega}
\end{figure}

Angular velocity boosts are centered on a factor of $3.5$, with each step above or below incurring an additional $5\%$ increase or decrease (i.e. two steps above $3.5$ is $3.5 \times 1.05 \times 1.1 =4.04 \approx 4$). By sampling more densely near the rotation rate where a resonance effect was reported by \citet{Cusinato_et_al:2025}, we hope to give our simulations the best chance to reproduce the phenomenon. Initial angular velocities are plotted in Figure \ref{fig:init_omega}. It is important to note that these are not hydrostatic rotation profiles, as no change is made to, e.g., the pressure gradient to offset the increase in centrifugal force. In the case of the most rapidly rotating model, this makes the star marginally rotationally supported near the equator at the shell interface at $r=6\times10^{8} \, \mathrm{cm}$, and at the outer boundary; on the timescale of these simulations, we would not expect to see signs of artificial stellar breakup. This does, however, emphasize a shortcoming of uniformly boosting the angular velocity by a constant factor, as the progenitor physics is less self-consistent. 

With the main set of $14$ models, we consider $3$ addition simulations for the aforementioned total of $17$. These supplementary simulations include one model (BF3.55) very similar to BF3.5, but which is configured to save hydrodynamic outputs with a frequency of $25 \, \mathrm{kHz}$ to enable tracking the dynamics which produce kHz-range GW signals. A final two models are run without magnetic fields to quantify their impact on GW emission; one model with rotation, and one model without.

\section{Explosion dynamics}
\label{sec:exp_dyn}
The usual diagnostics of CCSNe, e.g. explosion energies, remnant kicks and spins, etc., are likely to be strongly impacted by the artificial rotation of the progenitor (in all but the SR1 model). Thus, these measures are less physically interesting and will be omitted from our discussion.

\begin{figure*}
    \centering
    \includegraphics[width=1.0\textwidth]{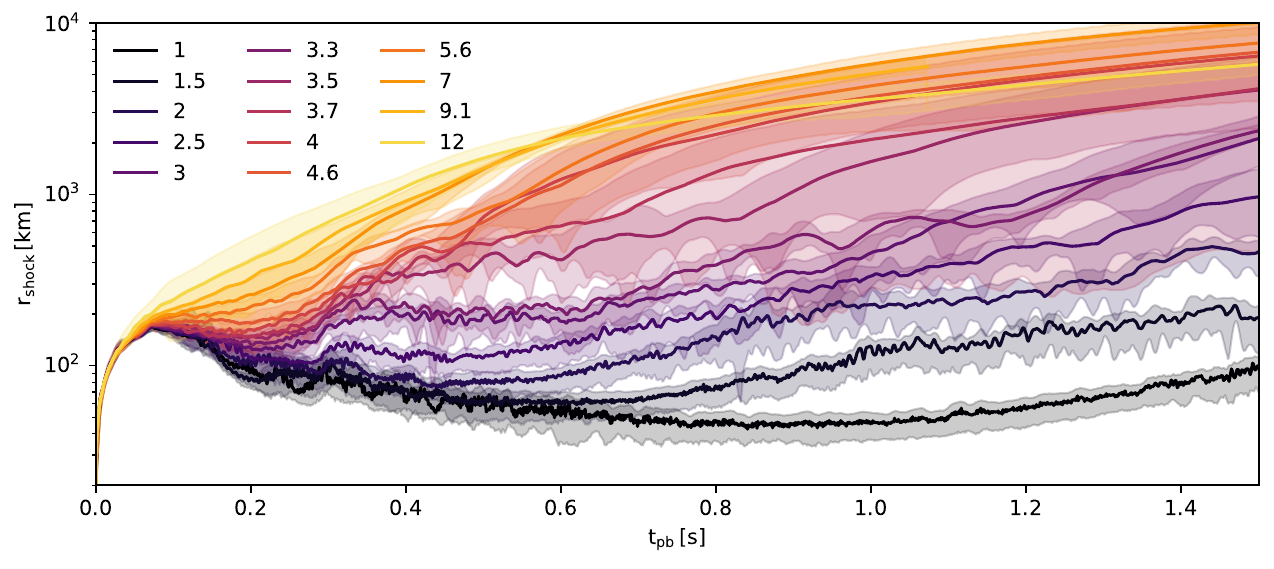}
    \caption{Shock radius as function of post-bounce time for simulations in the main set. The mean radius is shown as a solid line, and the minimum/maximum range as the shaded area. The minimum and maximum are smoothed within a $10 \, \mathrm{ms}$ window to improve clarity. Lines are labeled by their boost factors in the legend, excluding any prefixes.}
    \label{fig:shock_radius}
\end{figure*}

The trajectory of the shock remains relevant however, as the size of the shock bubble impacts the frequencies of, primarily, acoustic $p$-modes \citep{Torres-Forne_et_al:2018}; meanwhile, neutrino-driven convection in the gain region influences the forcing of these modes. For rapidly rotating models, the shock is deformed towards a toroidal configuration. Due to this asymmetry, the mean radius in Figure \ref{fig:shock_radius} is supplemented with the min/max range.

All models produce neutrino-driven explosions, with varying times of shock revival. There is minimal impact of magnetic fields on the dynamics of the shock, as will be discussed further in Section \ref{sec:magnetic}. Additionally, the ratio of kinetic to magnetic energies in the gain region remains significantly below equipartition \citep[c.p.][]{Varma_Mueller_Schneider:2023} before the shock is revived in all simulations, hinting that standard neutrino-driven convection is the dominant active process in the gain region. Typically, the magnetic energy is at least two orders of magnitude smaller than the turbulent kinetic energy when the shock starts expanding. Only at late times (i.e., $t_\mathrm{pb} \gtrsim 1 \, \mathrm{s}$) and in the most rapidly rotating models, do the energies become comparable, however by this point the neutrino mechanism has already precipitated an explosion. 

In all models of the main set, the shock expands rapidly past $100 \, \mathrm{km}$, with shock radii then diverging based on the rotation rate of the progenitor. Slower rotation results in the shock receding below $100 \, \mathrm{km}$ within ${\sim} 200 \, \mathrm{ms}$ post-bounce. The slower the rotation, the longer the shock spends in this contracted state, with the SR1 model taking almost a full second until the shock starts to expand again. The shock of the intermediate rotation model IR3.5, stays contracted for around $100 \, \mathrm{ms}$ before expanding. Models with fast rotation may not experience any contraction of the shock; model FR12 is the extreme example, with an unambiguously positive shock velocity at all times. Models BF4.6 and BF5.6 are more borderline, with the shock almost perfectly stalled (without contraction or expansion) for over $100 \, \mathrm{ms}$ before accretion of angular momentum drives equatorial expansion of a spindle-torus-shaped shock. Notably, the shocks of these rapidly rotating models reach many thousands of kilometers within the $1.5 \, \mathrm{s}$ of post-bounce simulation time. Such an expanded shock hints at very low $p$-mode frequencies. Rapid expansion of the shock also precludes SASI-driven GW emission \citep{Powell_Mueller:2024}.

With more rapid rotation, the mass accretion rate onto the shock decreases, thus reducing the ram pressure and promoting earlier shock revival. This may partially explain the trend in our models of earlier shock expansion with larger boost factors. However, we note that the reduction in accretion rate is significant only in the most rapidly rotating models: i.e., up to ${\sim}10\%$ in model FR12, and more typically less than $1\%$. On the other hand, the balance of forces on the shock is delicate, and a small adjustment from rotation may be enough to explain the trend. For moderate rotation, trends of the conditions for shock revival with rotation are less straightforward, and the monotonic trend seen in our set of models may not hold generally.
For example, the effect of rotation on SASI, and its subsequent impact on shock revival, may also play a role. As shown by, \citet[e.g.,][]{Yamasaki_Foglizzo:2008} and \citet{Blondin_Gipson_Harris_Mezzacappa:2017} rotation boosts the growth rate of SASI spiral modes, although of course these are absent in axisymmetric models such as ours. The impact of rotation on axisymmetric sloshing modes is weak \citep{Iwakami_et_al:2009}, and likely strongly subdominant to the reduction in ram pressure in producing earlier shock revival.
Other simulation sets tend to indicate a non-monotonic dependence of shock revival time on the rotation rate of the star \citep[e.g.][]{Marek_Janka:2009, Pajkos_et_al:2021, Pajkos_Boyeneni_EggenbergerAndersen:2026} (or even the opposite trend to our results, as in \citet{Fryer_Heger:2000}), suggesting the trend we observe may be confined to our limited sample.

\section{Gravitational waves}
\label{sec:gws}
Our discussion of the GWs from our simulations is organized into three parts, drawing variously from our simulation sets. We begin by presenting results from our main set of $14$ simulations with varying rotation boosts. This is aided by a comparison to the two non-MHD models. This part deals primarily with the long-term evolution of the GW signal, and the factors which influence this evolution. Subsequently, we analyze in detail the one remaining model, which aims to temporally resolve the PNS oscillations. The purpose of this part is to identify where GWs come from by directly examining the underlying fluid motions. Finally, we use one non-magnetic model, and one model from the main set, to assess the applicability of a linear eigenmode analysis to rotating stars.

\subsection{Evolution of GW signal}
\label{sec:gw_main_set}
In this section we present time series of the GW signals from our main set of simulations.
The GW signal is calculated from the axisymmetric quadrupole formula of \citet{Dimmelmeier_Font_Mueller:2002} and \citet{Mueller_Janka_Marek:2013} in the Newtonian limit,

\begin{align}
\label{eqn:quadrupole_formula}
    A^{\mathrm{E2}}_{20} = &\frac{32 \pi^{\frac{3}{2}} G }{\sqrt{15} c^{4}} \frac{\partial}{\partial t} \iint r^{3} \rho \big[ v_{r} (3 \cos^2 \theta - 1) \\ \nonumber
    &+ 3r^{-1} v_{\theta} \sin \theta \cos \theta \big] \sin \theta \ud \theta \ud r,
\end{align}
for a mass distribution of density $\rho$ and velocity $\mathbf{v}$. The contributions of neutrinos are small at frequencies of around $10 \, \mathrm{Hz}$ and above, and therefore neglected in the calculation for simplicity. As our simulations are axisymmetric, only the plus polarization of the GW is present, which we denote by \aplus. We assume an observer in the equatorial plane. 

\begin{figure*}
    \centering
    \includegraphics[width=1.0\textwidth]{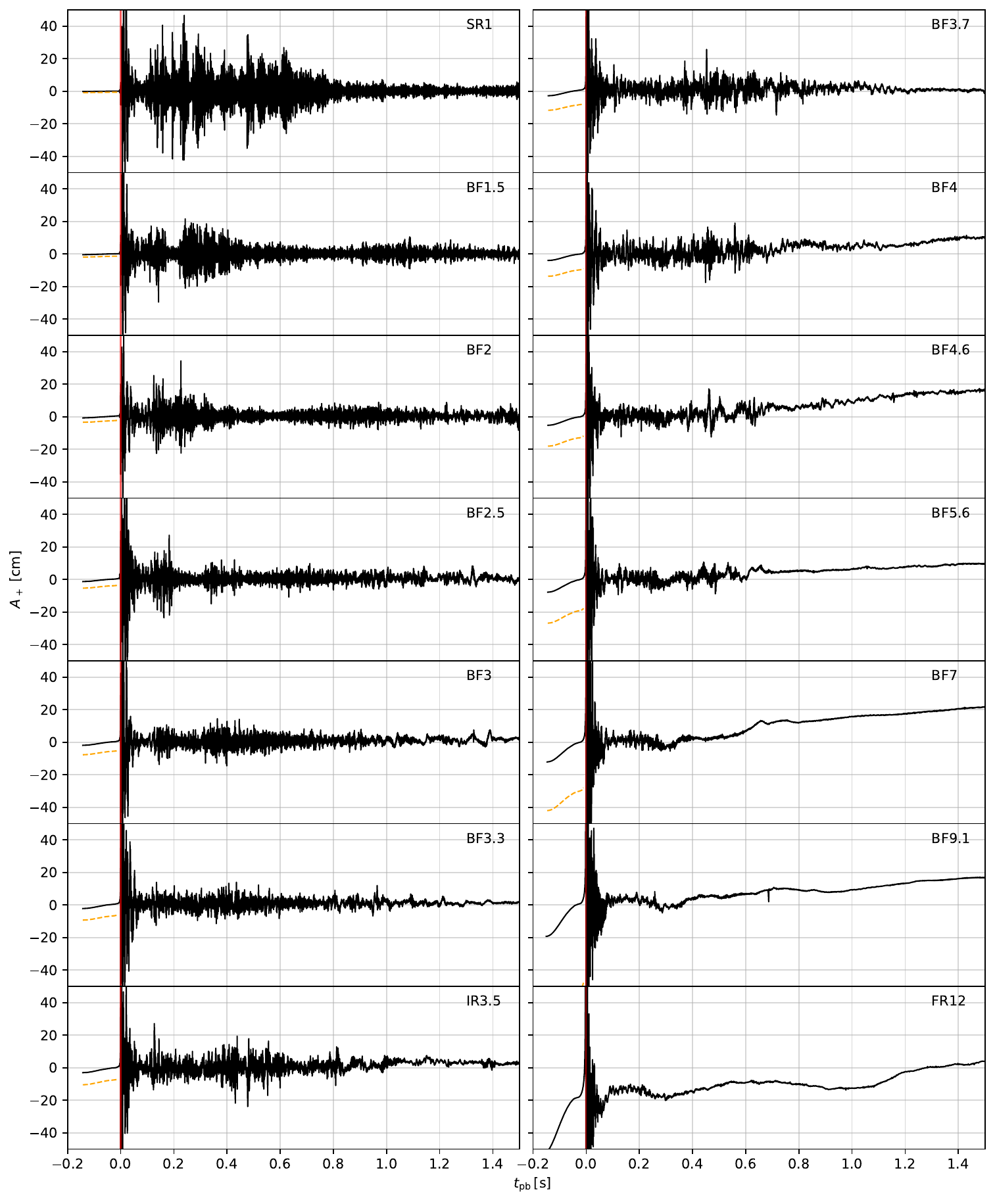}
    \caption{GW amplitudes, \aplus, for the 14 models of the main set. For models with large rotation boosts, a quadrupole moment is induced outside the core which produces low frequency ($f \ll 10 \, \mathrm{Hz}$) GWs. This causes \aplus to exhibit a secular offset from zero on the timescale of our simulations. Because we are interested in much higher frequency signals, it is neater to subtract this low frequency signal such that \aplus oscillates approximately around zero and each panel can have the same vertical scale. The low-frequency component is calculated in a post-processing step (since much lower time resolution is required) using the same time-integrated Newtonian quadrupole formula as the un-adjusted signal and restricting to the region $r > 10^{9} \, \mathrm{cm}$. We have confirmed by spectral analysis of both signals that this does not change the spectral properties of GWs in the region of interest. The  uncorrected pre-bounce amplitudes $A_+$ are provided as dashed orange lines for reference.}
    \label{fig:aplus}
\end{figure*}

Figure \ref{fig:aplus} shows the time series of the GW amplitudes, normalized to their respective times of bounce. Without boosting the rotation rate, the outer shells of the progenitor are, to good approximation, in hydrostatic equilibrium due to the self-consistent angular momentum transport of the \textsc{MESA} model. However, when the angular velocity of these shells is increased by a constant factor at the start of our simulations (except in model SR1), the centrifugal force is then insufficient to maintain a stable stellar structure and the star will bulge along its equator. This induces a drift in the quadrupole moment in these outer shells with a low frequency compared to typical PNS oscillation modes. On the timescale of these simulations, this low-frequency signal resembles a constant offset, most prominent at large rotation boosts and scaling in amplitude with $\Omega^2$. An immediate estimate of this offset may be obtained by isolating the (azimuthal) rotational terms in Equation (A14) of \citet{Mueller_Janka_Marek:2013}, i.e.,
\begin{equation}
\label{eqn:aetwo_azimuthal}
    A^{\mathrm{E2}}_{20} \simeq \frac{128 \pi^{\frac{3}{2}} G}{3\sqrt{15} c^{4}} \int r^{4} \rho \Omega^{2} \ud r,
\end{equation}
using the 1D progenitor data alone.

Because this effect is largely artificial -- a realistic star at the onset of core-collapse should not yet exhibit large non-equilibrium quadrupolar deformations -- and to ensure all panels of Figure \ref{fig:aplus} can share a consistent $y$-axis which is centered on zero, we post-process this low-frequency signal and subtract it from the time series. This is done by calculating the quadrupole amplitude only for the region with $r>10^{9} \, \mathrm{cm}$; however, if $r_{\mathrm{shock}} > 10^{9} \, \mathrm{cm}$, then the correction is extrapolated (as a constant) from previous times to avoid rapid changes to the quadrupole moment which can be produced by partial inclusion of the shock geometry (which is the case at late times in some rapidly rotating models). The threshold radius of $10^{9} \, \mathrm{cm}$ is justifiable by noting that the integrand in Equation \ref{eqn:aetwo_azimuthal} is heavily weighted towards large radii by the $r^{4}$ term, which more than compensates for the decay in $\rho$. In the case of model IR3.5, roughly $70\%$ of the initial secular signal comes from $r>10^{9} \, \mathrm{cm}$, and $99.99\%$ from above $10^{7} \, \mathrm{cm}$. Due to the scaling with $\Omega^{2}$, these fractions are even larger in more rapidly rotating models. While this method does not completely remove the secular contributions (e.g. ${\sim}30\%$ of the secular signal is between $10^{7}-10^{9} \, \mathrm{cm}$ in model IR3.5 and is not corrected for), it is sufficient for the purpose of simplifying Figure \ref{fig:aplus} to include only pertinent information.

All panels of Figure \ref{fig:aplus} show an initial period of quiescence during the collapse phase, during which some residue of the artificial quadrupole at $r<10^{9} \, \mathrm{cm}$ is most visible. At bounce, all models show similar characteristics, with a high amplitude burst lasting several milliseconds which exceeds $50 \, \mathrm{cm}$ in all models, and can be as large as $500 \, \mathrm{cm}$ for rapid rotators; in most cases, $A_{+}$ is largest at bounce, with the maximum (absolute) value listed in Table \ref{tab:run_summary}. This is followed by a short-lived ($\lesssim 100 \, \mathrm{ms}$) ringdown of the bounce, which decays exponentially with time. Higher amplitude oscillations associated with typical pressure and gravity modes ($p$- and $g$-modes) only emerge $50-100 \, \mathrm{ms}$ post-bounce, and also decay in amplitude but on longer timescales of $0.1-1 \, \mathrm{s}$. Furthermore, the amplitude of these modes decreases as the rotation rate of the model increases, with oscillations strongest in the SR1 model and weakest in FR12.

This trend of decreasing mode amplitude with increased rotation has been reported previously by \citet{Jardine_Powell_Mueller:2022}. \citet{Andresen_et_al:2019} also make note of the phenomenon, and discuss two mechanism by which high-frequency $g$-mode emission can be suppressed: a positive angular momentum gradient in the PNS convective region, and suppression of SASI-driven downflows. For our rapidly rotating models, both likely play a role. All models are stabilized against rotational instabilities in the PNS by $\partial \Omega / \partial r > 0$, which naturally fulfills Rayleigh's centrifugal stability criterion. Furthermore, as discussed in Section \ref{sec:exp_dyn}, the shock quickly expands in rapidly rotating models, inhibiting the growth of the SASI and effectively freezing-in any nascent convective plumes as the post-shock region expands. In essence, expansion of the shock cuts off the gain region as an energy reservoir for driving PNS oscillations. These stabilizing effects come with the natural caveat that they are most effective at the equator and weakest in the polar region. As will be discussed later in this section, the disparate dynamics of the equatorial and polar regions are critical to understanding the GW emission of rotational CCSNe.

\begin{figure*}
    \centering
    \includegraphics[width=0.9\textwidth]{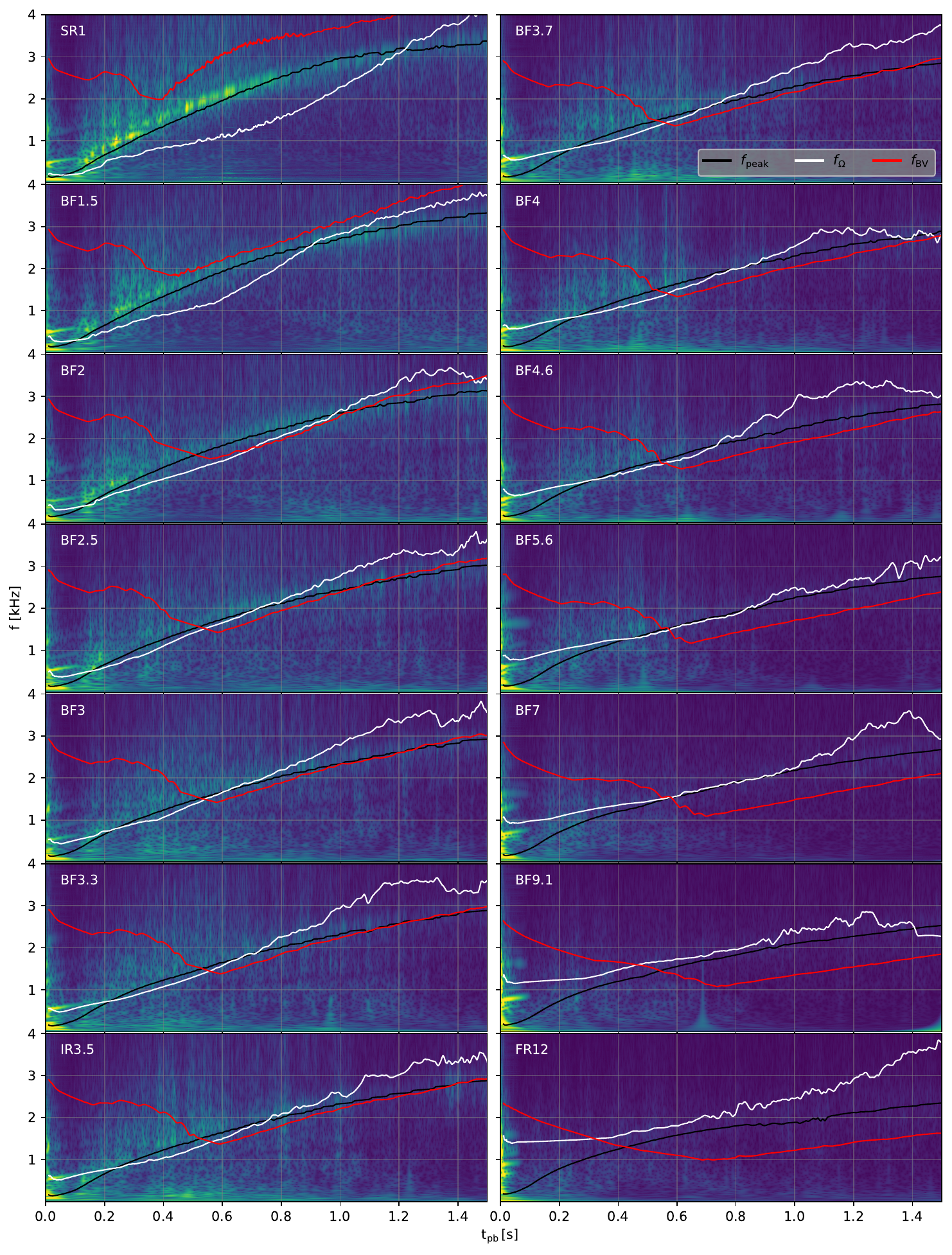}
    \caption{Normalized wavelet spectra of \aetwo for $1.5 \, \mathrm{s}$ post-bounce. The color scale is consistent between panels. The black line shows the predicted peak GW signal as computed from Equation \eqref{eqn:fpeak}. The white and red lines show, respectively, the maximum epicyclic and maximum Brunt-V\"ais\"al\"a frequencies inside a radius of $100 \, \mathrm{km}$ which are discussed further in Section \ref{sec:sh_crit}}. 
    \label{fig:dwt_grid}
\end{figure*}

Using a continuous wavelet transform of the amplitude time series \citep{Morlet_Arens_Forgeau_Giard:1982, Mueller_Janka_Marek:2013}, we visualize the frequency evolution of the GW signal in Figure \ref{fig:dwt_grid}. Looking first at the panel for the SR1 model, there are a few notable features. In the few milliseconds after bounce, there is evidence of high-amplitude, broad-spectrum emission consistent with the spike in the time series in Figure \ref{fig:aplus}. A slightly longer-lived bounce signal is found at around $500 \, \mathrm{Hz}$ which remains visible above the background for around $100 \, \mathrm{ms}$ and which slightly increases in frequency with time. This is consistent with other GW signal predictions of an extended ringdown phase following core bounce in rotating stars \citep{Dimmelmeier_Ott_Janka_Marek_Mueller:2007}, which appears to be triggered not so much by the initial bounce, but rather by the short period of prompt convection after the neutrino breakout burst. Although it is difficult to formally identify a mode corresponding to this feature (see Section \ref{sec:linear_mode}) its prompt emergence and frequency evolution suggest it is tied to $g$-mode oscillations in the PNS core \citep{Fuller_Klion_Abdikamalov_Ott:2015}; this is verified in Section \ref{sec:gw_spatial}. 

There is some emission of low-frequency GWs on the order of $100 \, \mathrm{Hz}$, however the majority of this signal is poorly resolved on this frequency scale. Signals in this frequency range have been associated with SASI activity \citep{Powell_Mueller:2024, Powell_Mueller:2025, Kuroda_Kotake_Takiwaki:2016} which is prominent in several of our models with slower rotation. The frequency of SASI motions in these models is $<100 \, \mathrm{Hz}$, which, to zeroth order, yield a GW frequency in the expected range once frequency doubling is taking into account.

The most significant feature of model SR1 is the dominant emission band which emerges after the ringdown of the PNS. The signal becomes prominent at a few hundred Hz and rapidly increases in frequency. It reaches $1 \, \mathrm{kHz}$ about $250 \, \mathrm{ms}$ after bounce and exceeds $3 \, \mathrm{kHz}$ by the end of the simulation. The amplitude of this emission band is strongest in the first ${\sim} 150 \, \mathrm{ms}$ after the ringdown of the PNS, but also varies in strength with time, with another bright patch around $500-600 \, \mathrm{ms}$ post-bounce. These periods of stronger GW emission are still significantly weaker than those noted by \citet{Cusinato_et_al:2025}. Unlike other studies, such as \citet{Jakobus_et_al:2023} who find a relatively strong $^{2}g_{1}$ mode with decreasing frequency, we see no clear evidence of other modes contributing significantly to the GW signal for simulations in main set. This is not true for all simulations considered in this work, as a core $g$-mode is visible in Figure \ref{fig:dwt_3d} for a 3D model, and additional modes (e.g. $p$-modes) are evident in the non-magnetic models of Section \ref{sec:magnetic}.

Moving now to the remaining models in the main set, a clear trend emerges with the dominant mode decreasing in amplitude as the rotation rate of the progenitor increases. At intermediate rotation (e.g. model IR3.5) the emission band is still visible, however it becomes effectively indistinguishable from the broadband stochastic emission in the fast rotating model FR12. Consistent with the finding of previous studies \citep{Dimmelmeier_Stergiouslas_Font:2006, Powell_Mueller:2020, Pajkos_Warren_Couch_OConnor_Pan:2021}, peak frequencies also decrease as the angular velocity of the star increases; this is also evident by examining $f_\mathrm{1s}$ -- the peak GW frequency $1 \, \mathrm{s}$ after bounce -- in Table \ref{tab:run_summary}. The bounce signal of all models show broad emission up to a few kHz, often with a strong low-frequency component. The prominent ringdown mode, active in the first $100 \, \mathrm{ms}$ post-bounce, also remains near $500 \, \mathrm{Hz}$ for most models, although increases slightly for very fast rotation.

\subsubsection{Understanding the dominant mode}
\label{sec:dom_mode}

\begin{figure}
    \centering
    \includegraphics[width=1.0\linewidth]{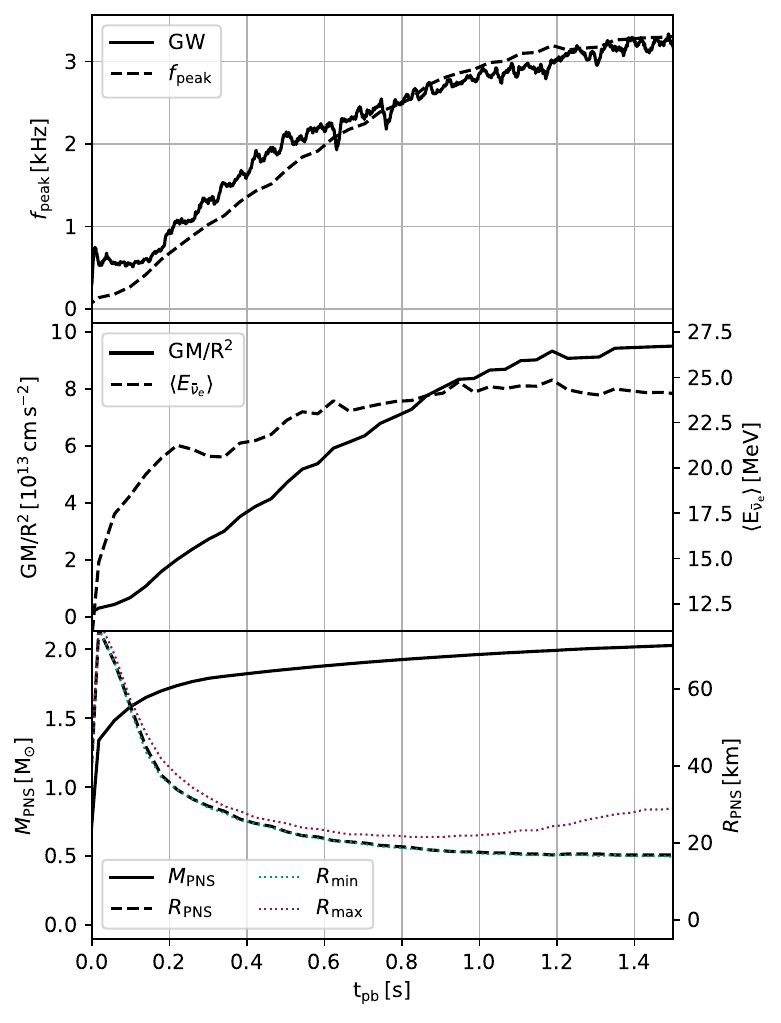}
    \caption{\textbf{Top:} Approximate frequency of the dominant GW emission band for the SR1 model (solid line) obtained by finding the maximum frequency in $12 \, \mathrm{ms}$ time windows and performing a moving average on the result to smooth the data. In essence this is a kind of line of best fit to the spectrograms in Figure \ref{fig:dwt_grid}. The analytic approximation, $f_\mathrm{peak}$, estimated from electron antineutrino mean energies is shown by the dashed line. \textbf{Middle:} Components of Equation \eqref{eqn:fpeak} as a function of time. PNS compactness is plotted with a solid line using the left axis while electron antineutrino mean energies are plotted with a dashed line and correspond to the right-hand axis. \textbf{Bottom:} PNS baryonic mass (solid line; left axis) and radius (dashed line; right axis), including the minimum and maximum radius satisfying $\rho > 10^{11} \, \mathrm{g \, cm^{-3}}$ (colored dotted lines).}
    \label{fig:sr1_fpeak}
\end{figure}

The peak frequencies of the GW emission in all our models, even if weak for rapid rotators, are high compared to the majority of existing simulations \citep[e.g.][]{Powell_Mueller:2024, Andresen_et_al:2019, Bugli_et_al:2023, Cusinato_et_al:2025, Cusinato_Obergaulinger_Aloy:2025} where the maximum frequency of the dominant mode is around $1 \, \mathrm{kHz}$, potentially up to $2 \, \mathrm{kHz}$ in certain cases. As in \citet{Powell_Mueller_AguileraDena_Langer:2023}, our use of a modified Newtonian gravitational potential will tend to shift peak GW frequencies upwards by ${\sim} 20\%$ compared to full general relativistic simulations \citep{Mueller_Janka_Marek:2013, Powell_Mueller:2020}, however this effect is insufficient to fully explain our high-frequency emission. Consequently, to explain the dominant mode frequency exceeding $3 \, \mathrm{kHz}$, we consider the analytic estimate of the dominant GW frequency, $f_\mathrm{peak}$, of \citet{Mueller_Janka_Marek:2013}:
\begin{equation}
    \label{eqn:fpeak}
    f_\mathrm{peak} \approx \frac{1}{2 \pi} \frac{GM}{R^{2}} \sqrt{\frac{1.1 m_{n}}{\langle E_{\bar{\nu}_\mathrm{e}} \rangle}},
\end{equation}
where $M$ and $R$ are the PNS mass and radius respectively, $m_n$ is the neutron mass, and $\langle E_{\bar{\nu}_\mathrm{e}} \rangle$ is the mean electron antineutrino energy in the PNS surface region (here defined as the surface for which $\rho = 10^{11} \, \mathrm{g \, cm^{-3}}$, although with a few caveats which will be discussed in due course). Equation \eqref{eqn:fpeak} predicts the dominant GW frequency of the SR1 model with remarkable accuracy -- within $5-10\%$ in the later half of the simulation, as shown in Figure \ref{fig:sr1_fpeak}. However, this varies depending on how the PNS radius is defined due to rotation-induced oblateness -- in this analysis we find that using the PNS radius near the \emph{polar axis} produces the best fit of $f_\mathrm{peak}$ to the simulated GW data. We stress that this is not a result that should necessarily be expected a priori, as Equation~(\ref{eqn:fpeak}) is formally valid only for a non-rotating, spherical PNS. However, the model may evidently be robust more generally; we motivate this observation in more detail in Section \ref{sec:sh_crit}.

The reasonable goodness-of-fit of the $f_\mathrm{peak}$ approximation allows us to attribute the high frequencies of the dominant GW emission in our models with physical properties of the PNS, specifically the PNS compactness term ($M/R^{2}$) computed \emph{at the poles}, and the electron antineutrino mean energies. Both are plotted in the middle panel of Figure \ref{fig:sr1_fpeak}. Comparing to \citet{Mueller_Janka_Marek:2013}, we find higher mean energies -- in the range of $20-25 \, \mathrm{MeV}$ compared to their ${\sim} 15 \, \mathrm{MeV}$; interestingly, on its own this would serve to reduce the frequency of the dominant GW emission, however it is overcompensated by the $M/R^{2}$ term being $2-4$ times that of \citet{Mueller_Janka_Marek:2013}, depending on the specific model.

\begin{figure}
    \centering
    \includegraphics[width=\linewidth]{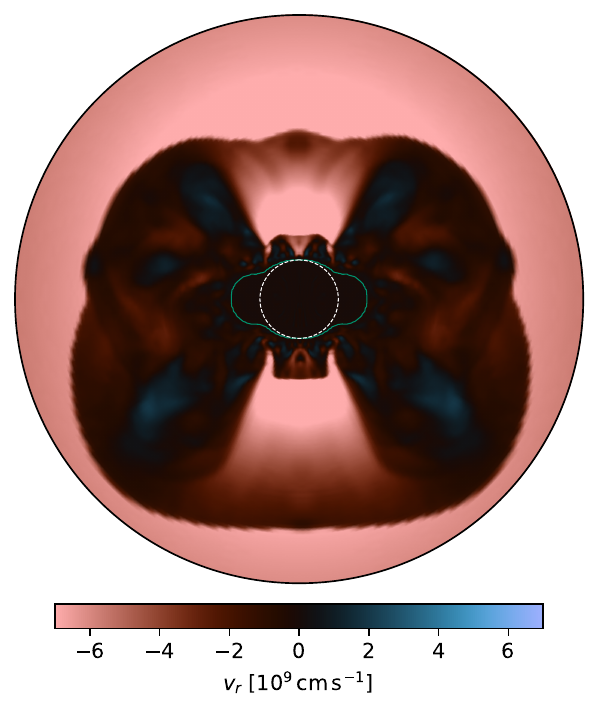}
    \caption{Slice of radial velocity of the SR1 model at $t_\mathrm{pb}=1.5 \, \mathrm{s}$, mirrored over the vertical polar axis. The radius of the outer boundary is $120 \, \mathrm{km}$. The nominal PNS boundary at $\rho = 10^{11} \, \mathrm{g \, cm^{-3}}$ is shown by a solid green line; for $t_\mathrm{pb} \gtrsim 1.0 \, \mathrm{s}$, this surface develops a prominent equatorial bulge. A dashed white circle with a radius equal to the minimum PNS radius is included for reference.}
    \label{fig:vr_slice_sr1}
\end{figure}

This high compactness of the PNS is the likely cause of the dominant GW frequency being so high. But why is the PNS so compact? The bottom panel of Figure \ref{fig:sr1_fpeak} shows the PNS baryonic mass and radius as a function of time. The PNS rapidly becomes quite massive and reaches $2.0 \, \msun$ by the end of the simulation. Because the core mass of this progenitor is quite large, the PNS also being quite massive is to be expected. This massive PNS is matched with a relatively small radius \citep[c.f.][]{Powell_Mueller:2024}, which drops below $20 \, \mathrm{km}$ within one second after bounce. This rapid contraction is achieved by efficient neutrino cooling (recalling that neutrino energies are also higher in our simulations), but is also influenced by the oblateness of the PNS.
The equilibrium configurations of rapidly rotating neutron stars have larger radii near the
equator, while the radius tends to shrink at the pole. Consequently, the maximum PNS radius at the equator is larger by a factor $1.7$ than the minimum radius at the pole by the end of the simulation.
The oblateness of the PNS is highlighted by the green line in Figure \ref{fig:vr_slice_sr1}. 

The reason why the dominant mode frequency is determined by the polar radius is found in the fast polar downflows seen in Figure \ref{fig:vr_slice_sr1}, which shows a slice of radial velocity up to a radius which encompasses the entire shock. If these downflows drive the oscillation mode responsible for the dominant GW emission of the model \citep{Andresen_Mueller_Mueller_Janka:2017,Vartanyan_et_al:2023}, then it is reasonable that Equation \eqref{eqn:fpeak} is most accurate using the PNS properties along the axis\footnote{Recall that Equation \eqref{eqn:fpeak} is essentially an approximation for the Brunt-V\"ais\"al\"a frequency, which is a locally defined quantity of the fluid, in terms of global PNS properties. In asymmetry, global PNS properties such as radius are ill-defined, so choosing to prioritize certain directions (in this case the polar region) by referring to the underlying physics is a robust approach for applying the approximation to non-spherical PNSs.}.

Our use of \textsc{CoCoNuT-FMT} may also explain some discrepancy when comparing to simulations performed with other CCSNe codes. As shown by \citet{Varma_Mueller_Obergaulinger:2021}, different simulation codes -- specifically \textsc{CoCoNuT} and \textsc{Aenus-ALCAR} -- are not necessarily in agreement with respect to the structure of rapidly rotating PNSs, particularly the radius and distribution of angular momentum to which the GW emission is sensitive. Thus, the PNS in our simulations (with \textsc{CoCoNuT}) may also be more compact than in the simulations of \citet{Cusinato_et_al:2025} (with \textsc{Aenus-ALCAR}), despite both sets of simulations using the same progenitor and including similar physics (i.e. both use 2D MHD with relativistic corrections, both use the SFHo EoS, and both apply the same method for boosting pre-collapse rotation rates). This would contribute the differences in dominant GW frequency between this study and theirs. Previous simulations of a $39 \, \msun$ progenitor from the same ``family'' as our $17 \, \msun$ model -- i.e., also within the set of \citet{Aguilera-Dena_Langer_Moriya_Schootemeijer:2018} -- by \citet{Powell_Mueller_AguileraDena_Langer:2023} using the \textsc{CoCoNuT} code, also found quite high peak frequencies of up to ${\sim} 2 \, \mathrm{kHz}$ despite the shorter $0.6 \, \mathrm{s}$ duration of their model. In fact, GWs at around $2 \, \mathrm{kHz}$ is quite consistent with the GW signal $0.6 \, \mathrm{s}$ into our slow-to-moderate rotation simulations. This could indicate that the stellar modeling of these stars tends to result, after collapse, in a PNS core structure which is prone to high frequency oscillations. In light of the ${\sim} 20\%$ increase from the modified Newtonian potential, the fact that this and similar progenitors have previously been found to produce high-frequency GWs during collapse, and the systemic differences known to exist between simulation codes, we conclude that it is reasonable for the dominant GW emission of these models to be above the typical range.

\begin{figure}
    \centering
    \includegraphics[width=\linewidth]{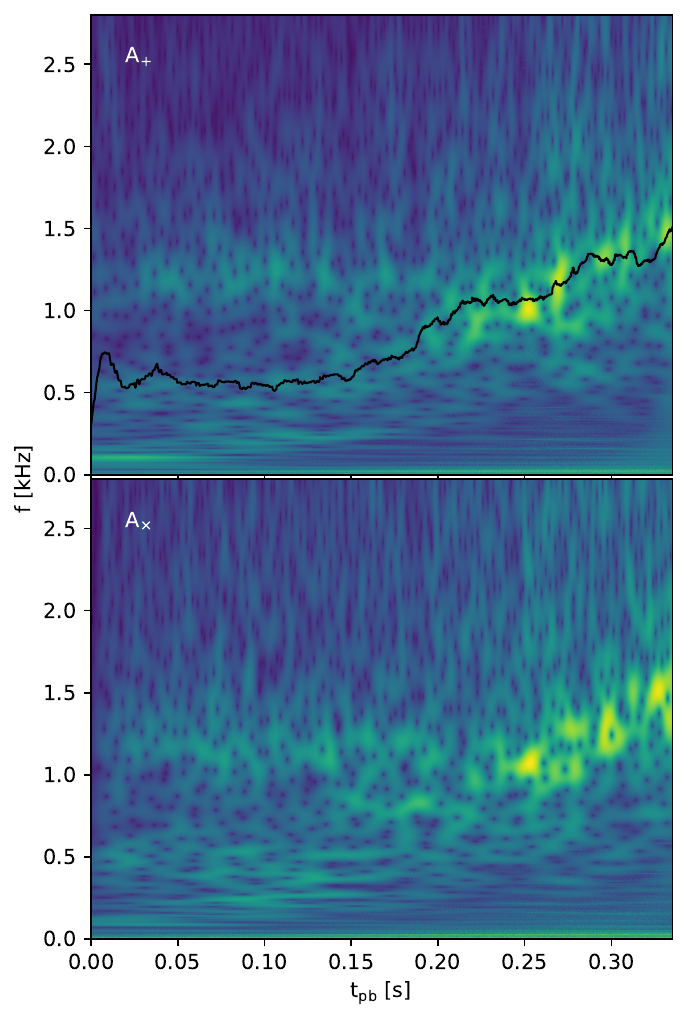}
    \caption{Normalized wavelet spectra of GWs from a simulation of model SR1 in 3D. Both plus and cross polarizations of the GWs are shown in the top and bottom panels respectively. For the plus polarization, the smoothed fit to the dominant emission band from Figure \ref{fig:sr1_fpeak} (solid black line in the top panel) is reproduced for comparison to the 2D results. The core $g$-mode is visible in both panels at around $1.2 \, \mathrm{kHz}$, starting from a few milliseconds after bounce.}
    \label{fig:dwt_3d}
\end{figure}

We further verify this result by comparing to a 3D simulation\footnote{This simulation is not part of our main set, but conveniently has a similar setup to our simulation of the SR1 model; its inclusion here is purely for verification and comparison with the 2D results. A more detailed analysis of the dynamics of this simulation will be the subject of a future study.} of model SR1. As shown by the spectrograms in Figure \ref{fig:dwt_3d}, we find a very similar frequency evolution of the dominant mode to the 2D simulation, despite the shorter time simulated in 3D. This is shown most clearly by the black line in the top panel, which shows the frequency of peak GW power in the 2D model as a function of time. It is clear that the line tracks the dominant emission band in the 3D simulation remarkably well. This confirms that the PNS structure, and particularly the transport of angular momentum within the PNS, is reliably captured in axisymmetry, and that the large GW frequencies of our main set are not an artifact the simulations being performed in 2D. 

It is also noteworthy that the 3D simulation includes a prominent, but sub-dominant, emission band at around $1.2-1.3 \, \mathrm{kHz}$, which decreases slightly in frequency with time. Owing to the prompt emergence of the mode, and its frequency evolution, we associate this emission band with a core $g$-mode \citep{Jakobus_et_al:2023}. We further note that this mode is either absent, or qualitatively different, in the spectrograms of 2D simulations, likely due to differences between convection in 2D and 3D \citep{Couch_OConnor:2014, Vartanyan_Burrows_Radice_Skinner_Dolence:2019} which impact the forcing motions of the modes.

In addition to the differences in convection, 3D simulations may also develop triaxial instabilities if the ratio of kinetic energy and gravitational energy, $T/|W|$, exceeds certain threshold values. These instabilities are suppressed in 2D, but could reasonably occur in real stars with very rapid rotation and may impact the GW signal produced \citep{Kuroda_Kawaguchi_Shibata:2025, Shibagaki_et_al:2026, Bugli_Guilet_Foglizzo_Obergaulinger:2023, Takiwaki_Kotake_Foglizzo:2021, Shibagaki_Kuroda_Kotake_Takiwaki:2020}. Thus, we compute this ratio for all models. We find that only the two most rapidly rotating models, BF9.1 and FR12, exceed the threshold for the onset of the low-$T/|W|$ instability, $T/|W| \approx 0.14$ \citep{Shapiro_Teukolsky:1983}\footnote{This threshold originates from a stability analysis of Maclaurin spheroids; different stellar structures may cause the precise value to change slightly (for instance, see \citet{Kuroda_Kawaguchi_Shibata:2025} for a numerical investigation in the context of white dwarfs), however this does not significantly impact our conclusion.}, and all models are significantly below the threshold for the dynamical instability. Additionally, for the two aforementioned models, the threshold for the low-$T/|W|$ instability is reached late in the simulation ($t \gtrsim 1 \, \mathrm{s}$), such that the earlier data is still reliable. While there may be other differences in 3D simulations of our models, such as non-axisymmetric modes of the magnetorotational instability \citep{Balbus_Hawley:1991, Akiyama_Wheeler_Meier_Lichtenstadt:2003, Obergaulinger_et_al:2009, Mosta_et_al:2015}, this check provides additional evidence that our simulations capture the most significant dynamics relevant to the GW signal.

Finally, while we have primarily focused on model SR1 in this discussion, Equation \eqref{eqn:fpeak} seems to hold reasonably well for all models SR1 through to FR12, despite the increasingly violated assumption of spherical symmetry. However, it is increasingly difficult to assess the validity of Equation \eqref{eqn:fpeak} for more rapidly rotating stars because the GW signal becomes less distinct from the background.

\subsubsection{Solberg-H{\o}iland criterion}
\label{sec:sh_crit}

In the previous section we showed that the dominant GW emission is well-modeled by an approximation of the Brunt-V\"ais\"al\"a frequency near the polar axis -- i.e. Equation \eqref{eqn:fpeak}.
While the local Brunt-V\"ais\"al\"a frequency at the pole is relevant to the deceleration of overshooting plumes near the axis, it is less intuitive why a local approximation for the Brunt-V\"ais\"al\"a frequency tracks the frequency of a global oscillation mode so well. One element of the explanation is the additional stabilization of the stratification by rotation, which increases the local characteristic frequency of fluid oscillations at lower latitudes, bringing them more in line with the higher Brunt-V\"ais\"al\"a frequency at the pole.
To further demonstrate that high GW frequencies naturally arise as a consequence of the hydrodynamic structure of the PNS, particularly where rotation modifies the structure, we investigate the two associated oscillation frequencies of the fluid -- namely the Brunt-V\"ais\"al\"a and epicyclic frequencies more closely. This also helps us to address the resonance condition of \citet{Cusinato_et_al:2025}, which depends on a crossing of the epicyclic frequency with the PNS $f$-mode.

In a rotating fluid, a displaced fluid bubble is subject to the Solberg-H{\o}iland stability criterion. In terms of the Brunt-V\"ais\"al\"a frequency, $f_\mathrm{BV}$, and epicyclic frequency, $f_\mathrm{\Omega}$, stable oscillations are obtained if,
\begin{equation}
\label{eqn:solhoi}
    f_\mathrm{SH}^{2} = f_\mathrm{BV}^{2} + f_{\Omega}^{2} \sin \theta > 0.
\end{equation}
Otherwise, the fluid is unstable and convection will develop. Following \citet{Maeder_Meynet_Lagarde_Charbonnel}, the $\sin \theta$ factor accounts for the restoring buoyancy and centrifugal forces not being coplanar, except at the equator. 
There is no trivial way to relate Equation~\eqref{eqn:solhoi} to the rotational shift in mode frequencies, but in the equatorial plane $f_\mathrm{SH}$ has a simple interpretation; it represents the local oscillation frequency of radially displaced blobs under the influence of stratification and rotation.

We take the Brunt-V\"ais\"al\"a frequency in its Newtonian form,
\begin{equation}
\label{eqn:bvfreq}
    f_\mathrm{BV}^{2} = \frac{1}{4 \pi^{2} \rho} \frac{\partial \Phi}{\partial r} \bigg(\frac{1}{c_{s}} \frac{\partial P}{\partial r} - \frac{\partial \rho}{\partial r} \bigg),
\end{equation}
where $\Phi$ is the gravitational potential, $\rho$ the fluid density, $P$ the pressure, and $c_{s}$ the local sound speed. The epicyclic frequency is here defined as,
\begin{equation}
    f_{\Omega}^{2} = \frac{\Omega}{2 \pi^{2}r} \frac{\partial (r^{2} \Omega)}{\partial r},
\end{equation}
with $\Omega$ being the spherically averaged angular velocity of the fluid. 

The components of $f_\mathrm{SH}$, the epicyclic and Brunt-V\"ais\"al\"a frequencies, are plotted in Figure \ref{fig:dwt_grid}. Specifically, we plot the maximum of these frequencies inside a radius of $100 \, \mathrm{km}$, containing the PNS and most of the gain region (until an explosion is well-underway and the shock quite advanced). For the epicyclic frequency, $f_\mathrm{\Omega}$, the maximum is typically an increasing function of time which, for most models, tracks the peak GW emission quite closely. The deviation is perhaps most obvious in the SR1 model, where $f_{\Omega}$ is ${\sim} 50\%$ lower than the GW signal $600 \, \mathrm{ms}$ post-bounce. For rapidly rotating models, the initial $f_\mathrm{\Omega}$ is quite large and difficult to associate with features of the spectrogram; however, this is primarily a result of low power in the spectrogram making a comparison more difficult. 

\begin{figure}
    \centering
    \includegraphics[width=\linewidth]{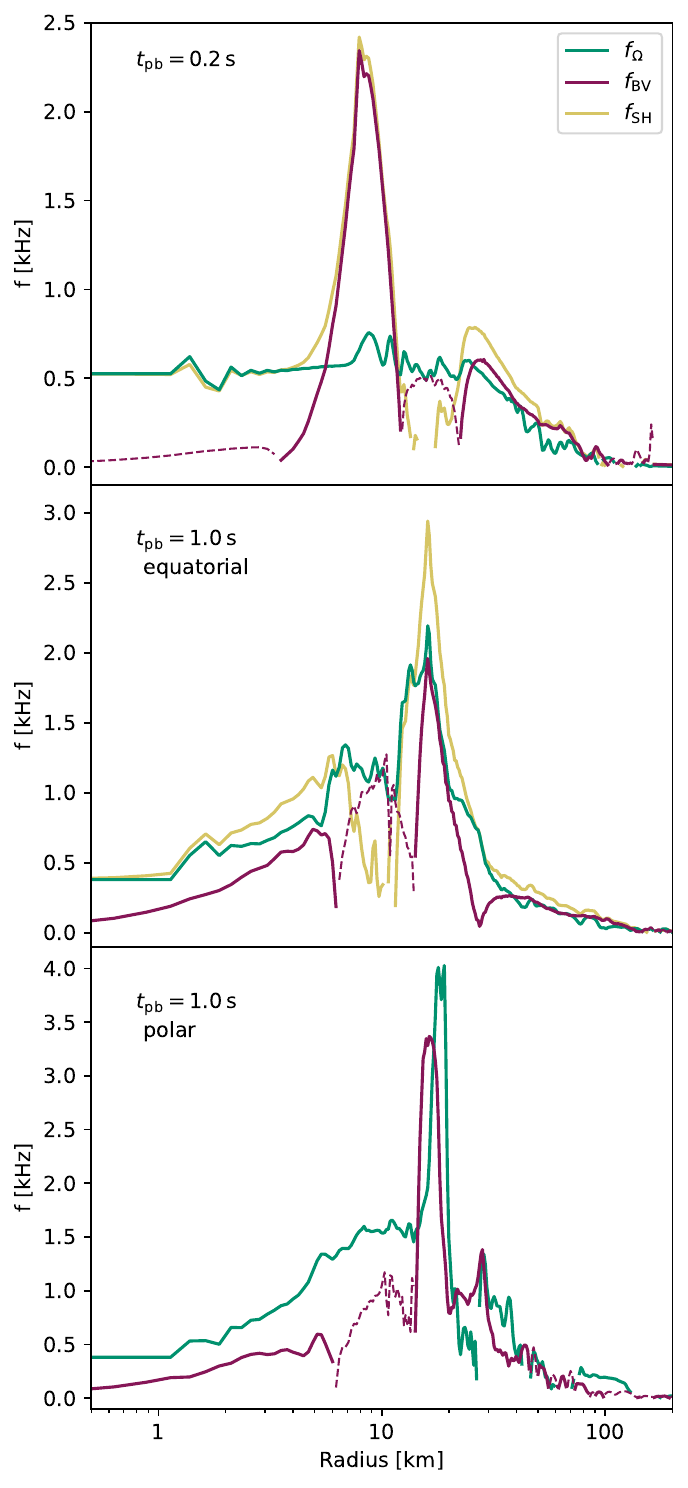}
    \caption{Exemplar radial profiles of the dynamical frequencies $f_\Omega$, $f_\mathrm{BV}$, and $f_\mathrm{SH}$ for model IR3.5. Dashed lines indicate imaginary Brunt-V\"ais\"al\"a frequencies, i.e. convectively unstable regions in the absence of rotational support. In the top panel, the PNS and surroundings are sufficiently close to spherical symmetry that a simple spherical average is a reliable measure of the frequencies. In the middle and bottom panels (both at $t_\mathrm{pb} = 1.0 \, \mathrm{s}$), the polar and equatorial regions produce quite different frequency profiles; thus the total volume is roughly split into a polar region ($0^{\circ} - 45^{\circ}$), and equatorial region ($45^{\circ} - 135^{\circ}$). The bottom panel does not include $f_\mathrm{SH}$ because it does not represent the local oscillation frequency of radially displaced blobs
    when $\sin \theta \not\approx 1$.}
    \label{fig:freqs_3.5}
\end{figure}

Figure \ref{fig:freqs_3.5} shows profiles of $f_\mathrm{SH}$, $f_\mathrm{BV}$, and $f_\Omega$ at $t_\mathrm{pb} = 0.2 \, \mathrm{s}$ and $t_\mathrm{pb} = 1.0 \, \mathrm{s}$, where the latter is decomposed into polar and equatorial regions. The evolution of the epicyclic frequency is largely driven by the growth of angular momentum gradients at the PNS surface. $f_\Omega$ is largest in the polar region, reaching around $4 \, \mathrm{kHz}$, while in the equatorial region it is just above $2 \, \mathrm{kHz}$. While this means that $f_\Omega$ in the equatorial region is, at first glance, more consistent with the GW emission ($f_\mathrm{1s} = 2.4 \, \mathrm{kHz}$ for model IR3.5, as per Table \ref{tab:run_summary}), the fluid is unlikely to oscillate at the peak frequency at a given point. Instead, the frequency of oscillations is influenced more broadly by the surrounding convectively stable region. We will expand on this point in due course.

Additionally, there is no guarantee that high frequency polar epicyclic modes are significantly excited due to the fact that accretion near the poles will tend to excite oscillations perpendicular to centrifugal forces.

As for the Brunt-V\"ais\"al\"a frequency, $f_\mathrm{BV}$ is consistently above the GW signal in Figure \ref{fig:dwt_grid} for the first few hundred milliseconds after bounce and tends to decrease with time.
The maximum of $f_\mathrm{BV}$ is generally expected to be higher than the mode frequency, which represents an effective average of $f_\mathrm{BV}$ over the stable region, as can be formally seen within the WKB approximation \citep[see, e.g.][]{Tassoul:1980}.
The initial decreasing trend in $f_\mathrm{BV}$ has another reason, however. It is simply due to our choice of detecting the maximum
of $f_\mathrm{BV}$  within the innermost $100\,\mathrm{km}$. The maximum Brunt-V\"ais\"al\"a frequency initially sits at the inner edge of the PNS convective zone (cp.\ top panel in Figure~\ref{fig:freqs_3.5}), and initially therefore tracks the trajectory of the core $g$-mode of \citet{Jakobus_et_al:2023} rather than the outer $f$- or $g$-mode represented by the dominant emission band; note that there is no feature from this core $g$-mode in the spectrogram at early times in 2D, though. The value of $f_\mathrm{BV}$ at the inner convective boundary in the PNS decreases with time \citep{Jakobus_Mueller_Heger:2025}. 
After several hundred milliseconds, the maximum moves to the outer edge of the convective region, and then tracks the increasing trend in the dominant mode frequency.

Because the mode frequency is sensitive to $f_\mathrm{BV}$ across the stable region and not just to its maximum value, the fact that the maximum of $f_\mathrm{BV}$ tracks the dominant frequency well is somewhat accidental. 
As shown in Figure~\ref{fig:freqs_3.5}, $f_\mathrm{BV}$ is not constant across the stable region at the PNS surface, in both the polar and equatorial directions. The maximum Brunt-V\"ais\"al\"a frequency along the polar direction is almost double that at the equator. This implies that, while an average over the stable region near the pole is of the order of the actual mode frequency, an average of $f_\mathrm{BV}$ in the equatorial plane would be somewhat lower than expected.

However, Figure~\ref{fig:freqs_3.5} shows that the lower Brunt-V\"ais\"al\"a frequency in the equatorial region is partially compensated for by a positive angular momentum gradient. 
Here, the epicyclic frequency $f_\mathrm{\Omega}$ reaches similar values to $f_\mathrm{BV}$, and hence $f_\mathrm{SH}$ -- as the relevant oscillation frequency for radially displaced blobs -- considerably exceeds $f_\mathrm{BV}$. 
For the late snapshot shown in the middle and bottom panels of Figure~\ref{fig:freqs_3.5}, the maximum of $f_\mathrm{SH}$ at the equator of about $2.9\,\mathrm{kHz}$ is in fact quite similar to the maximum of $f_\mathrm{BV}$ at the pole of
$3.4\,\mathrm{kHz}$.
Thus the local oscillation frequencies for radially displaced blobs do not vary excessively by latitude and are broadly consistent with the frequency of the dominant global oscillation mode. The small variation of the local radial oscillation frequency is likely coincidental, but possibly explains why the estimation of the mode frequency based on the PNS properties at the pole in Section~\ref{sec:dom_mode} works well. Since displaced fluid elements at the PNS surface are oscillating with a similar frequency, this frequency will also be similar to that of the corresponding lowest-order quadrupolar eigenmode.
This consistency across latitudes explains why the dominant mode effectively follows the Brunt-V\"ais\"al\"a frequency, and classical frequency relation~\eqref{eqn:fpeak}, at the pole. 

\begin{figure}
    \centering
    \includegraphics[width=\linewidth]{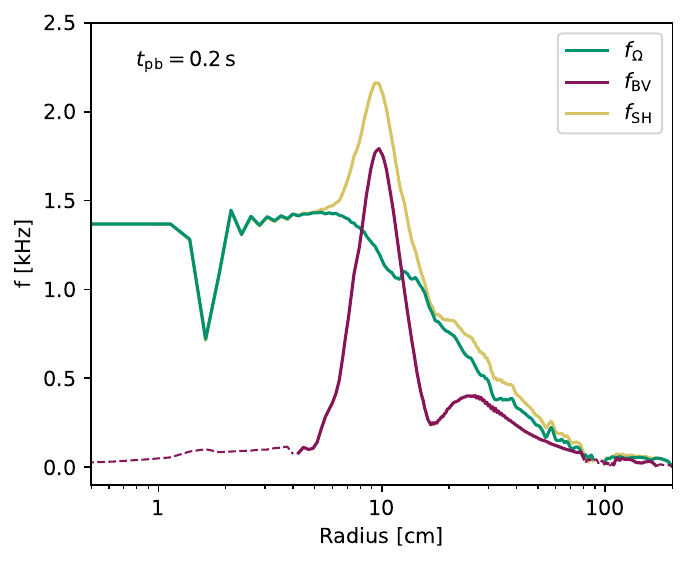}
    \caption{Same as the top panel of Figure \ref{fig:freqs_3.5}, but for model FR12.}
    \label{fig:freqs_12}
\end{figure}

Profiles for the more extreme rotating model FR12 at $t_\mathrm{pb} = 0.2 \, \mathrm{s}$ are shown in Figure \ref{fig:freqs_12}. These profiles contrast with those of model IR3.5 by the stronger contribution of $f_\Omega$ to $f_\mathrm{SH}$, especially inside the PNS. The net effect of this is to decisively stabilize the PNS surface region at most latitudes, thus reducing the prevalence of convective turbulence which can excite GW modes. This may partially explain why the dominant emission band of model FR12 in Figure \ref{fig:dwt_grid} is so weak.

\begin{figure}
    \centering
    \includegraphics[width=\linewidth]{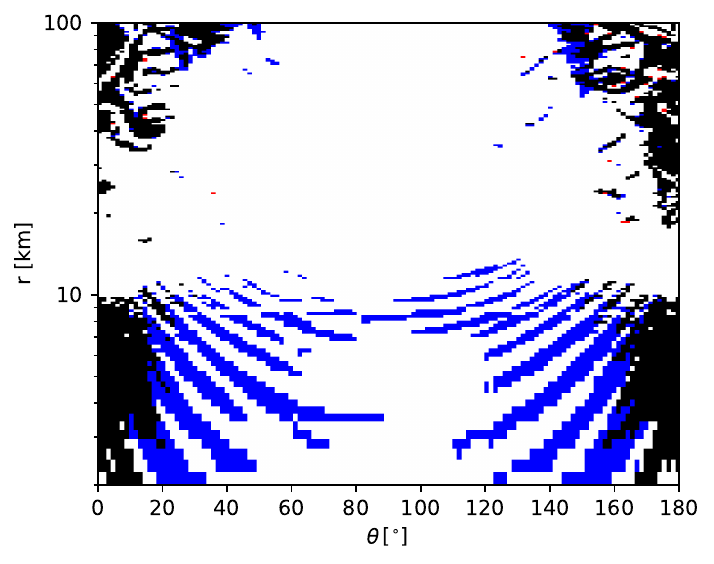}
    \caption{Slice of model FR12 at $t_\mathrm{pb} = 1 \, \mathrm{s}$ showing local stability conditions. Black regions denote locally unstable fluid parcels ($f_\mathrm{BV}^{2} < 0, \,f_\mathrm{SH}^{2} < 0$) while white regions are stable by either criterion ($f_\mathrm{BV}^{2} > 0, \, f_\mathrm{SH}^{2} > 0$). Blue regions are buoyancy-unstable but rotationally stabilized ($f_\mathrm{BV}^{2} < 0, f_\mathrm{SH}^{2} > 0$), while red regions -- which number few -- are the converse: buoyancy-stable, but overall destabilized by an angular momentum gradient ($f_\mathrm{BV}^{2} > 0, f_\mathrm{SH}^{2} < 0$).}
    \label{fig:fr12_stability}
\end{figure}

At later times, the more significant asymmetry in model FR12 makes a partitioning into equatorial and polar regions (as in Figure \ref{fig:freqs_3.5}) unreliable. Instead, we consider the local stability conditions of model FR12 in Figure \ref{fig:fr12_stability} where different colors are used to denote different regimes of stability and instability. The black (and red, although these are rare) regions are unstable regions where turbulence will likely develop. These are concentrated in the upper/lower ${\sim} 15^{\circ}$ near the poles inside the PNS core; this is the same region with $f_\mathrm{BV}^{2} < 0$ at the earlier snapshot in Figure \ref{fig:freqs_12}. There are filaments of instability in the post-shock region at tens of kilometers also, however they are produced by hydrodynamical gradients of material which is already turbulent. The large white region is convectively stable with $f_\mathrm{SH}^{2} > 0$. That most of the PNS and post-shock region is stable is unsurprising given the weak GW emission of this model. Most importantly, the blue regions are where $f_\mathrm{BV}^{2} < 0$ but $f_\mathrm{SH}^{2} > 0$, i.e., the fluid stratification is unstable, but stability is restored by a positive angular momentum gradient. In the PNS core, this limits the unstable regions to the poles, neatly demonstrating how rotation acts to stabilize the fluid.

Although rotation reduces the size of the unstable region, a local stability analysis with the spatial granularity of Figure~\ref{fig:fr12_stability} is not a perfect gauge of where the flow will be turbulent. Similar to convective overshoot entraining material over (typically spherical) convective boundaries, convection in the polar regions can be expected to spill over to lower latitudes, changing the stability conditions rather unpredictably.

What remains is to address the issue of resonance between the epicyclic and $f$-mode frequencies. Due to our (intentionally) similar simulation setup as \citet{Cusinato_et_al:2025} with the same progenitor and very similar input physics, it is first productive to compare our calculation of $f_\Omega$ to theirs\footnote{It is important to note that our method of taking the maximum of a spherical average differs to their (inverse) method of averaging radial maximums, avoiding polar and equatorial regions. We compute similar frequencies regardless of the method used.} in the case of the IR3.5 model. Initially, our $f_\Omega$ is consistent with theirs, at $600-700 \, \mathrm{Hz}$. However, their $f_\Omega$ evolves much slower than ours, reaching at most $1.2 \, \mathrm{kHz}$ versus our $>3 \, \mathrm{kHz}$. Significantly, they note resonant amplification of the GW signal at frequency crossings of $f_\Omega$ with the fundamental mode. In our simulation of model IR3.5, $f_\Omega$ is close to the dominant GW mode\footnote{We are unable to identify the modal character of the dominant emission band (see Section \ref{sec:linear_mode}) and hence will avoid describing it as the $f$-mode for internal consistency.} at numerous times throughout the simulation, however there is no obvious amplification of the GW emission as a result. The same holds for our other models where $f_\Omega$ tracks the GW emission band rather closely. 

It is not immediately obvious why our simulations do not reproduce the resonant amplification reported by \citet{Cusinato_et_al:2025}. Our fairly granular sweep over $\Omega_\mathrm{c}$ produces a range of $f_\Omega$ and GW emission curves, which undergo multiple crossings without amplification of the GW signal. The differences which arise in PNS structure discussed in Section \ref{sec:dom_mode} may play a role. Similarly, the angular variations of the epicyclic frequency found in this section may make the conditions for a resonance with the dominant mode more stringent. The problem may warrant additional exploration, especially in light of the recent simulations by \citet{Cusinato_Obergaulinger_Aloy:2025} where the phenomenon is present in the GW emission of a different progenitor.

\subsubsection{Impact of magnetic fields}
\label{sec:magnetic}
All simulations in our main set of $14$ are initialized with the same magnetic field, as described in Section \ref{sec:setup}. This choice has consequences for the PNS structure due to modified angular momentum transport, ultimately leading to an imprint of the magnetic field on the GW emission. This may contribute to the high GW frequencies in our simulations compared to others. It may also explain the lack of resonant amplification compared to \citet{Cusinato_et_al:2025, Cusinato_Obergaulinger_Aloy:2025} because our pre-collapse magnetic field is roughly ten times stronger than that of \citet{Cusinato_Obergaulinger_Aloy:2025}. To quantify the impact of magnetic fields, we perform two additional non-MHD simulations: one also without rotation, and one with a boost factor of $3.5$, similar to the IR3.5 model.

The non-rotating $\mathrm{BF0\_noB}$ model is unique among the models we consider in that it is the only one which does not explode, with no revival of the shock. However, the initial expansion and contraction of the shock after bounce follows a similar trajectory to the other models in Figure \ref{fig:shock_radius}. Meanwhile, the rotating $\mathrm{IR3.5\_noB}$ model exhibits very similar shock dynamics as the MHD-variant, model $\mathrm{IR3.5}$, with small differences which are easily attributed to stochasticity between simulations. This indicates that it is differences in rotation of the progenitor, not magnetic fields, which dictates the explodability in this set of simulations -- although we stress that this is likely not a general result.

\begin{figure}
    \centering
    \includegraphics[width=\linewidth]{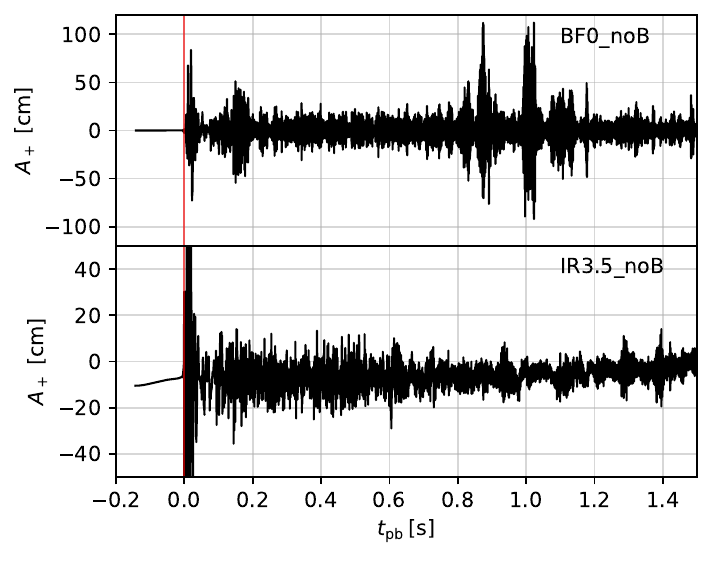}
    \caption{Time series of GW amplitudes, \aplus, for the two non-magnetic simulations. Note that, unlike in Figure \ref{fig:aplus}, the $y$-axis scales differ between top and bottom panels.}
    \label{fig:aplus_noB}
\end{figure}

\begin{figure*}
    \centering
    \includegraphics[width=\linewidth]{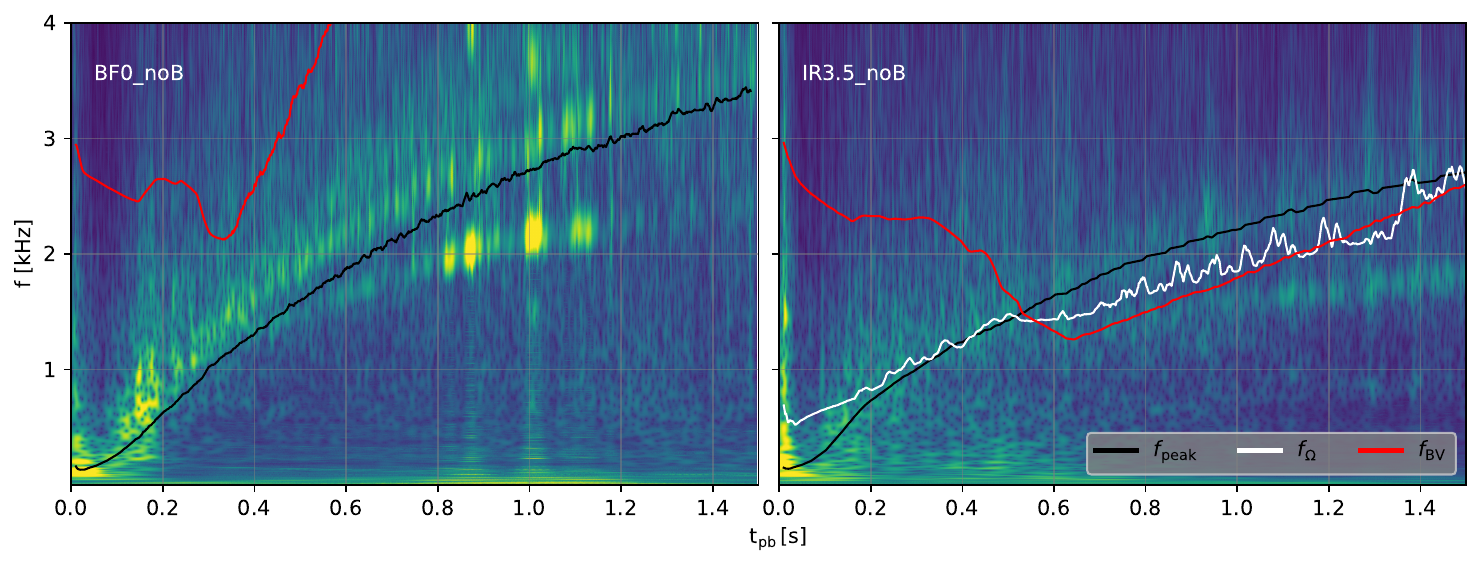}
    \caption{Same as Figure \ref{fig:dwt_grid} but for the two additional non-magnetic simulations. The epicyclic frequency, $f_\Omega$, does not exist for the non-rotating $\mathrm{BF0\_noB}$ model.}
    \label{fig:dwt_noB}
\end{figure*}

Time series of \aplus for the two non-MHD simulations are shown in Figure \ref{fig:aplus_noB} while spectrograms are given in Figure \ref{fig:dwt_noB}, again including the frequencies $f_\mathrm{peak}$, $f_\Omega$ and $f_\mathrm{BV}$. There are several notable differences between these spectrograms and those of Figure \ref{fig:dwt_grid}.

Looking first at the $\mathrm{BF0\_noB}$ model, there is evidently not a single dominant emission band visible in the spectrogram, as in the models of the main set, but multiple (three to be precise, as we will show in Section \ref{sec:linear_mode}). The lowest frequency band reaches ${\sim} 2.4 \, \mathrm{kHz}$ by the end of the simulation. The band with the next highest frequency is comparable to the dominant mode of models in the main set, surpassing $3 \, \mathrm{kHz}$ by the end of the simulation. The frequency evolution of this mode is quite similar to $f_\mathrm{peak}$ from Equation \ref{eqn:fpeak}, although slightly exceeds it. Another mode sits atop this one at slightly larger frequencies again, although it is less distinct in the spectrogram due to being quite broad; we find in Section \ref{sec:linear_mode} that this band corresponds to a $p$-mode, whose frequency is strongly coupled to the position of the shock. Thus, the frequency of the mode varies as the shock deforms, e.g., due to SASI. 

Unlike in the MHD models, the maximum Brunt-V\"ais\"al\"a frequency inside a radius of $100 \, \mathrm{km}$ does not follow a prominent emission band, instead, after the maximum of $f_\mathrm{BV}$ moves away from the inner boundary of the convective region, it becomes very large. This difference is due to the structure of the PNS in the non-rotating model where large gradients in density and pressure right at the PNS surface create a thin region (${\sim}2 \, \mathrm{km}$ thick) where $f_\mathrm{BV}$ reaches $5-6 \, \mathrm{kHz}$. As discussed in the previous section, the maximum Brunt-V\"ais\"al\"a frequency is not a good measure of the driven mode frequency when the peak in $f_\mathrm{BV}$ is narrow. Considering an average in the region around this peak brings the effective frequency down to around $4 \, \mathrm{kHz}$ by the end of the simulation -- comparable to the computed GW emission.

Model $\mathrm{BF0\_noB}$ is unique in that it undergoes brief periods of stronger GW emission, where the amplitude, \aplus, in Figure \ref{fig:aplus_noB} can exceed $100 \, \mathrm{cm}$ -- larger than the bounce signal for all but the most rapidly rotating models of the main set. These occur between $0.8-0.9 \, \mathrm{s}$, and at around $1 \, \mathrm{s}$ and $1.1 \, \mathrm{s}$ post-bounce, and primarily boost the power in the $f$-mode (as identified in Section \ref{sec:linear_mode}), however there is some broad-spectrum amplification too. These episodes are \textit{not} explained by phenomena such as increased PNS accretion or enhanced convection, and are also not evident in the neutrino luminosities for any neutrino species. SASI oscillations are generally strong during these phases and may play a role by their geometry preferentially exciting quadrupolar modes, although we do not quantify this.

The second non-magnetic model, $\mathrm{IR3.5\_noB}$, is more similar to those in the main set, although \aplus in Figure \ref{fig:aplus_noB} is slightly higher amplitude, indicating slightly stronger GW emission. Like the non-rotating model, the spectrogram for $\mathrm{IR3.5\_noB}$ also shows some splitting of the GW emission into two bands compared to its MHD counterpart, model IR3.5. The higher of the two frequency bands is a very good match to $f_\mathrm{peak}$ from Equation \eqref{eqn:fpeak}. The epicyclic and Brunt-V\"ais\"al\"a frequencies generally sit between the two bands, matching best with the lower band at earlier times, and the higher band at late times. This higher frequency band is similar to the dominant mode in the standard IR3.5 model, while the lower band has no obvious counterpart. The appearance of additional emission bands in the GW spectrum with or without magnetic fields and rotation may not be universal, with studies such as \citet{Jardine_Powell_Mueller:2022} seeing no such effect; however it has also been noted by \citet{Vartanyan_et_al:2023} that simulations extending less than a second after bounce may not capture the full GW signal. It is difficult to correlate specific hydrodynamical features induced by MHD (or a lack of MHD) with the excitation of specific emission bands/modes. Excitation of a particular mode (see Section \ref{sec:linear_mode}) requires driving forces (e.g. convection) in regions determined by the corresponding eigenfunction, which has a non-trivial spatial structure (see \citet{Torres-Forne_et_al:2018} for examples). Additionally, some modes are more difficult to excite or radiate with lower power, making them less obvious in spectrograms \citep{Torres-Forne_et_al:2018}.
Regardless, the presence of additional emission bands, or lack thereof, indicates that the effect of magnetic fields on the structure of the PNS can be significant enough to be potentially observable. While the effect of magnetic fields is also likely strong enough to impact the emergence of a resonance between PNS modes \citep{Cusinato_et_al:2025}, without obtaining such a resonance in our simulations, we cannot make a more conclusive statement to that effect.

\subsection{Spatially-resolved GW emission}
\label{sec:gw_spatial}
While the time-frequency structure of the GW signals provides clues
about the  emitting modes, it does not directly reveal the underlying fluid motions.
Complementary to our previous approach of explaining the PNS modes in terms of the frequencies which drive them, it is therefore beneficial to also develop an understanding of the actual fluid motions which are excited by, e.g., buoyancy and rotational pseudo-forces. This will be important for inferring PNS structure from GW observations of CCSNe. To this effect, we consider one additional model with a rotational boost factor of $3.55$, corresponding to a central angular velocity of $1.03 \ \mathrm{rad\, s^{-1}}$. Different to the previous simulations which produce output files every $2 \ \mathrm{ms}$, full output for this simulation is saved every $40 \ \mathrm{\text \textmu s}$. This corresponds to a sampling frequency of $25 \ \mathrm{kHz}$ and is sufficient to resolve temporo-spatial oscillations of the PNS corresponding to typical $p$- and $g$-modes, in a manner similar to recently developed approaches in the literature \citep{Jakobus_et_al:2023,Zha_EggenbergerAndersen_OConnor:2024,Cusinato_et_al:2025}.

\begin{figure*}
    \centering
    \includegraphics[width=\textwidth]{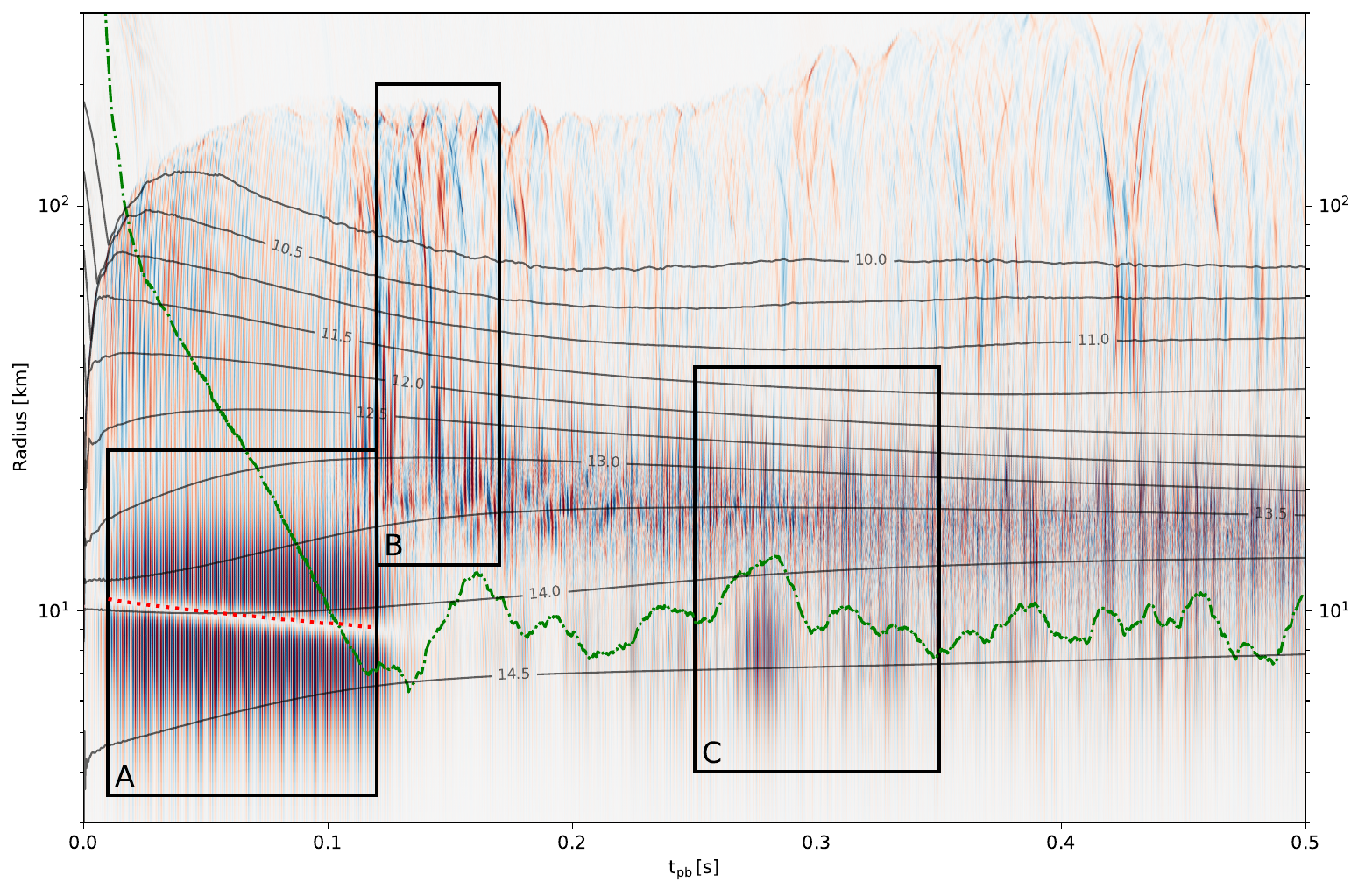}
    \caption{Normalized quadrupole amplitude $q(r,t)$ as function of radius and post-bounce time for model BF3.55. The amplitudes at a given time are normalized by the mean in a centered time window of $12 \ \mathrm{ms}$. The relative change of the normalization is shown by the dash-dot green line and shares the logarithmic scale of the left axis (excluding an -- in this context -- unimportant constant factor which sets the absolute scale of the signal). Solid gray lines mark isocontours of spherically-averaged density. Contour labels denote $\log_{10}$ of the density surface in units of $\mathrm{g \, cm^{-3}}$; i.e. a contour labeled $12.0$ marks a surface of $10^{12} \ \mathrm{g \, cm^{-3}}$. Three boxes enclose the time and radius ranges of several pertinent features. Box A includes a dotted red line, which tracks the radius of the $s=3 \, \mathrm{k_{B}/nucleon}$ entropy surface which approximately coincides with a node of the ringdown mode during the first $100 \ \mathrm{ms}$ post-bounce.}
    \label{fig:quad_tseries}
\end{figure*}

The time-integrated quadrupole formula \citep{Mueller_Janka_Marek:2013} is again used to compute the GW amplitude, \aetwo. Similar to \citet{Jakobus_et_al:2023}, by integrating in solid angle, but not in radius, we obtain a function of time and radius, $q(r,t)$\footnote{This is exactly Equation \eqref{eqn:quadrupole_formula} with the integration over $r$ omitted.}, which spatially localizes the source of GWs as the PNS and gain region evolve with time. This is plotted in Figure \ref{fig:quad_tseries} from bounce to $500 \ \mathrm{ms}$ later \citep[c.p.][]{Cusinato_Obergaulinger_Aloy:2025}.

Figure \ref{fig:quad_tseries} is not intended as an accurate representation of the changing strength of the GW signal produced over time, as the color scale at any given time is normalized by a smoothed amplitude curve (green dash-dot line) computed as a moving average in time of $q(r,t)$ with a boxcar window width of $12 \, \mathrm{ms}$. This removes most secular trends in the signal amplitude and normalizes the oscillations to the interval $[-1,1]$. A sense of the time evolution of the GW amplitude is preserved in the dash-dot amplitude line (up to a normalization factor containing unimportant physical and geometric constants which allow the line to share the scale of the y-axis), and one can immediately see a rapid decay of the mean amplitude in the first $100 \ \mathrm{ms}$ post-bounce. Following this is a prolonged period where the signal is relatively constant in amplitude, varying on the order of $20\%$ around its mean. The smoothed signal amplitude includes quadrupole contributions from radii above those shown in Figure \ref{fig:quad_tseries}, i.e. $r > 300 \, \mathrm{km}$, which may be artificially excited by boosting the rotation rate of the progenitor. These may present as low frequency fluctuations of the smoothed signal amplitude, or even as a secular drift, if the frequency is low enough compared to the smoothing window width of $12 \ \mathrm{ms}$.

\begin{figure}[h!]
    \centering
    \includegraphics[width=0.9\linewidth]{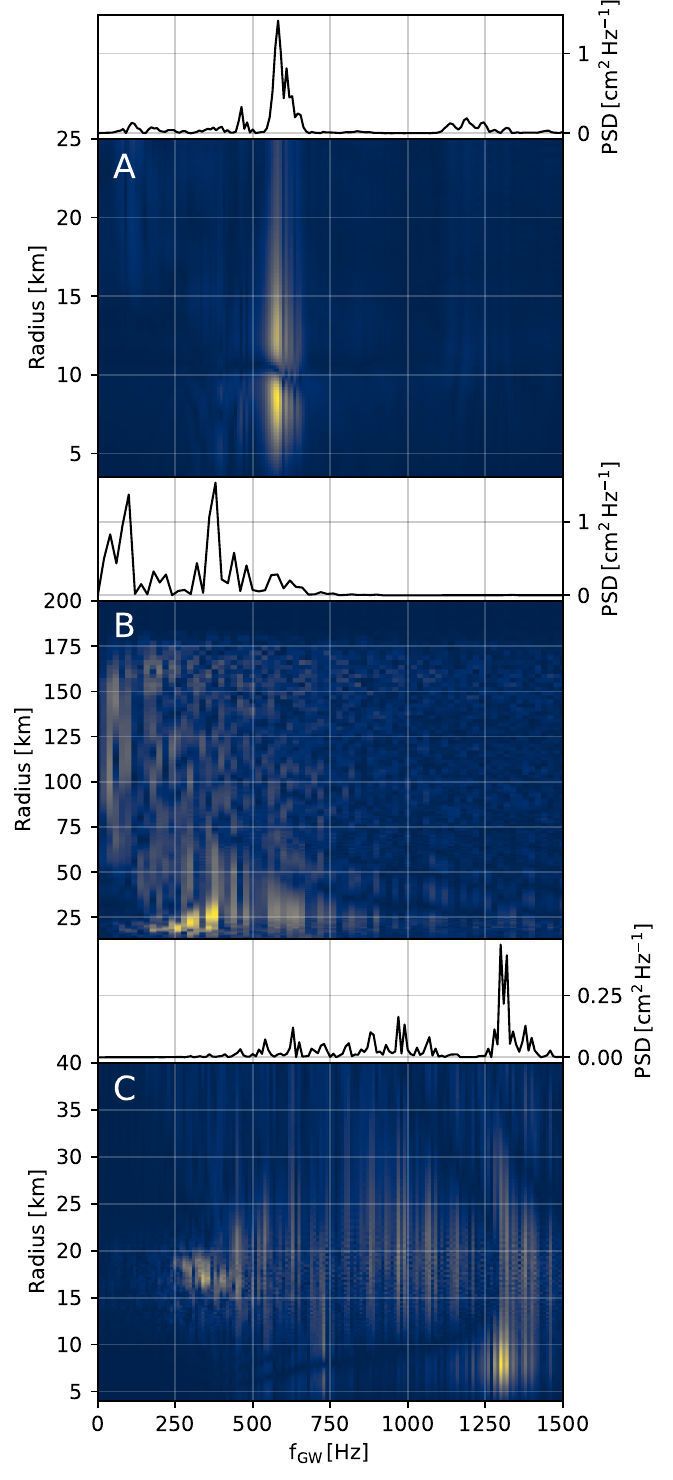}
    \caption{Fourier decompositions corresponding to boxes A-C in Figure \ref{fig:quad_tseries} corresponding to model BF3.55. The transform covers the entire time window of the box and is calculated independently for each radius. Note that the radius scale differs between panels. The frequency axis is truncated at $1.5 \ \mathrm{kHz}$; above this frequency we find only weak broadband emission at all radii. The PSD of \aetwo in each window is provided above the corresponding panel.}
    \label{fig:box_specs}
\end{figure}

The signal amplitude suggests the existence of two distinct phases: a ringdown of the PNS after bounce, and longer-duration oscillations driven by different processes -- namely accretion and convection, in the manner discussed in previous sections. Although Section \ref{sec:gw_main_set} established that the dominant GW emission evolves over several seconds, we find that the period of $0.5 \, \mathrm{s}$ presented in this section is sufficient to include a sampling of all the physical processes relevant to the GW signal. At later times, $q(r,t)$ looks very similar to the time window $300-500 \, \mathrm{ms}$, but with the strongest emission region oscillating at higher frequencies and contracting as the PNS cools. 
Three boxes in Figure \ref{fig:quad_tseries} contain prominent features in both phases which will be used to better characterize the GW emission. Fourier transforms of $q(r,t)$ over the full time window in each box are provided in Figure \ref{fig:box_specs} to show the frequencies which correspond to oscillations in Figure \ref{fig:quad_tseries}. The power spectral densities (PSD) within each box are also shown; unlike the Fourier transforms, these account for cancellation of quadrupolar perturbations when integrating $q(r,t)$ in radius to obtain \aetwo.

\subsubsection{Box A}

Box A features the ringdown of the PNS after bounce. The signal amplitude decays quite rapidly, halving every ${\sim}30 \ \mathrm{ms}$, which is consistent with previous studies on the ringdown of rotating PNSs \citep{Dimmelmeier_Ott_Marek_Janka:2008, Fuller_Klion_Abdikamalov_Ott:2015}. During the ringdown, gravitational radiation is dominated by narrowband emission centered around $580 \ \mathrm{Hz}$, as seen in Panel A of Figure \ref{fig:box_specs}. The frequency does not shift appreciably with time and the spatial decomposition shows a clear nodal structure which tracks the steep positive entropy gradient at the inner edge of the PNS convective zone, here identified by the $s=3 \, \mathrm{k_{B}/nucleon}$ entropy surface and shown by a red dotted line in Box A of Figure \ref{fig:quad_tseries}. 

\begin{figure}
    \centering
    \includegraphics[width=\linewidth]{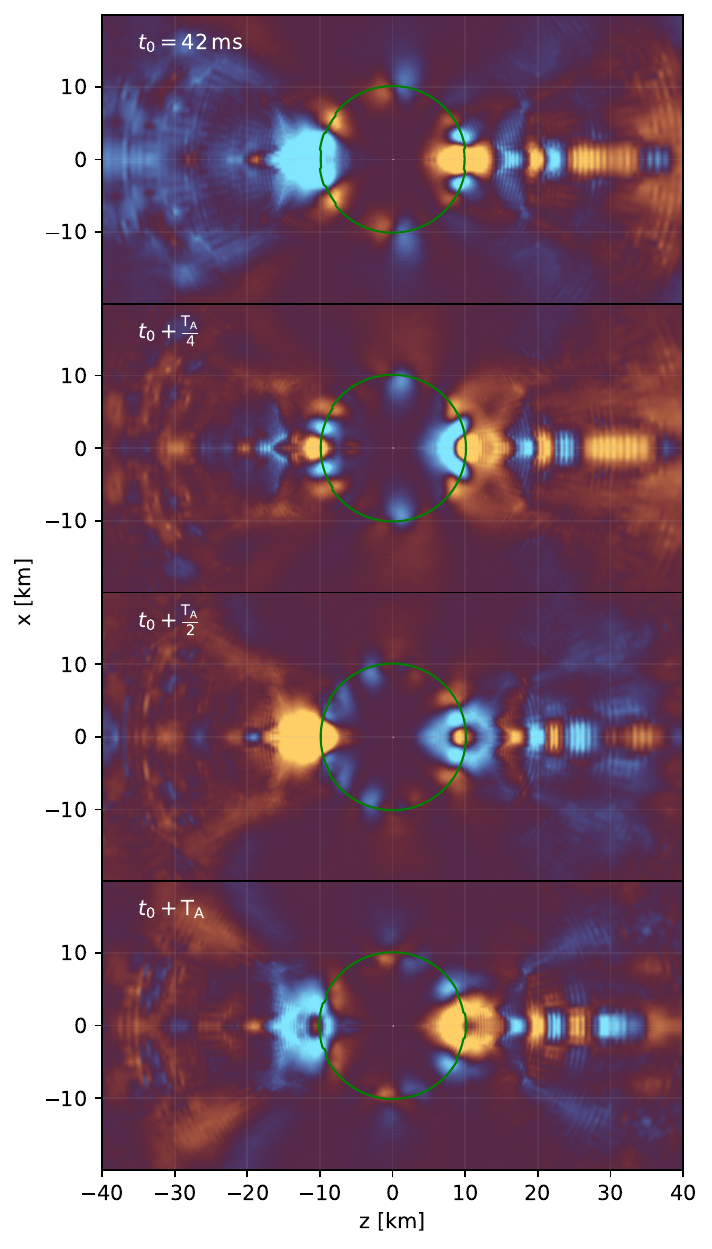}
    \caption{Slices of the time-derivative of the integrand in the quadrupole formula -- Equation \eqref{eqn:quadrupole_formula} -- for model BF3.55. The top panel is at an (arbitrary) start time of $42 \, \mathrm{ms}$ post-bounce; the subsequent panels are fractions of $\mathrm{T_{A}} \approx 1.7 \, \mathrm{ms}$ later, where $\mathrm{T_{A}}$ is the period corresponding to a frequency of $580 \, \mathrm{Hz}$. From the second panel down to the bottom panel, these fractions are a quarter, a half, and a full cycle respectively. Additionally, the integrand in Equation \eqref{eqn:quadrupole_formula} includes the usual $r^{2} \sin \theta$ volume element for integrating in spherical coordinates. The $r^{2}$ part is retained here to make the color scale easier to interpret, while the $\sin \theta$ is omitted to avoid suppressing features near the rotation axis. The $s=3 \, \mathrm{k_{B}/nucleon}$ entropy surface is plotted as a green line in each panel for reference. This early after bounce, the PNS is not strongly deformed by rotation and the entropy structure does not change appreciably over the sub-millisecond times between panels.}
    \label{fig:boxa_quad}
\end{figure}

Another view of these oscillations is given in Figure \ref{fig:boxa_quad} where the time-derivative of the integrand in the quadrupole formula is plotted (i.e. without any integration over spatial dimensions). The four panels show the evolution of the integrand over a period of $580^{-1} \, \mathrm{s}$ starting $42 \, \mathrm{ms}$ after bounce. Subsequent panels show how the integrand has evolved over the course of one oscillation at $580 \, \mathrm{Hz}$; i.e., a quarter, half, and full cycle later for the second, third, and forth panels respectively. It is important to note that, to aide the visualization, the $r^{2}$ term of the volume element is retained as part of the integrand, while the $\sin \theta$ term is omitted; this is done with the understanding that contributions along the axis (horizontal $y=0$ line) only contribute a small amount to the total integrated quadrupole. The $s=3 \, \mathrm{k_{B}/nucleon}$ entropy surface is marked by a green line. At this time it is a near-perfect sphere of radius $10 \, \mathrm{km}$.

All panels show similar features, primarily a set of oscillations roughly centered on the entropy isosurface. Consistent with a mode frequency of $580 \, \mathrm{Hz}$, these oscillations in the first and third panels are approximately $180^{\circ}$ out of phase, although the position of angular nodes may change due to the stochastic nature of accretion onto the PNS, and the impact of inertial modes. Because these oscillations track a positive entropy gradient, they are likely buoyancy-driven. However, Figure \ref{fig:boxa_quad} shows that the radial extent of perturbations can be several kilometers -- large enough that a perturbed fluid bubble is subject to a range of restoring buoyancy forces, again highlighting that the Brunt-V\"ais\"al\"a frequency at a single point -- e.g. the $s=3 \, \mathrm{k_{B}/nucleon}$ entropy surface -- is not directly relatable to the corresponding mode frequency. 

There is also a visible wave train along the (horizontal) rotation axis. Because this feature is strongly confined to the axis, and the waves are not massively energetic, it does not contribute significantly to the full volume-integrated quadrupole; i.e., the GW emissions from these waves are negligible. However, it is nonetheless a clear signal of sloshing motions of the PNS along the axis. While most prominent on the right-hand side in Figure \ref{fig:boxa_quad} at the time shown, both poles of the PNS generate outwards-traveling waves during the ringdown phase after bounce. 

\begin{figure}
    \centering
    \includegraphics[width=\linewidth]{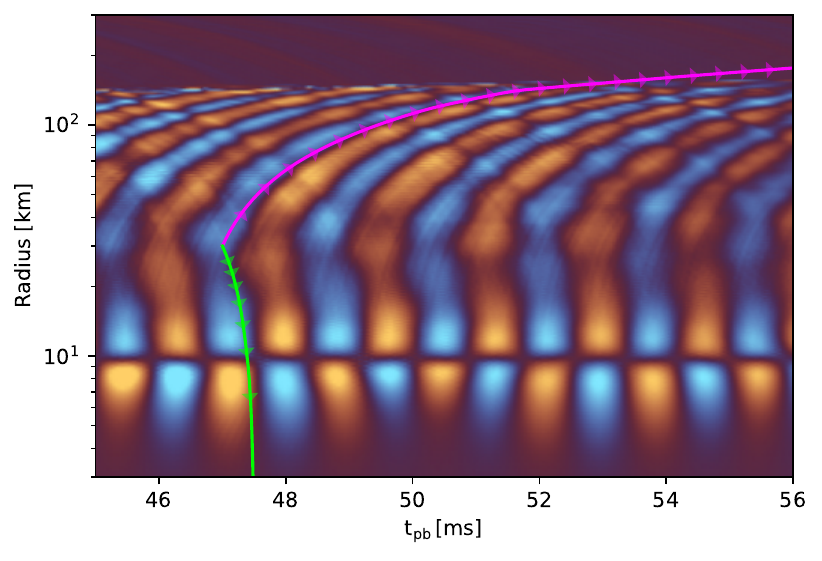}
    \caption{Close-up of a brief period in box A (although with different radial bounds) from Figure \ref{fig:quad_tseries} for model BF3.55. Sonic trajectories are over-plotted for both inwards (green) and outwards (magenta) directions. $q(r,t)$ is scaled by $r^{1/3}$ to achieve adequate color contrast at all radii.}
    \label{fig:box_a_zoom}
\end{figure}

Another type of wave is uncovered in Figure \ref{fig:box_a_zoom}, which shows ${\sim} 10 \, \mathrm{ms}$ of Box A in more detail. On these timescales, the oscillations of the quadrupole amplitude can be seen propagating out of the PNS and into the gain region -- similar to what is also seen by \citet{Cusinato_Obergaulinger_Aloy:2025}. 
We confirm that these are acoustic waves by also plotting the outwards trajectory of a tracer particle traveling outwards at the local sound speed, $c_{s}$; i.e., solutions to,
\begin{equation}
    r'(t)=\pm c_{s}(r,t), \ r(0)=30 \, \mathrm{km},
\end{equation}
where the initial radius of $30 \, \mathrm{km}$ is read off Figure \ref{fig:box_a_zoom} as the approximate radius of the source of the waves. The inwards solution from this radius is also shown and, up to the node at the entropy isosurface, is a reasonable fit for the wavefront. While this is a rather rough phenomenological fit, it suggests that the waves in quadrupole amplitude are produced somewhere near the PNS convective region, although they do not necessarily contribute strongly to the GW signal until they propagate deeper into the PNS. The interesting implication of this analysis is that it is possible to resolve the propagation of waves from the interior of the PNS into the gain region, which has been invoked as a means of energy transfer in the past \citep{burrows_06,gossan_20}.
It is also noteworthy that the propagation of waves into the gain
region is different in our case from \citet{Cusinato_Obergaulinger_Aloy:2025} in one important respect.
We find that waves in the gain region and within the PNS are of the same frequency, consistent with linear wave propagation, whereas they find waves 
of lower frequency (possibly subharmonics) in the gain region, which is indicative of non-linear wave interactions. This suggests subtle differences between simulation codes in regards to wave coupling in the PNS surface region, which may have considerable implications for energy transfer
into and out of GW-emitting modes.

\subsubsection{Box B}

There are two features of interest in Box B: first, the stronger oscillations between the PNS convective region and surface, peaking at around $20 \, \mathrm{km}$, and second around $100 \, \mathrm{km}$ in the gain region where sufficient time has passed since the shock stalled for large-scale convective plumes to develop. Both features persist outside Box B, but the gain region emission becomes much weaker in comparison, despite ongoing large-scale convection. The frequency of GW emissions inside the PNS are broad, but peak at several hundred Hz (panel B of Figure \ref{fig:box_specs}). This broadness is consistent with the frequency distributions implicit in Figures \ref{fig:dwt_grid} and \ref{fig:dwt_3d} around $150 \, \mathrm{ms}$ post-bounce. It is only after several hundred milliseconds that the dominant mode emerges and radiates GWs in a relatively narrow frequency band.

In the gain region, Figure \ref{fig:box_specs} shows that quadrupolar perturbations are not strongly peaked at any one location in frequency or radius. However, after integrating in radius to obtain the GW PSD, it is evident that GW power is concentrated at lower frequencies -- less than $100 \, \mathrm{Hz}$ -- in the gain region during this phase of the CCSNe evolution. This soon after the formation of the gain region, the spatial scale of convective plumes (which number few) is quite large, typically spanning the full range between the gain radius and the shock. These contribute a relatively low-frequency GW signal, in contrast to faster, smaller-scale eddies which develop later.

\subsubsection{Box C}

Box C and its corresponding panel in Figure \ref{fig:box_specs} show broadband contributions to \aetwo from the PNS convective region with some transient contributions from deeper in the core. There are hotspots in the range $300-400 \, \mathrm{Hz}$, and around $1.2 - 1.3 \, \mathrm{kHz}$ located in the PNS convective region and PNS core respectively. However, examining the PSD shows that the low-frequency components largely do not survive, and the emission peaks above $1.2 \, \mathrm{kHz}$ due to a combination of strong oscillations in the PNS core ($r < 10 \, \mathrm{km}$) and weaker oscillations towards the PNS surface. By examining spectrograms for the almost-identical IR3.5 model, we see that the high-frequency component likely corresponds to the emerging dominant emission mode at around $300\,\mathrm{ms}$ after bounce.

This analysis makes clear that there are at least two distinct processes (or combinations of processes) responsible for GW emissions in the early phase of rotational core bounce. First is the ringdown, which dominates immediately after bounce but decays rapidly. This process involves oscillations deep within the PNS and produces a short-lived feature on the GW spectrograms of rotating models, which is distinctive for its relatively narrow and steady frequency. The second process is excitation of the PNS surface region (down to the convective layer) through various forms of convection and accretion \citep[c.p.][]{Mezzacappa_Zanolin:2024}. These processes eventually drive the growth of the dominant GW mode.
The important insight from Box~C is that the excited modes sometimes entail motions deep inside the PNS, highlighting the importance of the PNS as an emission region, in agreement with other recent analyses \citep{Murphy_et_al:2025}.

\subsection{Linear mode analysis}
\label{sec:linear_mode}
By assuming perturbations producing GWs are small, the equations of motion of the fluid may be linearized, permitting an analysis of the stable oscillation modes of the PNS. Previous work on the PNS eigenvalue problem \citep{Torres-Forne_et_al:2018, Fuller_Klion_Abdikamalov_Ott:2015, Murphy_et_al:2025, WesternacherSchneider:2020} typically assume spherical, non-rotating stars for simplicity and are generally successful at matching eigenmodes to the predicted GW emission of CCSNe \citep[e.g.][]{Zha_EggenbergerAndersen_OConnor:2024}. The addition of rotation presents several complications, not only by directly modifying the fluid equations with rotational force terms, but also via the effects of mode mixing and introduction of inertial modes \citep{Fuller_Klion_Abdikamalov_Ott:2015}. It is currently not clear whether the assumptions of the analysis are valid in the context of rapidly rotating stars, or if there is an (as yet undetermined) upper bound on how fast the PNS can spin and still yield meaningful eigenmodes. Beyond this upper bound, it is not possible to rigorously identify the modes corresponding to GW emissions. It will be shown that this is particularly vexing in the case of our model $\mathrm{IR3.5\_noB}$ where multiple prominent emission bands exist (similarly for model $\mathrm{BF0\_noB}$, however we shall imminently see that this model easily permits mode identification).

\begin{figure*}
    \centering
    \includegraphics[width=\linewidth]{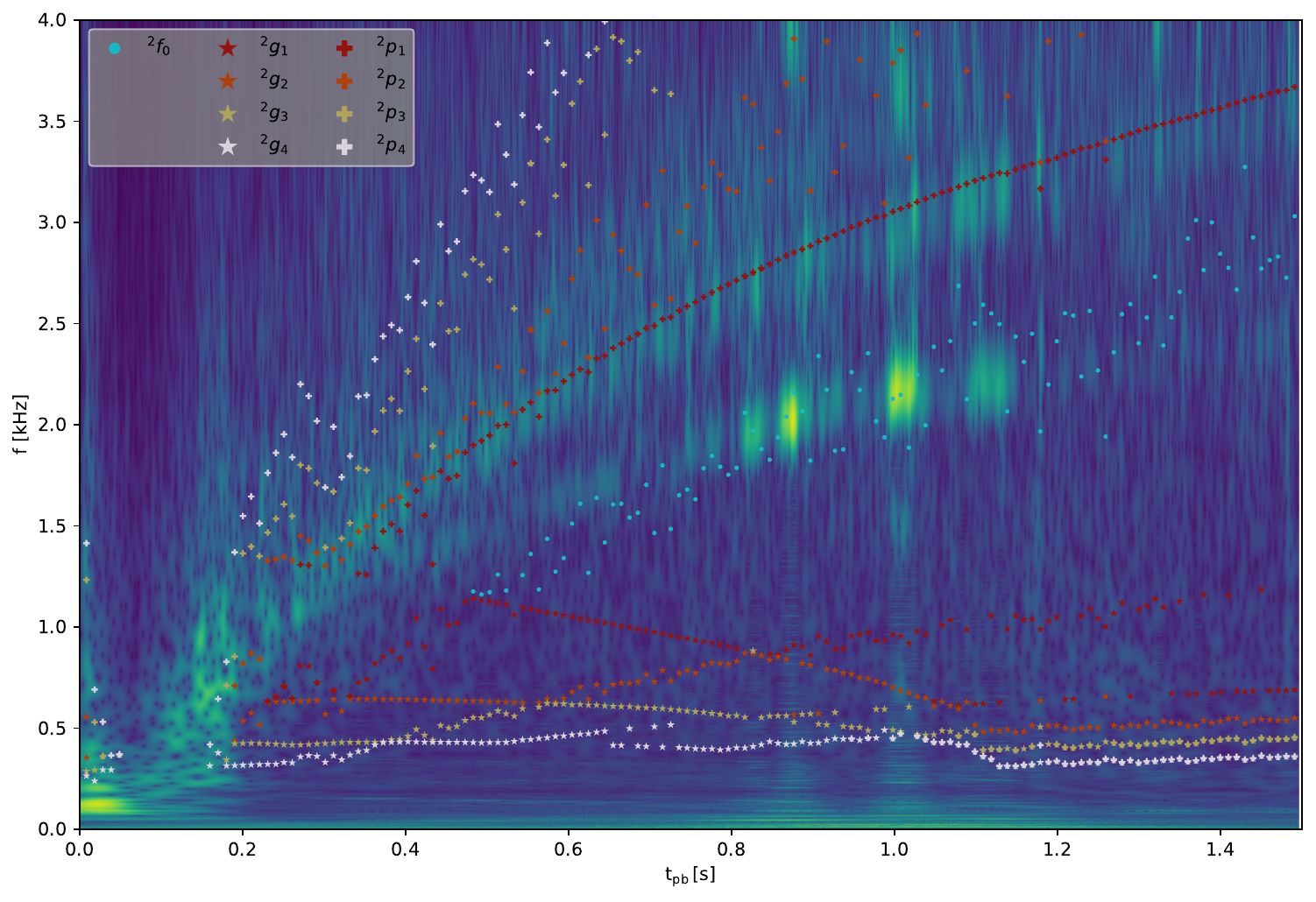}
    \caption{Frequency evolution of several low node count eigenmodes over-plotted on the spectrogram for $\mathrm{BF0\_noB}$ from the corresponding panel of Figure \ref{fig:dwt_noB}. The circle markers indicate the $f$-mode while $g$-mode and $p$-modes are denoted by a star and a plus, respectively.}
    \label{fig:mode_analysis_bf0}
\end{figure*}

\begin{figure*}
    \centering
    \includegraphics[width=\linewidth]{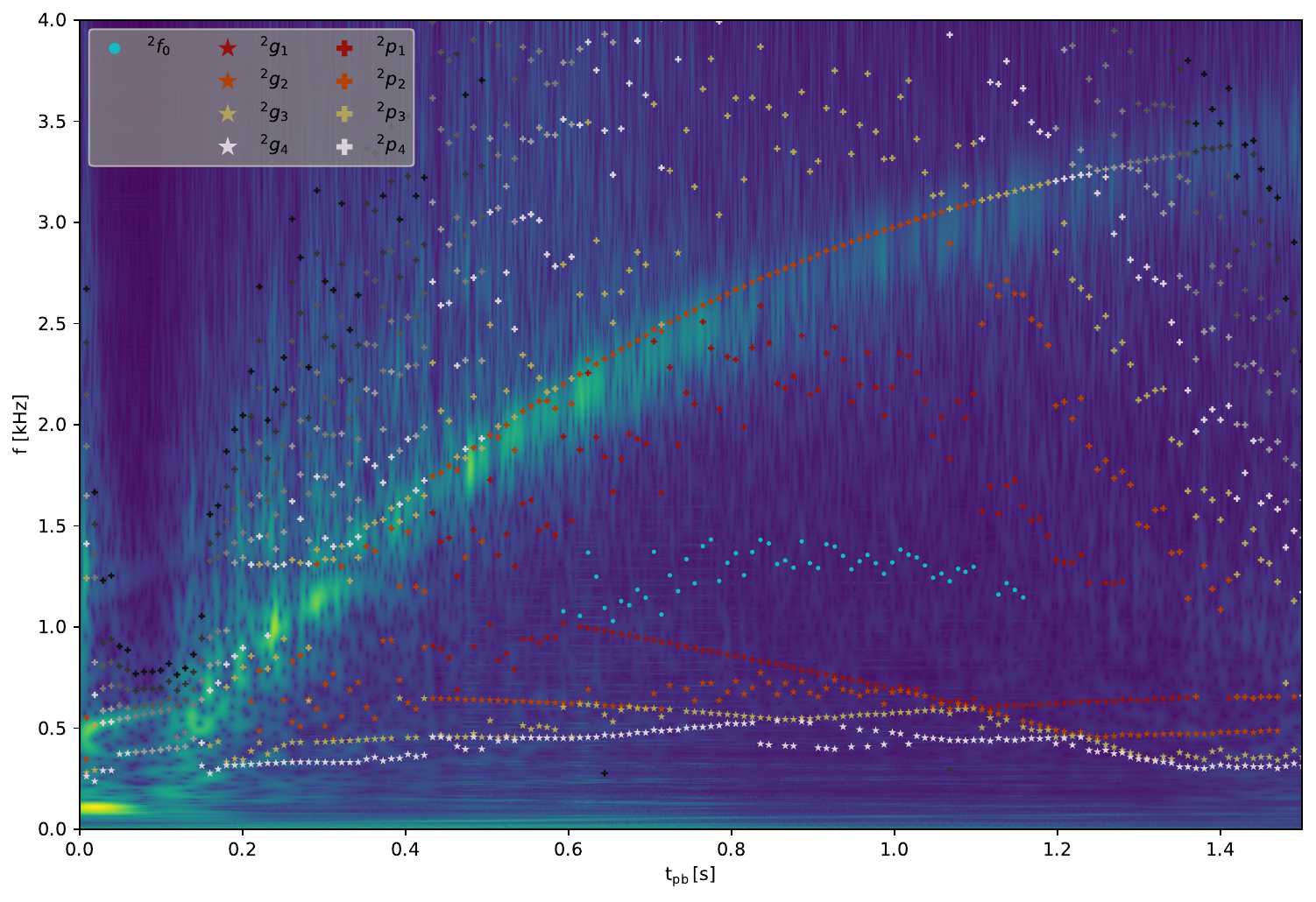}
    \caption{Same as Figure \ref{fig:mode_analysis_bf0} but for model SR1. The spectrogram is reproduced from the corresponding panel of Figure \ref{fig:dwt_grid}. Only the polar regions are used to compute the background hydrodynamic state for the linear mode analysis for this model. Additional $p$-modes $^{2}p_{5-9}$ are shown in shades of gray, but are not listed in the legend for brevity because the nominal mode identify is unimportant here.}
    \label{fig:mode_analysis_sr1}
\end{figure*}

We apply the eigenmode analysis framework of \citet{Torres-Forne_et_al:2018} to our $\mathrm{BF0\_noB}$ and SR1 models using the \textsc{great} code \citep{Torres-Forne_et_al:2018, Torres-Forne_et_al:2019}. The code is configured to assume Newtonian gravity with the Cowling approximation (static spacetime), which is appropriate given the use of monopole gravity in our simulations.
We reiterate that this analysis neglects the rotation of the star, and so is formally only valid for the $\mathrm{BF0\_noB}$ model. We also include the analysis of the SR1 model to assess the performance of the framework under the weakest violation of its assumptions, i.e., the slowest rotation model in our main set. 
The results of the linear mode analyses are summarized in Figures \ref{fig:mode_analysis_bf0} and \ref{fig:mode_analysis_sr1} for models $\mathrm{BF0\_noB}$ and SR1, respectively.

We adopt the same mode classification system as \citet{Torres-Forne_et_al:2018} based on a node counting method. The mode with the highest frequency out of all modes with $n$ nodes is labeled the $n$-th $p$-mode, while the mode with the lowest frequency with $n$ nodes is labeled the $n$-th $g$-mode. We use the notation $^{\ell}\!p_{n}$ and $^{\ell}\!g_{n}$ for $p$-modes and $g$-modes respectively, where $\ell = 2$ to isolate the quadrupole modes, and $n$ is the number of nodes of the eigenfunction between the grid center and the outer boundary at the shock. As usual, we label the mode with zero nodes the fundamental or $f$-mode, $^2f_0$. For simplicity, only modes with less than $5$ nodes are plotted; higher node counts only produce higher frequency $p$-modes and lower frequency $g$-modes, which is consistent with the trend established for $n<5$. To avoid overcrowded
plots, we do not include the hybrid modes in this classification scheme, which are identified as modes with frequencies between the $^{\ell}\!p_{n}$ and $^{\ell}\!g_{n}$ modes.
We note that these modes participate in avoided crossings \citep{Stergioulas_2003} with
both the $p$- and $g$-modes, e.g., with the $^2g_1$ mode in both models $\mathrm{BF0\_noB}$ and SR1 (at ${\sim} 440 \, \mathrm{ms}$ and ${\sim} 650 \, \mathrm{ms}$ post-bounce respectively), which causes a change in the mode character and the emergence of the fundamental mode.

We will first consider Figure \ref{fig:mode_analysis_bf0} and the linear mode analysis of the non-rotational model $\mathrm{BF0\_noB}$. For this model, aside from deviations from spherical symmetry caused by turbulence, the assumptions of the eigenmode analysis are met and we expect a close correspondence between the frequency evolution of the computed modes, and the GW emission bands. Indeed, the analysis performs well, with each band in the GW spectrum clearly associated with an eigenmode by around half a second after bounce. The lowest frequency band is associated with the $f$-mode which, as mentioned previously, emerges as the result of an avoided crossing between a hybrid mode and the $^{2}g_{1}$ mode. The remaining GW emission bands are found to be associated with the $^{2}p_{1}$ and $^{2}p_{2}$ modes (middle and top bands, respectively). The amount of scatter in the computed mode frequency over time varies between the three modes. This may occur due to variations in the shock radius \citep{Torres-Forne_et_al:2018}, however will also be impacted by the dynamics of different regions within the PNS and post-shock domain. The different eigenfunctions of each mode result in different sensitivities of the mode to changes in hydrodynamic conditions in a given region. For instance, the $^{2}p_{1}$ mode is likely sensitive to a stable region, while the $^{2}p_{2}$ mode may be significantly driven by a turbulent, or otherwise rapidly evolving region. 

For modes with larger node counts, the well-established trends apply: the frequency of the mode increases/decreases as the node count increases for subsequent $p$-modes/$g$-modes. Additionally, all $p$-modes increase in frequency with time -- largely because the shock in the non-rotating model does not expand on the timescale of our simulation \citep{Torres-Forne_et_al:2018} -- while $g$-modes are somewhat steady below $1 \, \mathrm{kHz}$ owing to the slow variation of PNS regions which generate them. Larger-n $p$- and $g$-modes are not associated with any notable GW emission.

The clear correspondence between modes of the linear analysis and the simulated GW signal gives us confidence that the existing eigenmode framework is robust for non-rotating models. Additionally, it independently corroborates the large GW frequencies found in our simulations.

Now considering the SR1 model, we first note that applying exactly the same analysis as for model $\mathrm{BF0\_noB}$ to model SR1 fails to produce meaningful results. Specifically, the dominant GW emission band is not associated with any of the computed modes after $800 \, \mathrm{ms}$, and prior to that is only coincidentally close in frequency to a few $p$-modes. The fundamental mode is far removed from any strong GW emission band.

The failure of the linear analysis for model SR1 has a number of reasons. First and foremost, the analysis, as previously mentioned, is formulated assuming no rotation. It neglects inertial forces as restoring forces for oscillations, and becomes inaccurate due to the spherical averaging of the rotationally-deformed structure of the PNS and post-shock region. Second, rotation strongly influences the dynamics of the shock, which in turn heavily influence the frequency of $p$-modes, with the frequency of a given $p$-mode tending to decrease as the (average) shock radius expands at late times. To sidestep some of these issues, we restrict the calculation of the hydrodynamic background state to the polar regions (cones along the north and south poles with a half-angle of $23^{\circ}$) where rotation minimally impacts the PNS structure.  Note, however, that such an analysis is mostly a provisional tool to gain more insight into the mode structure than can be obtained by considering the profiles of the Brunt-V\"ais\"al\"a frequency and epicyclic frequency. It cannot replace a fully multi-dimensional mode analysis including rotation.

The resulting modes shown in Figure~\ref{fig:mode_analysis_sr1} clearly show an evolving mode which tracks the dominant GW emission for the entire simulation, however the character of this mode changes frequently. Particularly for $t_\mathrm{pb} > 1 \, \mathrm{s}$, the ``emitting'' mode progresses from $^{2}p_{2}$ through to $^{2}p_{9}$, changing as the trajectory of these $p$-modes -- which are decreasing in frequency as the shock expands -- brings them close to the GW emission band. There is no obvious physical interpretation for this phenomenon, and instead it is more likely to be a failure of the mode identification to ascribe a persistent ``name'' for this mode over time.
We additionally note that mode classification based on a linear eigenmode analysis on polar data must be approached with caution.

Of additional interest is the behavior of the $f$-mode, which is not associated with any significant GW emission (which is also true without restricting the hydrodynamic inputs of the linear analysis to the polar regions), and which disappears again after some time due to an avoided crossing with a hybrid mode. Similar avoided crossings terminate the $^{2}p_{1}$ and $^{2}p_{2}$ modes when their respective frequencies drop below ${\sim} 1 \, \mathrm{kHz}$. The $f$-mode is often assumed to be the mode associated with the dominant GW emission.

While the failure of the mode analysis to conclusively identify the dominant mode is unsatisfactory, the mere presence of an evolving mode structure, which only appears when considering perturbations of the PNS structure in the polar region, is of some interest. The usefulness of this finding is somewhat limited however, since no definitive information about the modes is provided by the linear analysis without a more robust identification scheme. The partial success of the analysis when restricted to the poles mirrors the finding of Section \ref{sec:dom_mode} that the analytical $f_\mathrm{peak}$ approximation is robust when using the polar PNS structure, even for rotating models. However, more rapidly rotating models remain unsuitable for the linear mode analysis because rotation has too strong of an effect on their structure, even near the poles.

\section{Conclusions}
\label{sec:conclusions}

The search for GWs from CCSNe continues to be supported by the development of physically-informed GW waveforms from hydrodynamic simulations of core-collapse. While the emerging picture of ``standard'' CCSNe waveforms, i.e. without significant rotation, is powering more advanced searches for burst signals in GW detector data, the signals from rarer rapidly-rotating stars is less constrained. High-quality 3D simulations remain the ideal method for generating sample GW waveforms in these cases, however 2D simulations offer a compelling trade-off between physical accuracy and computational cost and are consequently well suited to exploring emerging trends with varying stellar rotation rates.

We have here computed a set of $17$ simulations of rotational core-collapse from a single progenitor, each differing only by a constant factor in their rotation profiles (except two non-magnetic models). Within the main set, these factors range from $1$ (i.e. model SR1 with no artificial boost to the rotation rate) to $12$, and result in varied explosion dynamics and GW emission. While slower rotation produces a fairly typical bounce followed by contraction of the shock, which later expands after a period of heating in the gain region, the most rapidly rotating models shocks' never stall, and instead produce a prompt, asymmetric explosio.

The GW signals show a prominent trend towards lower GW amplitudes as the progenitor rotation is increased. The stabilizing effect of a positive angular velocity gradient in the PNS convective region, and suppression of SASI-driven downflows by increasingly rapid shock expansion are likely to contribute to this trend. Notably, all models still show a strong GW burst at bounce, followed by a ringdown phase, which is consistent with previous models of rotational core-collapse. On longer timescales, all models also show a dominant mode which increases in frequency with time, although the frequency evolution of models with low GW amplitudes is more ambiguous (i.e. due to low signal-to-noise). Interestingly, a complementary
3D model showed another mode that is not regularly seen in simulations: the core $g$-mode band found by \citet{Jakobus_et_al:2023}.
The two non-MHD simulations exhibit qualitatively different GW morphology, with multiple prominent GW emission bands, including $p$-modes
above the familiar dominant $f$-mode or $^{2}g_1$-mode. In the magnetized cases, the $p$-modes are suppressed even in the non-exploding models where we expect similar conditions for $p$-modes
as in the non-MHD case.
Evidently, there is a fairly significant imprint of PNS magnetic fields on the GW emissions of CCSNe.

The GW frequencies from these simulations are notable for being rather high compared to simulations with other codes, and simulations of the collapse of other progenitors. We find that the structure of the PNS which emerges in our simulations naturally explains these high frequencies, although there remains some artificial increase due to the use of a modified Newtonian gravitational potential instead of general relativistic gravity. The unusually high frequencies
were also confirmed by a 3D simulation.
Furthermore, the approximate formula for peak gravitational wave frequencies of \citet{Mueller_Janka_Marek:2013}, despite not including rotational effects, remains valid for rapidly rotating stars provided that the PNS properties which enter into the formula are computed in the polar region of the PNS. As the polar radius of the PNS shrinks for very rapid rotation, the dominant mode frequency is shifted upwards compared to the non-rotating case. At lower latitudes, centrifugal
support expands the PNS, resulting in a lower Brunt-V\"ais\"al\"a frequency than at the poles, but centrifugal stabilization almost
compensates for this and results in similarly high local oscillation
frequencies as in the polar region.

Our simulations did not reproduce the resonant amplification effect noted by \citet{Cusinato_et_al:2025} despite achieving a crossing of the nominal $f$-mode frequency with the epicyclic frequency in a number of cases. The effect consequently may require more precise tuning of the simulation to appear, and/or may have a dependence on the details of numerical implementations within different CCSNe simulation codes.

We additionally perform a spatial decomposition of the GW signal by radius. We identify the regions and frequencies primarily associated with the post-bounce ringdown phase, as well as later, once nascent convection and accretion start exciting $p$-modes and $g$-modes in the PNS and gain region. The oscillations associated with ringdown in particular possess a distinctive structure, with standing waves oscillating at $580 \, \mathrm{Hz}$ around a node which closely tracks the large positive entropy gradient at the edge of the PNS convective region. These oscillations appear to be at least partially driven by forcing motions in the PNS surface region. In about $100 \, \mathrm{ms}$ the ringdown decays away, revealing more stochastic motions in a layer between the PNS convective and surface regions. Some quadrupolar activity is present in the gain region (in the form of convective plumes, etc.), however as a direct contribution to the GW signal, it is subdominant to the contribution from the PNS. The primary role of these convective plumes in generating GWs is to accrete onto the PNS to excite PNS oscillation modes.

Finally, we applied linear mode analysis to two of our simulations ($\mathrm{BF0\_noB}$ and SR1), to determine its suitability for rotating stars. 
While the analysis predicts mode frequency trajectories which match the respective GW emissions of models $\mathrm{BF0\_noB}$ and SR1, the emitting modes can only be reliably identified for the former, where the assumptions of spherical symmetry and staticity are fulfilled. For moderate rotation, only the mode frequency can be recovered, not the character of the mode, while for even faster rotation, no results can be obtained.

Our results underscore that there are still significant open questions about the  structure, excitation and coupling of PNS oscillation modes, especially in the case of rapid rotation. While rapidly rotating core-collapse supernova progenitors are rare, further research on these questions is required to prepare for a potential event in the Milky Way or its vicinity. The increased frequencies of the dominant high-frequency emission for the rotating case are plausible and strengthen the case for high-frequency GW detectors
\citep{ackley_20} as probes of supernova physics in the event of a detection. However, the frequency structure of rotating PNSs
clearly needs to be analyzed rigorously in the future by generalizing the techniques of the eigenmode analysis to the 2D case.
This needs to be combined with a better understanding of mode coupling and mode excitation to resolve when and if more exciting features in the GW signal, such as emission bursts due to resonant coupling, can occur.

\section*{Acknowledgements}
We thank the anonymous referee for their useful comments and suggestions.
BM acknowledges support from the Australian Research Council through Discovery Project DP240101786. The authors acknowledge computer time allocations from Astronomy Australia Limited's ASTAC scheme, the National Computational Merit Allocation Scheme (NCMAS), and from an Australasian Leadership Computing Grant. Some of this work was performed on the Gadi supercomputer with the assistance of resources and services from the National Computational Infrastructure (NCI), which is supported by the Australian Government.  Some of this work was performed on the OzSTAR national facility at Swinburne University of Technology. The OzSTAR program receives funding in part from the Astronomy National Collaborative Research Infrastructure Strategy (NCRIS) allocation provided by the Australian Government, and from the Victorian Higher Education State Investment Fund (VHESIF) provided by the Victorian Government.

%%%%%%%%%%%%%%%%%%%%%%%%%%%%%%%%%%%%%%%%%%%%%%%%%%
\section*{Data Availability}
The data from our simulations will be made available upon reasonable requests made to the authors.

\bibliographystyle{apsrev4-2-truc_auth}
\bibliography{bibliography}

%apsrev4-2.bst 2019-01-14 (MD) hand-edited version of apsrev4-1.bst
%Control: key (0)
%Control: author (72) initials jnrlst
%Control: editor formatted (1) identically to author
%Control: production of article title (-1) disabled
%Control: page (0) single
%Control: year (1) truncated
%Control: production of eprint (0) enabled
\begin{thebibliography}{81}%
\makeatletter
\providecommand \@ifxundefined [1]{%
 \@ifx{#1\undefined}
}%
\providecommand \@ifnum [1]{%
 \ifnum #1\expandafter \@firstoftwo
 \else \expandafter \@secondoftwo
 \fi
}%
\providecommand \@ifx [1]{%
 \ifx #1\expandafter \@firstoftwo
 \else \expandafter \@secondoftwo
 \fi
}%
\providecommand \natexlab [1]{#1}%
\providecommand \enquote  [1]{``#1''}%
\providecommand \bibnamefont  [1]{#1}%
\providecommand \bibfnamefont [1]{#1}%
\providecommand \citenamefont [1]{#1}%
\providecommand \href@noop [0]{\@secondoftwo}%
\providecommand \href [0]{\begingroup \@sanitize@url \@href}%
\providecommand \@href[1]{\@@startlink{#1}\@@href}%
\providecommand \@@href[1]{\endgroup#1\@@endlink}%
\providecommand \@sanitize@url [0]{\catcode `\\12\catcode `\$12\catcode `\&12\catcode `\#12\catcode `\^12\catcode `\_12\catcode `\%12\relax}%
\providecommand \@@startlink[1]{}%
\providecommand \@@endlink[0]{}%
\providecommand \url  [0]{\begingroup\@sanitize@url \@url }%
\providecommand \@url [1]{\endgroup\@href {#1}{\urlprefix }}%
\providecommand \urlprefix  [0]{URL }%
\providecommand \Eprint [0]{\href }%
\providecommand \doibase [0]{https://doi.org/}%
\providecommand \selectlanguage [0]{\@gobble}%
\providecommand \bibinfo  [0]{\@secondoftwo}%
\providecommand \bibfield  [0]{\@secondoftwo}%
\providecommand \translation [1]{[#1]}%
\providecommand \BibitemOpen [0]{}%
\providecommand \bibitemStop [0]{}%
\providecommand \bibitemNoStop [0]{.\EOS\space}%
\providecommand \EOS [0]{\spacefactor3000\relax}%
\providecommand \BibitemShut  [1]{\csname bibitem#1\endcsname}%
\let\auto@bib@innerbib\@empty
%</preamble>
\bibitem [{\citenamefont {{Hyper-Kamiokande Proto-Collaboration}}\ \emph {et~al.}(2018)\citenamefont {{Hyper-Kamiokande Proto-Collaboration}}, \citenamefont {{Abe}}, \citenamefont {{Abe}}, \citenamefont {{Aihara}}, \citenamefont {{Aimi}}, \citenamefont {{Akutsu}}, \citenamefont {{Andreopoulos}}, \citenamefont {{Anghel}}, \citenamefont {{Anthony}}, \citenamefont {{Antonova}}, \citenamefont {{Ashida}}, \citenamefont {{Aushev}}, \citenamefont {{Barbi}}, \citenamefont {{Barker}}, \citenamefont {{Barr}}, \citenamefont {{Beltrame}}, \citenamefont {{Berardi}}, \citenamefont {{Bergevin}}, \citenamefont {{Berkman}}, \citenamefont {{Berns}}, \citenamefont {{Berry}}, \citenamefont {{Bhadra}}, \citenamefont {{Bravo-Bergu{\~n}o}}, \citenamefont {{Blaszczyk}}, \citenamefont {{Blondel}}, \citenamefont {{Bolognesi}}, \citenamefont {{Boyd}}, \citenamefont {{Bravar}}, \citenamefont {{Bronner}}, \citenamefont {{Buizza Avanzini}}, \citenamefont {{Cafagna}}, \citenamefont {{Cole}}, \citenamefont {{Calland}}, \citenamefont
  {{Cao}}, \citenamefont {{Cartwright}}, \citenamefont {{Catanesi}}, \citenamefont {{Checchia}}, \citenamefont {{Chen-Wishart}}, \citenamefont {{Choi}}, \citenamefont {{Choi}}, \citenamefont {{Coleman}}, \citenamefont {{Collazuol}}, \citenamefont {{Cowan}}, \citenamefont {{Cremonesi}}, \citenamefont {{Dealtry}}, \citenamefont {{De Rosa}}, \citenamefont {{Densham}}, \citenamefont {{Dewhurst}}, \citenamefont {{Drakopoulou}}, \citenamefont {{Di Lodovico}}, \citenamefont {{Drapier}}, \citenamefont {{Dumarchez}}, \citenamefont {{Dunne}}, \citenamefont {{Dziewiecki}}, \citenamefont {{Emery}}, \citenamefont {{Esmaili}}, \citenamefont {{Evangelisti}}, \citenamefont {{Fernandez-Martinez}}, \citenamefont {{Feusels}}, \citenamefont {{Finch}}, \citenamefont {{Fiorentini}}, \citenamefont {{Fiorillo}}, \citenamefont {{Fitton}}, \citenamefont {{Frankiewicz}}, \citenamefont {{Friend}}, \citenamefont {{Fujii}}, \citenamefont {{Fukuda}}, \citenamefont {{Fukuda}}, \citenamefont {{Ganezer}}, \citenamefont {{Giganti}},
  \citenamefont {{Gonin}}, \citenamefont {{Grant}}, \citenamefont {{Gumplinger}}, \citenamefont {{Hadley}}, \citenamefont {{Hartfiel}}, \citenamefont {{Hartz}}, \citenamefont {{Hayato}}, \citenamefont {{Hayrapetyan}}, \citenamefont {{Hill}}, \citenamefont {{Hirota}}, \citenamefont {{Horiuchi}}, \citenamefont {{Ichikawa}}, \citenamefont {{Iijima}}, \citenamefont {{Ikeda}}, \citenamefont {{Imber}}, \citenamefont {{Inoue}}, \citenamefont {{Insler}}, \citenamefont {{Intonti}}, \citenamefont {{Ioannisian}}, \citenamefont {{Ishida}}, \citenamefont {{Ishino}}, \citenamefont {{Ishitsuka}}, \citenamefont {{Itow}}, \citenamefont {{Iwamoto}}, \citenamefont {{Izmaylov}}, \citenamefont {{Jamieson}}, \citenamefont {{Jang}}, \citenamefont {{Jang}}, \citenamefont {{Jeon}}, \citenamefont {{Jiang}}, \citenamefont {{Jonsson}}, \citenamefont {{Joo}}, \citenamefont {{Kaboth}}, \citenamefont {{Kachulis}}, \citenamefont {{Kajita}}, \citenamefont {{Kameda}}, \citenamefont {{Kataoka}}, \citenamefont {{Katori}}, \citenamefont
  {{Kayrapetyan}}, \citenamefont {{Kearns}}, \citenamefont {{Khabibullin}}, \citenamefont {{Khotjantsev}}, \citenamefont {{Kim}}, \citenamefont {{Kim}}, \citenamefont {{Kim}}, \citenamefont {{Kim}}, \citenamefont {{King}}, \citenamefont {{Kishimoto}}, \citenamefont {{Kobayashi}}, \citenamefont {{Koga}}, \citenamefont {{Konaka}}, \citenamefont {{Kormos}}, \citenamefont {{Koshio}}, \citenamefont {{Korzenev}}, \citenamefont {{Kowalik}}, \citenamefont {{Kropp}}, \citenamefont {{Kudenko}}, \citenamefont {{Kurjata}}, \citenamefont {{Kutter}}, \citenamefont {{Kuze}}, \citenamefont {{Labarga}}, \citenamefont {{Lagoda}}, \citenamefont {{Lasorak}}, \citenamefont {{Laveder}}, \citenamefont {{Lawe}}, \citenamefont {{Learned}}, \citenamefont {{Lim}}, \citenamefont {{Lindner}}, \citenamefont {{Litchfield}}, \citenamefont {{Longhin}}, \citenamefont {{Loverre}}, \citenamefont {{Lou}}, \citenamefont {{Ludovici}}, \citenamefont {{Ma}}, \citenamefont {{Magaletti}}, \citenamefont {{Mahn}}, \citenamefont {{Malek}}, \citenamefont
  {{Maret}}, \citenamefont {{Mariani}}, \citenamefont {{Martens}}, \citenamefont {{Marti}}, \citenamefont {{Martin}}, \citenamefont {{Marzec}}, \citenamefont {{Matsuno}}, \citenamefont {{Mazzucato}}, \citenamefont {{McCarthy}}, \citenamefont {{McCauley}}, \citenamefont {{McFarland}}, \citenamefont {{McGrew}}, \citenamefont {{Mefodiev}}, \citenamefont {{Mermod}}, \citenamefont {{Metelko}}, \citenamefont {{Mezzetto}}, \citenamefont {{Migenda}}, \citenamefont {{Mijakowski}}, \citenamefont {{Minakata}}, \citenamefont {{Minamino}}, \citenamefont {{Mine}}, \citenamefont {{Mineev}}, \citenamefont {{Mitra}}, \citenamefont {{Miura}}, \citenamefont {{Mochizuki}}, \citenamefont {{Monroe}}, \citenamefont {{Moon}}, \citenamefont {{Moriyama}}, \citenamefont {{Mueller}}, \citenamefont {{Muheim}}, \citenamefont {{Murase}}, \citenamefont {{Muto}}, \citenamefont {{Nakahata}}, \citenamefont {{Nakajima}}, \citenamefont {{Nakamura}}, \citenamefont {{Nakaya}}, \citenamefont {{Nakayama}}, \citenamefont {{Nantais}}, \citenamefont
  {{Needham}}, \citenamefont {{Nicholls}}, \citenamefont {{Nishimura}}, \citenamefont {{Noah}}, \citenamefont {{Nova}}, \citenamefont {{Nowak}}, \citenamefont {{Nunokawa}}, \citenamefont {{Obayashi}}, \citenamefont {{O'Keeffe}}, \citenamefont {{Okajima}}, \citenamefont {{Okumura}}, \citenamefont {{Onishchuk}}, \citenamefont {{O'Sullivan}},\ and\ \citenamefont {{O'Sullivan}}}]{HyperK:2018}%
  \BibitemOpen
  \bibfield  {author} {\bibinfo {author} {\bibnamefont {{Hyper-Kamiokande Proto-Collaboration}}}, \bibinfo {author} {\bibfnamefont {K.}~\bibnamefont {{Abe}}}, \bibinfo {author} {\bibfnamefont {K.}~\bibnamefont {{Abe}}}, \bibinfo {author} {\bibfnamefont {H.}~\bibnamefont {{Aihara}}}, \bibinfo {author} {\bibfnamefont {A.}~\bibnamefont {{Aimi}}}, \bibinfo {author} {\bibfnamefont {R.}~\bibnamefont {{Akutsu}}}, \bibinfo {author} {\bibfnamefont {C.}~\bibnamefont {{Andreopoulos}}}, \bibinfo {author} {\bibfnamefont {I.}~\bibnamefont {{Anghel}}}, \bibinfo {author} {\bibfnamefont {L.~H.~V.}\ \bibnamefont {{Anthony}}}, \bibinfo {author} {\bibfnamefont {M.}~\bibnamefont {{Antonova}}}, \bibinfo {author} {\bibfnamefont {Y.}~\bibnamefont {{Ashida}}}, \bibinfo {author} {\bibfnamefont {V.}~\bibnamefont {{Aushev}}}, \bibinfo {author} {\bibfnamefont {M.}~\bibnamefont {{Barbi}}}, \bibinfo {author} {\bibfnamefont {G.~J.}\ \bibnamefont {{Barker}}}, \emph {et~al.},\ }\href {https://doi.org/10.48550/arXiv.1805.04163} {\bibfield
  {journal} {\bibinfo  {journal} {arXiv e-prints}\ ,\ \bibinfo {eid} {arXiv:1805.04163}} (\bibinfo {year} {2018})}\BibitemShut {NoStop}%
\bibitem [{\citenamefont {{LIGO Scientific Collaboration}}\ \emph {et~al.}(2015)\citenamefont {{LIGO Scientific Collaboration}}, \citenamefont {{Aasi}}, \citenamefont {{Abbott}}, \citenamefont {{Abbott}}, \citenamefont {{Abbott}}, \citenamefont {{Abernathy}}, \citenamefont {{Ackley}}, \citenamefont {{Adams}}, \citenamefont {{Adams}}, \citenamefont {{Addesso}}, \citenamefont {{Adhikari}}, \citenamefont {{Adya}}, \citenamefont {{Affeldt}}, \citenamefont {{Aggarwal}}, \citenamefont {{Aguiar}}, \citenamefont {{Ain}}, \citenamefont {{Ajith}}, \citenamefont {{Alemic}}, \citenamefont {{Allen}}, \citenamefont {{Amariutei}}, \citenamefont {{Anderson}}, \citenamefont {{Anderson}}, \citenamefont {{Arai}}, \citenamefont {{Araya}}, \citenamefont {{Arceneaux}}, \citenamefont {{Areeda}}, \citenamefont {{Ashton}}, \citenamefont {{Ast}}, \citenamefont {{Aston}}, \citenamefont {{Aufmuth}}, \citenamefont {{Aulbert}}, \citenamefont {{Aylott}}, \citenamefont {{Babak}}, \citenamefont {{Baker}}, \citenamefont {{Ballmer}},
  \citenamefont {{Barayoga}}, \citenamefont {{Barbet}}, \citenamefont {{Barclay}}, \citenamefont {{Barish}}, \citenamefont {{Barker}}, \citenamefont {{Barr}}, \citenamefont {{Barsotti}}, \citenamefont {{Bartlett}}, \citenamefont {{Barton}}, \citenamefont {{Bartos}}, \citenamefont {{Bassiri}}, \citenamefont {{Batch}}, \citenamefont {{Baune}}, \citenamefont {{Behnke}}, \citenamefont {{Bell}}, \citenamefont {{Bell}}, \citenamefont {{Benacquista}}, \citenamefont {{Bergman}}, \citenamefont {{Bergmann}}, \citenamefont {{Berry}}, \citenamefont {{Betzwieser}}, \citenamefont {{Bhagwat}}, \citenamefont {{Bhandare}}, \citenamefont {{Bilenko}}, \citenamefont {{Billingsley}}, \citenamefont {{Birch}}, \citenamefont {{Biscans}}, \citenamefont {{Biwer}}, \citenamefont {{Blackburn}}, \citenamefont {{Blackburn}}, \citenamefont {{Blair}}, \citenamefont {{Blair}}, \citenamefont {{Bock}}, \citenamefont {{Bodiya}}, \citenamefont {{Bojtos}}, \citenamefont {{Bond}}, \citenamefont {{Bork}}, \citenamefont {{Born}}, \citenamefont
  {{Bose}}, \citenamefont {{Brady}}, \citenamefont {{Braginsky}}, \citenamefont {{Brau}}, \citenamefont {{Bridges}}, \citenamefont {{Brinkmann}}, \citenamefont {{Brooks}}, \citenamefont {{Brown}}, \citenamefont {{Brown}}, \citenamefont {{Brown}}, \citenamefont {{Buchman}}, \citenamefont {{Buikema}}, \citenamefont {{Buonanno}}, \citenamefont {{Cadonati}}, \citenamefont {{Calder{\'o}n Bustillo}}, \citenamefont {{Camp}}, \citenamefont {{Cannon}}, \citenamefont {{Cao}}, \citenamefont {{Capano}}, \citenamefont {{Caride}}, \citenamefont {{Caudill}}, \citenamefont {{Cavagli{\`a}}}, \citenamefont {{Cepeda}}, \citenamefont {{Chakraborty}}, \citenamefont {{Chalermsongsak}}, \citenamefont {{Chamberlin}}, \citenamefont {{Chao}}, \citenamefont {{Charlton}}, \citenamefont {{Chen}}, \citenamefont {{Cho}}, \citenamefont {{Cho}}, \citenamefont {{Chow}}, \citenamefont {{Christensen}}, \citenamefont {{Chu}}, \citenamefont {{Chung}}, \citenamefont {{Ciani}}, \citenamefont {{Clara}}, \citenamefont {{Clark}}, \citenamefont
  {{Collette}}, \citenamefont {{Cominsky}}, \citenamefont {{Constancio}}, \citenamefont {{Cook}}, \citenamefont {{Corbitt}}, \citenamefont {{Cornish}}, \citenamefont {{Corsi}}, \citenamefont {{Costa}}, \citenamefont {{Coughlin}}, \citenamefont {{Countryman}}, \citenamefont {{Couvares}}, \citenamefont {{Coward}}, \citenamefont {{Cowart}}, \citenamefont {{Coyne}}, \citenamefont {{Coyne}}, \citenamefont {{Craig}}, \citenamefont {{Creighton}}, \citenamefont {{Creighton}}, \citenamefont {{Cripe}}, \citenamefont {{Crowder}}, \citenamefont {{Cumming}}, \citenamefont {{Cunningham}}, \citenamefont {{Cutler}}, \citenamefont {{Dahl}}, \citenamefont {{Dal Canton}}, \citenamefont {{Damjanic}}, \citenamefont {{Danilishin}}, \citenamefont {{Danzmann}}, \citenamefont {{Dartez}}, \citenamefont {{Dave}}, \citenamefont {{Daveloza}}, \citenamefont {{Davies}}, \citenamefont {{Daw}}, \citenamefont {{DeBra}}, \citenamefont {{Del Pozzo}}, \citenamefont {{Denker}}, \citenamefont {{Dent}}, \citenamefont {{Dergachev}}, \citenamefont
  {{DeRosa}}, \citenamefont {{DeSalvo}}, \citenamefont {{Dhurandhar}}, \citenamefont {{D{\textasciiacute}{\i}az}}, \citenamefont {{Di Palma}}, \citenamefont {{Dojcinoski}}, \citenamefont {{Dominguez}}, \citenamefont {{Donovan}}, \citenamefont {{Dooley}}, \citenamefont {{Doravari}}, \citenamefont {{Douglas}}, \citenamefont {{Downes}}, \citenamefont {{Driggers}}, \citenamefont {{Du}}, \citenamefont {{Dwyer}}, \citenamefont {{Eberle}}, \citenamefont {{Edo}}, \citenamefont {{Edwards}}, \citenamefont {{Edwards}}, \citenamefont {{Effler}}, \citenamefont {{Eggenstein}}, \citenamefont {{Ehrens}}, \citenamefont {{Eichholz}}, \citenamefont {{Eikenberry}}, \citenamefont {{Essick}}, \citenamefont {{Etzel}}, \citenamefont {{Evans}}, \citenamefont {{Evans}}, \citenamefont {{Factourovich}}, \citenamefont {{Fairhurst}}, \citenamefont {{Fan}}, \citenamefont {{Fang}}, \citenamefont {{Farr}}, \citenamefont {{Farr}}, \citenamefont {{Favata}}, \citenamefont {{Fays}}, \citenamefont {{Fehrmann}}, \citenamefont {{Fejer}},
  \citenamefont {{Feldbaum}}, \citenamefont {{Ferreira}}, \citenamefont {{Fisher}}, \citenamefont {{Frei}}, \citenamefont {{Freise}}, \citenamefont {{Frey}}, \citenamefont {{Fricke}}, \citenamefont {{Fritschel}}, \citenamefont {{Frolov}}, \citenamefont {{Fuentes-Tapia}}, \citenamefont {{Fulda}}, \citenamefont {{Fyffe}},\ and\ \citenamefont {{Gair}}}]{aLIGO:2015}%
  \BibitemOpen
  \bibfield  {author} {\bibinfo {author} {\bibnamefont {{LIGO Scientific Collaboration}}}, \bibinfo {author} {\bibfnamefont {J.}~\bibnamefont {{Aasi}}}, \bibinfo {author} {\bibfnamefont {B.~P.}\ \bibnamefont {{Abbott}}}, \bibinfo {author} {\bibfnamefont {R.}~\bibnamefont {{Abbott}}}, \bibinfo {author} {\bibfnamefont {T.}~\bibnamefont {{Abbott}}}, \bibinfo {author} {\bibfnamefont {M.~R.}\ \bibnamefont {{Abernathy}}}, \bibinfo {author} {\bibfnamefont {K.}~\bibnamefont {{Ackley}}}, \bibinfo {author} {\bibfnamefont {C.}~\bibnamefont {{Adams}}}, \bibinfo {author} {\bibfnamefont {T.}~\bibnamefont {{Adams}}}, \bibinfo {author} {\bibfnamefont {P.}~\bibnamefont {{Addesso}}}, \bibinfo {author} {\bibfnamefont {R.~X.}\ \bibnamefont {{Adhikari}}}, \bibinfo {author} {\bibfnamefont {V.}~\bibnamefont {{Adya}}}, \bibinfo {author} {\bibfnamefont {C.}~\bibnamefont {{Affeldt}}}, \bibinfo {author} {\bibfnamefont {N.}~\bibnamefont {{Aggarwal}}}, \emph {et~al.},\ }\href {https://doi.org/10.1088/0264-9381/32/7/074001} {\bibfield
  {journal} {\bibinfo  {journal} {Classical and Quantum Gravity}\ }\textbf {\bibinfo {volume} {32}},\ \bibinfo {eid} {074001} (\bibinfo {year} {2015})}\BibitemShut {NoStop}%
\bibitem [{\citenamefont {{Powell}}\ and\ \citenamefont {{Lasky}}(2025)}]{Powell_Lasky:2025}%
  \BibitemOpen
  \bibfield  {author} {\bibinfo {author} {\bibfnamefont {J.}~\bibnamefont {{Powell}}}\ and\ \bibinfo {author} {\bibfnamefont {P.~D.}\ \bibnamefont {{Lasky}}},\ }\href {https://doi.org/10.1017/pasa.2025.10} {\bibfield  {journal} {\bibinfo  {journal} {\pasa}\ }\textbf {\bibinfo {volume} {42}},\ \bibinfo {eid} {e030} (\bibinfo {year} {2025})}\BibitemShut {NoStop}%
\bibitem [{\citenamefont {{Abbott}}\ \emph {et~al.}(2023)\citenamefont {{Abbott}}, \citenamefont {{Abbott}}, \citenamefont {{Acernese}}, \citenamefont {{Ackley}}, \citenamefont {{Adams}}, \citenamefont {{Adhikari}}, \citenamefont {{Adhikari}}, \citenamefont {{Adya}}, \citenamefont {{Affeldt}}, \citenamefont {{Agarwal}}, \citenamefont {{Agathos}}, \citenamefont {{Agatsuma}}, \citenamefont {{Aggarwal}}, \citenamefont {{Aguiar}}, \citenamefont {{Aiello}}, \citenamefont {{Ain}}, \citenamefont {{Ajith}}, \citenamefont {{Akcay}}, \citenamefont {{Akutsu}}, \citenamefont {{Albanesi}}, \citenamefont {{Allocca}}, \citenamefont {{Altin}}, \citenamefont {{Amato}}, \citenamefont {{Anand}}, \citenamefont {{Anand}}, \citenamefont {{Ananyeva}}, \citenamefont {{Anderson}}, \citenamefont {{Anderson}}, \citenamefont {{Ando}}, \citenamefont {{Andrade}}, \citenamefont {{Andres}}, \citenamefont {{Andri{\'c}}}, \citenamefont {{Angelova}}, \citenamefont {{Ansoldi}}, \citenamefont {{Antelis}}, \citenamefont {{Antier}}, \citenamefont
  {{Appert}}, \citenamefont {{Arai}}, \citenamefont {{Arai}}, \citenamefont {{Arai}}, \citenamefont {{Araki}}, \citenamefont {{Araya}}, \citenamefont {{Araya}}, \citenamefont {{Areeda}}, \citenamefont {{Ar{\`e}ne}}, \citenamefont {{Aritomi}}, \citenamefont {{Arnaud}}, \citenamefont {{Arogeti}}, \citenamefont {{Aronson}}, \citenamefont {{Arun}}, \citenamefont {{Asada}}, \citenamefont {{Asali}}, \citenamefont {{Ashton}}, \citenamefont {{Aso}}, \citenamefont {{Assiduo}}, \citenamefont {{Aston}}, \citenamefont {{Astone}}, \citenamefont {{Aubin}}, \citenamefont {{Austin}}, \citenamefont {{Babak}}, \citenamefont {{Badaracco}}, \citenamefont {{Bader}}, \citenamefont {{Badger}}, \citenamefont {{Bae}}, \citenamefont {{Bae}}, \citenamefont {{Baer}}, \citenamefont {{Bagnasco}}, \citenamefont {{Bai}}, \citenamefont {{Baiotti}}, \citenamefont {{Baird}}, \citenamefont {{Bajpai}}, \citenamefont {{Ball}}, \citenamefont {{Ballardin}}, \citenamefont {{Ballmer}}, \citenamefont {{Balsamo}}, \citenamefont {{Baltus}},
  \citenamefont {{Banagiri}}, \citenamefont {{Bankar}}, \citenamefont {{Barayoga}}, \citenamefont {{Barbieri}}, \citenamefont {{Barish}}, \citenamefont {{Barker}}, \citenamefont {{Barneo}}, \citenamefont {{Barone}}, \citenamefont {{Barr}}, \citenamefont {{Barsotti}}, \citenamefont {{Barsuglia}}, \citenamefont {{Barta}}, \citenamefont {{Bartlett}}, \citenamefont {{Barton}}, \citenamefont {{Bartos}}, \citenamefont {{Bassiri}}, \citenamefont {{Basti}}, \citenamefont {{Bawaj}}, \citenamefont {{Bayley}}, \citenamefont {{Baylor}}, \citenamefont {{Bazzan}}, \citenamefont {{B{\'e}csy}}, \citenamefont {{Bedakihale}}, \citenamefont {{Bejger}}, \citenamefont {{Belahcene}}, \citenamefont {{Benedetto}}, \citenamefont {{Beniwal}}, \citenamefont {{Bennett}}, \citenamefont {{Bentley}}, \citenamefont {{Benyaala}}, \citenamefont {{Bergamin}}, \citenamefont {{Berger}}, \citenamefont {{Bernuzzi}}, \citenamefont {{Berry}}, \citenamefont {{Bersanetti}}, \citenamefont {{Bertolini}}, \citenamefont {{Betzwieser}}, \citenamefont
  {{Beveridge}}, \citenamefont {{Bhandare}}, \citenamefont {{Bhardwaj}}, \citenamefont {{Bhattacharjee}}, \citenamefont {{Bhaumik}}, \citenamefont {{Bilenko}}, \citenamefont {{Billingsley}}, \citenamefont {{Bini}}, \citenamefont {{Birney}}, \citenamefont {{Birnholtz}}, \citenamefont {{Biscans}}, \citenamefont {{Bischi}}, \citenamefont {{Biscoveanu}}, \citenamefont {{Bisht}}, \citenamefont {{Biswas}}, \citenamefont {{Bitossi}}, \citenamefont {{Bizouard}}, \citenamefont {{Blackburn}}, \citenamefont {{Blair}}, \citenamefont {{Blair}}, \citenamefont {{Blair}}, \citenamefont {{Bobba}}, \citenamefont {{Bode}}, \citenamefont {{Boer}}, \citenamefont {{Bogaert}}, \citenamefont {{Boldrini}}, \citenamefont {{Bonavena}}, \citenamefont {{Bondu}}, \citenamefont {{Bonilla}}, \citenamefont {{Bonnand}}, \citenamefont {{Booker}}, \citenamefont {{Boom}}, \citenamefont {{Bork}}, \citenamefont {{Boschi}}, \citenamefont {{Bose}}, \citenamefont {{Bose}}, \citenamefont {{Bossilkov}}, \citenamefont {{Boudart}}, \citenamefont
  {{Bouffanais}}, \citenamefont {{Bozzi}}, \citenamefont {{Bradaschia}}, \citenamefont {{Brady}}, \citenamefont {{Bramley}}, \citenamefont {{Branch}}, \citenamefont {{Branchesi}}, \citenamefont {{Brandt}}, \citenamefont {{Brau}}, \citenamefont {{Breschi}}, \citenamefont {{Briant}}, \citenamefont {{Briggs}}, \citenamefont {{Brillet}}, \citenamefont {{Brinkmann}}, \citenamefont {{Brockill}}, \citenamefont {{Brooks}}, \citenamefont {{Brooks}}, \citenamefont {{Brown}}, \citenamefont {{Brunett}}, \citenamefont {{Bruno}}, \citenamefont {{Bruntz}}, \citenamefont {{Bryant}}, \citenamefont {{Bulik}}, \citenamefont {{Bulten}}, \citenamefont {{Buonanno}}, \citenamefont {{Buscicchio}}, \citenamefont {{Buskulic}}, \citenamefont {{Buy}}, \citenamefont {{Byer}}, \citenamefont {{Davies}}, \citenamefont {{Cadonati}}, \citenamefont {{Cagnoli}}, \citenamefont {{Cahillane}}, \citenamefont {{Bustillo}}, \citenamefont {{Callaghan}}, \citenamefont {{Callister}}, \citenamefont {{Calloni}}, \citenamefont {{Cameron}}, \citenamefont
  {{Camp}}, \citenamefont {{Canepa}}, \citenamefont {{Canevarolo}}, \citenamefont {{Cannavacciuolo}}, \citenamefont {{Cannon}}, \citenamefont {{Cao}}, \citenamefont {{Cao}}, \citenamefont {{Capocasa}}, \citenamefont {{Capote}}, \citenamefont {{Carapella}},\ and\ \citenamefont {{Carbognani}}}]{gwtc3:2023}%
  \BibitemOpen
  \bibfield  {author} {\bibinfo {author} {\bibfnamefont {R.}~\bibnamefont {{Abbott}}}, \bibinfo {author} {\bibfnamefont {T.~D.}\ \bibnamefont {{Abbott}}}, \bibinfo {author} {\bibfnamefont {F.}~\bibnamefont {{Acernese}}}, \bibinfo {author} {\bibfnamefont {K.}~\bibnamefont {{Ackley}}}, \bibinfo {author} {\bibfnamefont {C.}~\bibnamefont {{Adams}}}, \bibinfo {author} {\bibfnamefont {N.}~\bibnamefont {{Adhikari}}}, \bibinfo {author} {\bibfnamefont {R.~X.}\ \bibnamefont {{Adhikari}}}, \bibinfo {author} {\bibfnamefont {V.~B.}\ \bibnamefont {{Adya}}}, \bibinfo {author} {\bibfnamefont {C.}~\bibnamefont {{Affeldt}}}, \bibinfo {author} {\bibfnamefont {D.}~\bibnamefont {{Agarwal}}}, \bibinfo {author} {\bibfnamefont {M.}~\bibnamefont {{Agathos}}}, \bibinfo {author} {\bibfnamefont {K.}~\bibnamefont {{Agatsuma}}}, \bibinfo {author} {\bibfnamefont {N.}~\bibnamefont {{Aggarwal}}}, \bibinfo {author} {\bibfnamefont {O.~D.}\ \bibnamefont {{Aguiar}}}, \emph {et~al.},\ }\href {https://doi.org/10.1103/PhysRevX.13.041039} {\bibfield
   {journal} {\bibinfo  {journal} {Physical Review X}\ }\textbf {\bibinfo {volume} {13}},\ \bibinfo {eid} {041039} (\bibinfo {year} {2023})}\BibitemShut {NoStop}%
\bibitem [{\citenamefont {{Murphy}}\ \emph {et~al.}(2009)\citenamefont {{Murphy}}, \citenamefont {{Ott}},\ and\ \citenamefont {{Burrows}}}]{Murphy_Ott_Burrows:2009}%
  \BibitemOpen
  \bibfield  {author} {\bibinfo {author} {\bibfnamefont {J.~W.}\ \bibnamefont {{Murphy}}}, \bibinfo {author} {\bibfnamefont {C.~D.}\ \bibnamefont {{Ott}}},\ and\ \bibinfo {author} {\bibfnamefont {A.}~\bibnamefont {{Burrows}}},\ }\href {https://doi.org/10.1088/0004-637X/707/2/1173} {\bibfield  {journal} {\bibinfo  {journal} {\apj}\ }\textbf {\bibinfo {volume} {707}},\ \bibinfo {pages} {1173} (\bibinfo {year} {2009})}\BibitemShut {NoStop}%
\bibitem [{\citenamefont {{M{\"u}ller}}\ \emph {et~al.}(2013)\citenamefont {{M{\"u}ller}}, \citenamefont {{Janka}},\ and\ \citenamefont {{Marek}}}]{Mueller_Janka_Marek:2013}%
  \BibitemOpen
  \bibfield  {author} {\bibinfo {author} {\bibfnamefont {B.}~\bibnamefont {{M{\"u}ller}}}, \bibinfo {author} {\bibfnamefont {H.-T.}\ \bibnamefont {{Janka}}},\ and\ \bibinfo {author} {\bibfnamefont {A.}~\bibnamefont {{Marek}}},\ }\href {https://doi.org/10.1088/0004-637X/766/1/43} {\bibfield  {journal} {\bibinfo  {journal} {\apj}\ }\textbf {\bibinfo {volume} {766}},\ \bibinfo {eid} {43} (\bibinfo {year} {2013})}\BibitemShut {NoStop}%
\bibitem [{\citenamefont {{Jardine}}\ \emph {et~al.}(2022)\citenamefont {{Jardine}}, \citenamefont {{Powell}},\ and\ \citenamefont {{M{\"u}ller}}}]{Jardine_Powell_Mueller:2022}%
  \BibitemOpen
  \bibfield  {author} {\bibinfo {author} {\bibfnamefont {R.}~\bibnamefont {{Jardine}}}, \bibinfo {author} {\bibfnamefont {J.}~\bibnamefont {{Powell}}},\ and\ \bibinfo {author} {\bibfnamefont {B.}~\bibnamefont {{M{\"u}ller}}},\ }\href {https://doi.org/10.1093/mnras/stab3763} {\bibfield  {journal} {\bibinfo  {journal} {\mnras}\ }\textbf {\bibinfo {volume} {510}},\ \bibinfo {pages} {5535} (\bibinfo {year} {2022})}\BibitemShut {NoStop}%
\bibitem [{\citenamefont {{Jakobus}}\ \emph {et~al.}(2023)\citenamefont {{Jakobus}}, \citenamefont {{M{\"u}ller}}, \citenamefont {{Heger}}, \citenamefont {{Zha}}, \citenamefont {{Powell}}, \citenamefont {{Motornenko}}, \citenamefont {{Steinheimer}},\ and\ \citenamefont {{St{\"o}cker}}}]{Jakobus_et_al:2023}%
  \BibitemOpen
  \bibfield  {author} {\bibinfo {author} {\bibfnamefont {P.}~\bibnamefont {{Jakobus}}}, \bibinfo {author} {\bibfnamefont {B.}~\bibnamefont {{M{\"u}ller}}}, \bibinfo {author} {\bibfnamefont {A.}~\bibnamefont {{Heger}}}, \bibinfo {author} {\bibfnamefont {S.}~\bibnamefont {{Zha}}}, \bibinfo {author} {\bibfnamefont {J.}~\bibnamefont {{Powell}}}, \bibinfo {author} {\bibfnamefont {A.}~\bibnamefont {{Motornenko}}}, \bibinfo {author} {\bibfnamefont {J.}~\bibnamefont {{Steinheimer}}},\ and\ \bibinfo {author} {\bibfnamefont {H.}~\bibnamefont {{St{\"o}cker}}},\ }\href {https://doi.org/10.1103/PhysRevLett.131.191201} {\bibfield  {journal} {\bibinfo  {journal} {\prl}\ }\textbf {\bibinfo {volume} {131}},\ \bibinfo {eid} {191201} (\bibinfo {year} {2023})}\BibitemShut {NoStop}%
\bibitem [{\citenamefont {{Yakunin}}\ \emph {et~al.}(2015)\citenamefont {{Yakunin}}, \citenamefont {{Mezzacappa}}, \citenamefont {{Marronetti}}, \citenamefont {{Yoshida}}, \citenamefont {{Bruenn}}, \citenamefont {{Hix}}, \citenamefont {{Lentz}}, \citenamefont {{Bronson Messer}}, \citenamefont {{Harris}}, \citenamefont {{Endeve}}, \citenamefont {{Blondin}},\ and\ \citenamefont {{Lingerfelt}}}]{Yakunin_et_al:2015}%
  \BibitemOpen
  \bibfield  {author} {\bibinfo {author} {\bibfnamefont {K.~N.}\ \bibnamefont {{Yakunin}}}, \bibinfo {author} {\bibfnamefont {A.}~\bibnamefont {{Mezzacappa}}}, \bibinfo {author} {\bibfnamefont {P.}~\bibnamefont {{Marronetti}}}, \bibinfo {author} {\bibfnamefont {S.}~\bibnamefont {{Yoshida}}}, \bibinfo {author} {\bibfnamefont {S.~W.}\ \bibnamefont {{Bruenn}}}, \bibinfo {author} {\bibfnamefont {W.~R.}\ \bibnamefont {{Hix}}}, \bibinfo {author} {\bibfnamefont {E.~J.}\ \bibnamefont {{Lentz}}}, \bibinfo {author} {\bibfnamefont {O.~E.}\ \bibnamefont {{Bronson Messer}}}, \bibinfo {author} {\bibfnamefont {J.~A.}\ \bibnamefont {{Harris}}}, \bibinfo {author} {\bibfnamefont {E.}~\bibnamefont {{Endeve}}}, \bibinfo {author} {\bibfnamefont {J.~M.}\ \bibnamefont {{Blondin}}},\ and\ \bibinfo {author} {\bibfnamefont {E.~J.}\ \bibnamefont {{Lingerfelt}}},\ }\href {https://doi.org/10.1103/PhysRevD.92.084040} {\bibfield  {journal} {\bibinfo  {journal} {\prd}\ }\textbf {\bibinfo {volume} {92}},\ \bibinfo {eid} {084040} (\bibinfo
  {year} {2015})}\BibitemShut {NoStop}%
\bibitem [{\citenamefont {{Richers}}\ \emph {et~al.}(2017)\citenamefont {{Richers}}, \citenamefont {{Ott}}, \citenamefont {{Abdikamalov}}, \citenamefont {{O'Connor}},\ and\ \citenamefont {{Sullivan}}}]{Richers_et_al:2017}%
  \BibitemOpen
  \bibfield  {author} {\bibinfo {author} {\bibfnamefont {S.}~\bibnamefont {{Richers}}}, \bibinfo {author} {\bibfnamefont {C.~D.}\ \bibnamefont {{Ott}}}, \bibinfo {author} {\bibfnamefont {E.}~\bibnamefont {{Abdikamalov}}}, \bibinfo {author} {\bibfnamefont {E.}~\bibnamefont {{O'Connor}}},\ and\ \bibinfo {author} {\bibfnamefont {C.}~\bibnamefont {{Sullivan}}},\ }\href {https://doi.org/10.1103/PhysRevD.95.063019} {\bibfield  {journal} {\bibinfo  {journal} {\prd}\ }\textbf {\bibinfo {volume} {95}},\ \bibinfo {eid} {063019} (\bibinfo {year} {2017})}\BibitemShut {NoStop}%
\bibitem [{\citenamefont {{Andersen}}\ \emph {et~al.}(2021)\citenamefont {{Andersen}}, \citenamefont {{Zha}}, \citenamefont {{da Silva Schneider}}, \citenamefont {{Betranhandy}}, \citenamefont {{Couch}},\ and\ \citenamefont {{O'Connor}}}]{EggenbergerAndersen_et_al:2021}%
  \BibitemOpen
  \bibfield  {author} {\bibinfo {author} {\bibfnamefont {O.}~\bibnamefont {{Andersen}}}, \bibinfo {author} {\bibfnamefont {S.}~\bibnamefont {{Zha}}}, \bibinfo {author} {\bibfnamefont {A.}~\bibnamefont {{da Silva Schneider}}}, \bibinfo {author} {\bibfnamefont {A.}~\bibnamefont {{Betranhandy}}}, \bibinfo {author} {\bibfnamefont {S.~M.}\ \bibnamefont {{Couch}}},\ and\ \bibinfo {author} {\bibfnamefont {E.~P.}\ \bibnamefont {{O'Connor}}},\ }\href {https://doi.org/10.3847/1538-4357/ac294c} {\bibfield  {journal} {\bibinfo  {journal} {\apj}\ }\textbf {\bibinfo {volume} {923}},\ \bibinfo {eid} {201} (\bibinfo {year} {2021})}\BibitemShut {NoStop}%
\bibitem [{\citenamefont {{Ott}}\ \emph {et~al.}(2013)\citenamefont {{Ott}}, \citenamefont {{Abdikamalov}}, \citenamefont {{M{\"o}sta}}, \citenamefont {{Haas}}, \citenamefont {{Drasco}}, \citenamefont {{O'Connor}}, \citenamefont {{Reisswig}}, \citenamefont {{Meakin}},\ and\ \citenamefont {{Schnetter}}}]{Ott_et_al:2013}%
  \BibitemOpen
  \bibfield  {author} {\bibinfo {author} {\bibfnamefont {C.~D.}\ \bibnamefont {{Ott}}}, \bibinfo {author} {\bibfnamefont {E.}~\bibnamefont {{Abdikamalov}}}, \bibinfo {author} {\bibfnamefont {P.}~\bibnamefont {{M{\"o}sta}}}, \bibinfo {author} {\bibfnamefont {R.}~\bibnamefont {{Haas}}}, \bibinfo {author} {\bibfnamefont {S.}~\bibnamefont {{Drasco}}}, \bibinfo {author} {\bibfnamefont {E.~P.}\ \bibnamefont {{O'Connor}}}, \bibinfo {author} {\bibfnamefont {C.}~\bibnamefont {{Reisswig}}}, \bibinfo {author} {\bibfnamefont {C.~A.}\ \bibnamefont {{Meakin}}},\ and\ \bibinfo {author} {\bibfnamefont {E.}~\bibnamefont {{Schnetter}}},\ }\href {https://doi.org/10.1088/0004-637X/768/2/115} {\bibfield  {journal} {\bibinfo  {journal} {\apj}\ }\textbf {\bibinfo {volume} {768}},\ \bibinfo {eid} {115} (\bibinfo {year} {2013})}\BibitemShut {NoStop}%
\bibitem [{\citenamefont {{Kuroda}}\ \emph {et~al.}(2016)\citenamefont {{Kuroda}}, \citenamefont {{Kotake}},\ and\ \citenamefont {{Takiwaki}}}]{Kuroda_Kotake_Takiwaki:2016}%
  \BibitemOpen
  \bibfield  {author} {\bibinfo {author} {\bibfnamefont {T.}~\bibnamefont {{Kuroda}}}, \bibinfo {author} {\bibfnamefont {K.}~\bibnamefont {{Kotake}}},\ and\ \bibinfo {author} {\bibfnamefont {T.}~\bibnamefont {{Takiwaki}}},\ }\href {https://doi.org/10.3847/2041-8205/829/1/L14} {\bibfield  {journal} {\bibinfo  {journal} {\apjl}\ }\textbf {\bibinfo {volume} {829}},\ \bibinfo {eid} {L14} (\bibinfo {year} {2016})}\BibitemShut {NoStop}%
\bibitem [{\citenamefont {{Andresen}}\ \emph {et~al.}(2017)\citenamefont {{Andresen}}, \citenamefont {{M{\"u}ller}}, \citenamefont {{M{\"u}ller}},\ and\ \citenamefont {{Janka}}}]{Andresen_Mueller_Mueller_Janka:2017}%
  \BibitemOpen
  \bibfield  {author} {\bibinfo {author} {\bibfnamefont {H.}~\bibnamefont {{Andresen}}}, \bibinfo {author} {\bibfnamefont {B.}~\bibnamefont {{M{\"u}ller}}}, \bibinfo {author} {\bibfnamefont {E.}~\bibnamefont {{M{\"u}ller}}},\ and\ \bibinfo {author} {\bibfnamefont {H.~T.}\ \bibnamefont {{Janka}}},\ }\href {https://doi.org/10.1093/mnras/stx618} {\bibfield  {journal} {\bibinfo  {journal} {\mnras}\ }\textbf {\bibinfo {volume} {468}},\ \bibinfo {pages} {2032} (\bibinfo {year} {2017})}\BibitemShut {NoStop}%
\bibitem [{\citenamefont {{Andresen}}\ \emph {et~al.}(2019)\citenamefont {{Andresen}}, \citenamefont {{M{\"u}ller}}, \citenamefont {{Janka}}, \citenamefont {{Summa}}, \citenamefont {{Gill}},\ and\ \citenamefont {{Zanolin}}}]{Andresen_et_al:2019}%
  \BibitemOpen
  \bibfield  {author} {\bibinfo {author} {\bibfnamefont {H.}~\bibnamefont {{Andresen}}}, \bibinfo {author} {\bibfnamefont {E.}~\bibnamefont {{M{\"u}ller}}}, \bibinfo {author} {\bibfnamefont {H.~T.}\ \bibnamefont {{Janka}}}, \bibinfo {author} {\bibfnamefont {A.}~\bibnamefont {{Summa}}}, \bibinfo {author} {\bibfnamefont {K.}~\bibnamefont {{Gill}}},\ and\ \bibinfo {author} {\bibfnamefont {M.}~\bibnamefont {{Zanolin}}},\ }\href {https://doi.org/10.1093/mnras/stz990} {\bibfield  {journal} {\bibinfo  {journal} {\mnras}\ }\textbf {\bibinfo {volume} {486}},\ \bibinfo {pages} {2238} (\bibinfo {year} {2019})}\BibitemShut {NoStop}%
\bibitem [{\citenamefont {{O'Connor}}\ and\ \citenamefont {{Couch}}(2018)}]{oConnor_Couch:2018}%
  \BibitemOpen
  \bibfield  {author} {\bibinfo {author} {\bibfnamefont {E.~P.}\ \bibnamefont {{O'Connor}}}\ and\ \bibinfo {author} {\bibfnamefont {S.~M.}\ \bibnamefont {{Couch}}},\ }\href {https://doi.org/10.3847/1538-4357/aadcf7} {\bibfield  {journal} {\bibinfo  {journal} {\apj}\ }\textbf {\bibinfo {volume} {865}},\ \bibinfo {eid} {81} (\bibinfo {year} {2018})}\BibitemShut {NoStop}%
\bibitem [{\citenamefont {{Powell}}\ and\ \citenamefont {{M{\"u}ller}}(2020)}]{Powell_Mueller:2020}%
  \BibitemOpen
  \bibfield  {author} {\bibinfo {author} {\bibfnamefont {J.}~\bibnamefont {{Powell}}}\ and\ \bibinfo {author} {\bibfnamefont {B.}~\bibnamefont {{M{\"u}ller}}},\ }\href {https://doi.org/10.1093/mnras/staa1048} {\bibfield  {journal} {\bibinfo  {journal} {\mnras}\ }\textbf {\bibinfo {volume} {494}},\ \bibinfo {pages} {4665} (\bibinfo {year} {2020})}\BibitemShut {NoStop}%
\bibitem [{\citenamefont {{Powell}}\ and\ \citenamefont {{M{\"u}ller}}(2024)}]{Powell_Mueller:2024}%
  \BibitemOpen
  \bibfield  {author} {\bibinfo {author} {\bibfnamefont {J.}~\bibnamefont {{Powell}}}\ and\ \bibinfo {author} {\bibfnamefont {B.}~\bibnamefont {{M{\"u}ller}}},\ }\href {https://doi.org/10.1093/mnras/stae1731} {\bibfield  {journal} {\bibinfo  {journal} {\mnras}\ }\textbf {\bibinfo {volume} {532}},\ \bibinfo {pages} {4326} (\bibinfo {year} {2024})}\BibitemShut {NoStop}%
\bibitem [{\citenamefont {{Radice}}\ \emph {et~al.}(2019)\citenamefont {{Radice}}, \citenamefont {{Morozova}}, \citenamefont {{Burrows}}, \citenamefont {{Vartanyan}},\ and\ \citenamefont {{Nagakura}}}]{Radice_et_al:2019}%
  \BibitemOpen
  \bibfield  {author} {\bibinfo {author} {\bibfnamefont {D.}~\bibnamefont {{Radice}}}, \bibinfo {author} {\bibfnamefont {V.}~\bibnamefont {{Morozova}}}, \bibinfo {author} {\bibfnamefont {A.}~\bibnamefont {{Burrows}}}, \bibinfo {author} {\bibfnamefont {D.}~\bibnamefont {{Vartanyan}}},\ and\ \bibinfo {author} {\bibfnamefont {H.}~\bibnamefont {{Nagakura}}},\ }\href {https://doi.org/10.3847/2041-8213/ab191a} {\bibfield  {journal} {\bibinfo  {journal} {\apjl}\ }\textbf {\bibinfo {volume} {876}},\ \bibinfo {eid} {L9} (\bibinfo {year} {2019})}\BibitemShut {NoStop}%
\bibitem [{\citenamefont {{Vartanyan}}\ \emph {et~al.}(2023)\citenamefont {{Vartanyan}}, \citenamefont {{Burrows}}, \citenamefont {{Wang}}, \citenamefont {{Coleman}},\ and\ \citenamefont {{White}}}]{Vartanyan_et_al:2023}%
  \BibitemOpen
  \bibfield  {author} {\bibinfo {author} {\bibfnamefont {D.}~\bibnamefont {{Vartanyan}}}, \bibinfo {author} {\bibfnamefont {A.}~\bibnamefont {{Burrows}}}, \bibinfo {author} {\bibfnamefont {T.}~\bibnamefont {{Wang}}}, \bibinfo {author} {\bibfnamefont {M.~S.~B.}\ \bibnamefont {{Coleman}}},\ and\ \bibinfo {author} {\bibfnamefont {C.~J.}\ \bibnamefont {{White}}},\ }\href {https://doi.org/10.1103/PhysRevD.107.103015} {\bibfield  {journal} {\bibinfo  {journal} {\prd}\ }\textbf {\bibinfo {volume} {107}},\ \bibinfo {eid} {103015} (\bibinfo {year} {2023})}\BibitemShut {NoStop}%
\bibitem [{\citenamefont {{Andresen}}\ \emph {et~al.}(2021)\citenamefont {{Andresen}}, \citenamefont {{Glas}},\ and\ \citenamefont {{Janka}}}]{Andresen_Glas_Janka:2021}%
  \BibitemOpen
  \bibfield  {author} {\bibinfo {author} {\bibfnamefont {H.}~\bibnamefont {{Andresen}}}, \bibinfo {author} {\bibfnamefont {R.}~\bibnamefont {{Glas}}},\ and\ \bibinfo {author} {\bibfnamefont {H.~T.}\ \bibnamefont {{Janka}}},\ }\href {https://doi.org/10.1093/mnras/stab675} {\bibfield  {journal} {\bibinfo  {journal} {\mnras}\ }\textbf {\bibinfo {volume} {503}},\ \bibinfo {pages} {3552} (\bibinfo {year} {2021})}\BibitemShut {NoStop}%
\bibitem [{\citenamefont {{Mezzacappa}}\ \emph {et~al.}(2020)\citenamefont {{Mezzacappa}}, \citenamefont {{Marronetti}}, \citenamefont {{Landfield}}, \citenamefont {{Lentz}}, \citenamefont {{Yakunin}}, \citenamefont {{Bruenn}}, \citenamefont {{Hix}}, \citenamefont {{Messer}}, \citenamefont {{Endeve}}, \citenamefont {{Blondin}},\ and\ \citenamefont {{Harris}}}]{Mezzacappa_et_al:2020}%
  \BibitemOpen
  \bibfield  {author} {\bibinfo {author} {\bibfnamefont {A.}~\bibnamefont {{Mezzacappa}}}, \bibinfo {author} {\bibfnamefont {P.}~\bibnamefont {{Marronetti}}}, \bibinfo {author} {\bibfnamefont {R.~E.}\ \bibnamefont {{Landfield}}}, \bibinfo {author} {\bibfnamefont {E.~J.}\ \bibnamefont {{Lentz}}}, \bibinfo {author} {\bibfnamefont {K.~N.}\ \bibnamefont {{Yakunin}}}, \bibinfo {author} {\bibfnamefont {S.~W.}\ \bibnamefont {{Bruenn}}}, \bibinfo {author} {\bibfnamefont {W.~R.}\ \bibnamefont {{Hix}}}, \bibinfo {author} {\bibfnamefont {O.~E.~B.}\ \bibnamefont {{Messer}}}, \bibinfo {author} {\bibfnamefont {E.}~\bibnamefont {{Endeve}}}, \bibinfo {author} {\bibfnamefont {J.~M.}\ \bibnamefont {{Blondin}}},\ and\ \bibinfo {author} {\bibfnamefont {J.~A.}\ \bibnamefont {{Harris}}},\ }\href {https://doi.org/10.1103/PhysRevD.102.023027} {\bibfield  {journal} {\bibinfo  {journal} {\prd}\ }\textbf {\bibinfo {volume} {102}},\ \bibinfo {eid} {023027} (\bibinfo {year} {2020})}\BibitemShut {NoStop}%
\bibitem [{\citenamefont {{Powell}}\ and\ \citenamefont {{M{\"u}ller}}(2025)}]{Powell_Mueller:2025}%
  \BibitemOpen
  \bibfield  {author} {\bibinfo {author} {\bibfnamefont {J.}~\bibnamefont {{Powell}}}\ and\ \bibinfo {author} {\bibfnamefont {B.}~\bibnamefont {{M{\"u}ller}}},\ }\href {https://doi.org/10.48550/arXiv.2506.03581} {\bibfield  {journal} {\bibinfo  {journal} {arXiv e-prints}\ ,\ \bibinfo {eid} {arXiv:2506.03581}} (\bibinfo {year} {2025})}\BibitemShut {NoStop}%
\bibitem [{\citenamefont {{Powell}}\ and\ \citenamefont {{M{\"u}ller}}(2022)}]{Powell_Mueller:2022}%
  \BibitemOpen
  \bibfield  {author} {\bibinfo {author} {\bibfnamefont {J.}~\bibnamefont {{Powell}}}\ and\ \bibinfo {author} {\bibfnamefont {B.}~\bibnamefont {{M{\"u}ller}}},\ }\href {https://doi.org/10.1103/PhysRevD.105.063018} {\bibfield  {journal} {\bibinfo  {journal} {\prd}\ }\textbf {\bibinfo {volume} {105}},\ \bibinfo {eid} {063018} (\bibinfo {year} {2022})}\BibitemShut {NoStop}%
\bibitem [{\citenamefont {{Bruel}}\ \emph {et~al.}(2023)\citenamefont {{Bruel}}, \citenamefont {{Bizouard}}, \citenamefont {{Obergaulinger}}, \citenamefont {{Maturana-Russel}}, \citenamefont {{Torres-Forn{\'e}}}, \citenamefont {{Cerd{\'a}-Dur{\'a}n}}, \citenamefont {{Christensen}}, \citenamefont {{Font}},\ and\ \citenamefont {{Meyer}}}]{Bruel_et_al:2023}%
  \BibitemOpen
  \bibfield  {author} {\bibinfo {author} {\bibfnamefont {T.}~\bibnamefont {{Bruel}}}, \bibinfo {author} {\bibfnamefont {M.-A.}\ \bibnamefont {{Bizouard}}}, \bibinfo {author} {\bibfnamefont {M.}~\bibnamefont {{Obergaulinger}}}, \bibinfo {author} {\bibfnamefont {P.}~\bibnamefont {{Maturana-Russel}}}, \bibinfo {author} {\bibfnamefont {A.}~\bibnamefont {{Torres-Forn{\'e}}}}, \bibinfo {author} {\bibfnamefont {P.}~\bibnamefont {{Cerd{\'a}-Dur{\'a}n}}}, \bibinfo {author} {\bibfnamefont {N.}~\bibnamefont {{Christensen}}}, \bibinfo {author} {\bibfnamefont {J.~A.}\ \bibnamefont {{Font}}},\ and\ \bibinfo {author} {\bibfnamefont {R.}~\bibnamefont {{Meyer}}},\ }\href {https://doi.org/10.1103/PhysRevD.107.083029} {\bibfield  {journal} {\bibinfo  {journal} {\prd}\ }\textbf {\bibinfo {volume} {107}},\ \bibinfo {eid} {083029} (\bibinfo {year} {2023})}\BibitemShut {NoStop}%
\bibitem [{\citenamefont {{Andresen}}\ and\ \citenamefont {{Finkel}}(2024)}]{Andresen_Finkel:2024}%
  \BibitemOpen
  \bibfield  {author} {\bibinfo {author} {\bibfnamefont {H.}~\bibnamefont {{Andresen}}}\ and\ \bibinfo {author} {\bibfnamefont {B.}~\bibnamefont {{Finkel}}},\ }\href {https://doi.org/10.48550/arXiv.2411.12524} {\bibfield  {journal} {\bibinfo  {journal} {arXiv e-prints}\ ,\ \bibinfo {eid} {arXiv:2411.12524}} (\bibinfo {year} {2024})}\BibitemShut {NoStop}%
\bibitem [{\citenamefont {{Torres-Forn{\'e}}}\ \emph {et~al.}(2018)\citenamefont {{Torres-Forn{\'e}}}, \citenamefont {{Cerd{\'a}-Dur{\'a}n}}, \citenamefont {{Passamonti}},\ and\ \citenamefont {{Font}}}]{Torres-Forne_et_al:2018}%
  \BibitemOpen
  \bibfield  {author} {\bibinfo {author} {\bibfnamefont {A.}~\bibnamefont {{Torres-Forn{\'e}}}}, \bibinfo {author} {\bibfnamefont {P.}~\bibnamefont {{Cerd{\'a}-Dur{\'a}n}}}, \bibinfo {author} {\bibfnamefont {A.}~\bibnamefont {{Passamonti}}},\ and\ \bibinfo {author} {\bibfnamefont {J.~A.}\ \bibnamefont {{Font}}},\ }\href {https://doi.org/10.1093/mnras/stx3067} {\bibfield  {journal} {\bibinfo  {journal} {\mnras}\ }\textbf {\bibinfo {volume} {474}},\ \bibinfo {pages} {5272} (\bibinfo {year} {2018})}\BibitemShut {NoStop}%
\bibitem [{\citenamefont {{Sotani}}\ \emph {et~al.}(2021)\citenamefont {{Sotani}}, \citenamefont {{Takiwaki}},\ and\ \citenamefont {{Togashi}}}]{Sotani_Takiwaki_Togashi:2021}%
  \BibitemOpen
  \bibfield  {author} {\bibinfo {author} {\bibfnamefont {H.}~\bibnamefont {{Sotani}}}, \bibinfo {author} {\bibfnamefont {T.}~\bibnamefont {{Takiwaki}}},\ and\ \bibinfo {author} {\bibfnamefont {H.}~\bibnamefont {{Togashi}}},\ }\href {https://doi.org/10.1103/PhysRevD.104.123009} {\bibfield  {journal} {\bibinfo  {journal} {\prd}\ }\textbf {\bibinfo {volume} {104}},\ \bibinfo {eid} {123009} (\bibinfo {year} {2021})}\BibitemShut {NoStop}%
\bibitem [{\citenamefont {{Fuller}}\ \emph {et~al.}(2015)\citenamefont {{Fuller}}, \citenamefont {{Klion}}, \citenamefont {{Abdikamalov}},\ and\ \citenamefont {{Ott}}}]{Fuller_Klion_Abdikamalov_Ott:2015}%
  \BibitemOpen
  \bibfield  {author} {\bibinfo {author} {\bibfnamefont {J.}~\bibnamefont {{Fuller}}}, \bibinfo {author} {\bibfnamefont {H.}~\bibnamefont {{Klion}}}, \bibinfo {author} {\bibfnamefont {E.}~\bibnamefont {{Abdikamalov}}},\ and\ \bibinfo {author} {\bibfnamefont {C.~D.}\ \bibnamefont {{Ott}}},\ }\href {https://doi.org/10.1093/mnras/stv698} {\bibfield  {journal} {\bibinfo  {journal} {\mnras}\ }\textbf {\bibinfo {volume} {450}},\ \bibinfo {pages} {414} (\bibinfo {year} {2015})}\BibitemShut {NoStop}%
\bibitem [{\citenamefont {{Dimmelmeier}}\ \emph {et~al.}(2008)\citenamefont {{Dimmelmeier}}, \citenamefont {{Ott}}, \citenamefont {{Marek}},\ and\ \citenamefont {{Janka}}}]{Dimmelmeier_Ott_Marek_Janka:2008}%
  \BibitemOpen
  \bibfield  {author} {\bibinfo {author} {\bibfnamefont {H.}~\bibnamefont {{Dimmelmeier}}}, \bibinfo {author} {\bibfnamefont {C.~D.}\ \bibnamefont {{Ott}}}, \bibinfo {author} {\bibfnamefont {A.}~\bibnamefont {{Marek}}},\ and\ \bibinfo {author} {\bibfnamefont {H.~T.}\ \bibnamefont {{Janka}}},\ }\href {https://doi.org/10.1103/PhysRevD.78.064056} {\bibfield  {journal} {\bibinfo  {journal} {\prd}\ }\textbf {\bibinfo {volume} {78}},\ \bibinfo {eid} {064056} (\bibinfo {year} {2008})}\BibitemShut {NoStop}%
\bibitem [{\citenamefont {{Cusinato}}\ \emph {et~al.}(2026{\natexlab{a}})\citenamefont {{Cusinato}}, \citenamefont {{Obergaulinger}}, \citenamefont {{Aloy}},\ and\ \citenamefont {{Font}}}]{Cusinato_et_al:2025}%
  \BibitemOpen
  \bibfield  {author} {\bibinfo {author} {\bibfnamefont {M.}~\bibnamefont {{Cusinato}}}, \bibinfo {author} {\bibfnamefont {M.}~\bibnamefont {{Obergaulinger}}}, \bibinfo {author} {\bibfnamefont {M.~{\'A}.}\ \bibnamefont {{Aloy}}},\ and\ \bibinfo {author} {\bibfnamefont {J.~A.}\ \bibnamefont {{Font}}},\ }\href {https://doi.org/10.1103/6wp3-sw9y} {\bibfield  {journal} {\bibinfo  {journal} {Physical Review Research}\ }\textbf {\bibinfo {volume} {8}},\ \bibinfo {eid} {013180} (\bibinfo {year} {2026}{\natexlab{a}})}\BibitemShut {NoStop}%
\bibitem [{\citenamefont {{Cusinato}}\ \emph {et~al.}(2026{\natexlab{b}})\citenamefont {{Cusinato}}, \citenamefont {{Obergaulinger}},\ and\ \citenamefont {{Aloy}}}]{Cusinato_Obergaulinger_Aloy:2025}%
  \BibitemOpen
  \bibfield  {author} {\bibinfo {author} {\bibfnamefont {M.}~\bibnamefont {{Cusinato}}}, \bibinfo {author} {\bibfnamefont {M.}~\bibnamefont {{Obergaulinger}}},\ and\ \bibinfo {author} {\bibfnamefont {M.~{\'A}.}\ \bibnamefont {{Aloy}}},\ }\href {https://doi.org/10.1051/0004-6361/202556543} {\bibfield  {journal} {\bibinfo  {journal} {\aap}\ }\textbf {\bibinfo {volume} {705}},\ \bibinfo {eid} {A179} (\bibinfo {year} {2026}{\natexlab{b}})}\BibitemShut {NoStop}%
\bibitem [{\citenamefont {{M{\"u}ller}}\ \emph {et~al.}(2010)\citenamefont {{M{\"u}ller}}, \citenamefont {{Janka}},\ and\ \citenamefont {{Dimmelmeier}}}]{Mueller_Janka_Dimmelmeier:2010}%
  \BibitemOpen
  \bibfield  {author} {\bibinfo {author} {\bibfnamefont {B.}~\bibnamefont {{M{\"u}ller}}}, \bibinfo {author} {\bibfnamefont {H.-T.}\ \bibnamefont {{Janka}}},\ and\ \bibinfo {author} {\bibfnamefont {H.}~\bibnamefont {{Dimmelmeier}}},\ }\href {https://doi.org/10.1088/0067-0049/189/1/104} {\bibfield  {journal} {\bibinfo  {journal} {\apjs}\ }\textbf {\bibinfo {volume} {189}},\ \bibinfo {pages} {104} (\bibinfo {year} {2010})}\BibitemShut {NoStop}%
\bibitem [{\citenamefont {{M{\"u}ller}}\ \emph {et~al.}(2012)\citenamefont {{M{\"u}ller}}, \citenamefont {{Janka}},\ and\ \citenamefont {{Marek}}}]{Mueller_Janka_Marek:2012}%
  \BibitemOpen
  \bibfield  {author} {\bibinfo {author} {\bibfnamefont {B.}~\bibnamefont {{M{\"u}ller}}}, \bibinfo {author} {\bibfnamefont {H.-T.}\ \bibnamefont {{Janka}}},\ and\ \bibinfo {author} {\bibfnamefont {A.}~\bibnamefont {{Marek}}},\ }\href {https://doi.org/10.1088/0004-637X/756/1/84} {\bibfield  {journal} {\bibinfo  {journal} {\apj}\ }\textbf {\bibinfo {volume} {756}},\ \bibinfo {eid} {84} (\bibinfo {year} {2012})}\BibitemShut {NoStop}%
\bibitem [{\citenamefont {{M{\"u}ller}}\ and\ \citenamefont {{Varma}}(2020)}]{Mueller_Varma:2020}%
  \BibitemOpen
  \bibfield  {author} {\bibinfo {author} {\bibfnamefont {B.}~\bibnamefont {{M{\"u}ller}}}\ and\ \bibinfo {author} {\bibfnamefont {V.}~\bibnamefont {{Varma}}},\ }\href {https://doi.org/10.1093/mnrasl/slaa137} {\bibfield  {journal} {\bibinfo  {journal} {\mnras}\ }\textbf {\bibinfo {volume} {498}},\ \bibinfo {pages} {L109} (\bibinfo {year} {2020})}\BibitemShut {NoStop}%
\bibitem [{\citenamefont {{M{\"u}ller}}\ and\ \citenamefont {{Janka}}(2015)}]{Mueller_Janka:2015}%
  \BibitemOpen
  \bibfield  {author} {\bibinfo {author} {\bibfnamefont {B.}~\bibnamefont {{M{\"u}ller}}}\ and\ \bibinfo {author} {\bibfnamefont {H.~T.}\ \bibnamefont {{Janka}}},\ }\href {https://doi.org/10.1093/mnras/stv101} {\bibfield  {journal} {\bibinfo  {journal} {\mnras}\ }\textbf {\bibinfo {volume} {448}},\ \bibinfo {pages} {2141} (\bibinfo {year} {2015})}\BibitemShut {NoStop}%
\bibitem [{\citenamefont {{Steiner}}\ \emph {et~al.}(2013)\citenamefont {{Steiner}}, \citenamefont {{Hempel}},\ and\ \citenamefont {{Fischer}}}]{Steiner_Hempel_Fischer:2013}%
  \BibitemOpen
  \bibfield  {author} {\bibinfo {author} {\bibfnamefont {A.~W.}\ \bibnamefont {{Steiner}}}, \bibinfo {author} {\bibfnamefont {M.}~\bibnamefont {{Hempel}}},\ and\ \bibinfo {author} {\bibfnamefont {T.}~\bibnamefont {{Fischer}}},\ }\href {https://doi.org/10.1088/0004-637X/774/1/17} {\bibfield  {journal} {\bibinfo  {journal} {\apj}\ }\textbf {\bibinfo {volume} {774}},\ \bibinfo {eid} {17} (\bibinfo {year} {2013})}\BibitemShut {NoStop}%
\bibitem [{\citenamefont {{Aguilera-Dena}}\ \emph {et~al.}(2018)\citenamefont {{Aguilera-Dena}}, \citenamefont {{Langer}}, \citenamefont {{Moriya}},\ and\ \citenamefont {{Schootemeijer}}}]{Aguilera-Dena_Langer_Moriya_Schootemeijer:2018}%
  \BibitemOpen
  \bibfield  {author} {\bibinfo {author} {\bibfnamefont {D.~R.}\ \bibnamefont {{Aguilera-Dena}}}, \bibinfo {author} {\bibfnamefont {N.}~\bibnamefont {{Langer}}}, \bibinfo {author} {\bibfnamefont {T.~J.}\ \bibnamefont {{Moriya}}},\ and\ \bibinfo {author} {\bibfnamefont {A.}~\bibnamefont {{Schootemeijer}}},\ }\href {https://doi.org/10.3847/1538-4357/aabfc1} {\bibfield  {journal} {\bibinfo  {journal} {\apj}\ }\textbf {\bibinfo {volume} {858}},\ \bibinfo {eid} {115} (\bibinfo {year} {2018})}\BibitemShut {NoStop}%
\bibitem [{\citenamefont {{Paxton}}\ \emph {et~al.}(2011)\citenamefont {{Paxton}}, \citenamefont {{Bildsten}}, \citenamefont {{Dotter}}, \citenamefont {{Herwig}}, \citenamefont {{Lesaffre}},\ and\ \citenamefont {{Timmes}}}]{Paxton_et_al:2011}%
  \BibitemOpen
  \bibfield  {author} {\bibinfo {author} {\bibfnamefont {B.}~\bibnamefont {{Paxton}}}, \bibinfo {author} {\bibfnamefont {L.}~\bibnamefont {{Bildsten}}}, \bibinfo {author} {\bibfnamefont {A.}~\bibnamefont {{Dotter}}}, \bibinfo {author} {\bibfnamefont {F.}~\bibnamefont {{Herwig}}}, \bibinfo {author} {\bibfnamefont {P.}~\bibnamefont {{Lesaffre}}},\ and\ \bibinfo {author} {\bibfnamefont {F.}~\bibnamefont {{Timmes}}},\ }\href {https://doi.org/10.1088/0067-0049/192/1/3} {\bibfield  {journal} {\bibinfo  {journal} {\apjs}\ }\textbf {\bibinfo {volume} {192}},\ \bibinfo {eid} {3} (\bibinfo {year} {2011})}\BibitemShut {NoStop}%
\bibitem [{\citenamefont {{Sykes}}\ and\ \citenamefont {{M{\"u}ller}}(2025)}]{Sykes_Mueller:2025}%
  \BibitemOpen
  \bibfield  {author} {\bibinfo {author} {\bibfnamefont {B.}~\bibnamefont {{Sykes}}}\ and\ \bibinfo {author} {\bibfnamefont {B.}~\bibnamefont {{M{\"u}ller}}},\ }\href {https://doi.org/10.1103/PhysRevD.111.063042} {\bibfield  {journal} {\bibinfo  {journal} {\prd}\ }\textbf {\bibinfo {volume} {111}},\ \bibinfo {eid} {063042} (\bibinfo {year} {2025})}\BibitemShut {NoStop}%
\bibitem [{\citenamefont {{Varma}}\ \emph {et~al.}(2021)\citenamefont {{Varma}}, \citenamefont {{M{\"u}ller}},\ and\ \citenamefont {{Obergaulinger}}}]{Varma_Mueller_Obergaulinger:2021}%
  \BibitemOpen
  \bibfield  {author} {\bibinfo {author} {\bibfnamefont {V.}~\bibnamefont {{Varma}}}, \bibinfo {author} {\bibfnamefont {B.}~\bibnamefont {{M{\"u}ller}}},\ and\ \bibinfo {author} {\bibfnamefont {M.}~\bibnamefont {{Obergaulinger}}},\ }\href {https://doi.org/10.1093/mnras/stab2983} {\bibfield  {journal} {\bibinfo  {journal} {\mnras}\ }\textbf {\bibinfo {volume} {508}},\ \bibinfo {pages} {6033} (\bibinfo {year} {2021})}\BibitemShut {NoStop}%
\bibitem [{\citenamefont {{Obergaulinger}}\ \emph {et~al.}(2018)\citenamefont {{Obergaulinger}}, \citenamefont {{Just}},\ and\ \citenamefont {{Aloy}}}]{Obergaulinger_Just_Aloy:2018}%
  \BibitemOpen
  \bibfield  {author} {\bibinfo {author} {\bibfnamefont {M.}~\bibnamefont {{Obergaulinger}}}, \bibinfo {author} {\bibfnamefont {O.}~\bibnamefont {{Just}}},\ and\ \bibinfo {author} {\bibfnamefont {M.~A.}\ \bibnamefont {{Aloy}}},\ }\href {https://doi.org/10.1088/1361-6471/aac982} {\bibfield  {journal} {\bibinfo  {journal} {Journal of Physics G Nuclear Physics}\ }\textbf {\bibinfo {volume} {45}},\ \bibinfo {pages} {084001} (\bibinfo {year} {2018})}\BibitemShut {NoStop}%
\bibitem [{\citenamefont {{Varma}}\ \emph {et~al.}(2023)\citenamefont {{Varma}}, \citenamefont {{M{\"u}ller}},\ and\ \citenamefont {{Schneider}}}]{Varma_Mueller_Schneider:2023}%
  \BibitemOpen
  \bibfield  {author} {\bibinfo {author} {\bibfnamefont {V.}~\bibnamefont {{Varma}}}, \bibinfo {author} {\bibfnamefont {B.}~\bibnamefont {{M{\"u}ller}}},\ and\ \bibinfo {author} {\bibfnamefont {F.~R.~N.}\ \bibnamefont {{Schneider}}},\ }\href {https://doi.org/10.1093/mnras/stac3247} {\bibfield  {journal} {\bibinfo  {journal} {\mnras}\ }\textbf {\bibinfo {volume} {518}},\ \bibinfo {pages} {3622} (\bibinfo {year} {2023})}\BibitemShut {NoStop}%
\bibitem [{\citenamefont {{Yamasaki}}\ and\ \citenamefont {{Foglizzo}}(2008)}]{Yamasaki_Foglizzo:2008}%
  \BibitemOpen
  \bibfield  {author} {\bibinfo {author} {\bibfnamefont {T.}~\bibnamefont {{Yamasaki}}}\ and\ \bibinfo {author} {\bibfnamefont {T.}~\bibnamefont {{Foglizzo}}},\ }\href {https://doi.org/10.1086/587732} {\bibfield  {journal} {\bibinfo  {journal} {\apj}\ }\textbf {\bibinfo {volume} {679}},\ \bibinfo {pages} {607} (\bibinfo {year} {2008})}\BibitemShut {NoStop}%
\bibitem [{\citenamefont {{Blondin}}\ \emph {et~al.}(2017)\citenamefont {{Blondin}}, \citenamefont {{Gipson}}, \citenamefont {{Harris}},\ and\ \citenamefont {{Mezzacappa}}}]{Blondin_Gipson_Harris_Mezzacappa:2017}%
  \BibitemOpen
  \bibfield  {author} {\bibinfo {author} {\bibfnamefont {J.~M.}\ \bibnamefont {{Blondin}}}, \bibinfo {author} {\bibfnamefont {E.}~\bibnamefont {{Gipson}}}, \bibinfo {author} {\bibfnamefont {S.}~\bibnamefont {{Harris}}},\ and\ \bibinfo {author} {\bibfnamefont {A.}~\bibnamefont {{Mezzacappa}}},\ }\href {https://doi.org/10.3847/1538-4357/835/2/170} {\bibfield  {journal} {\bibinfo  {journal} {\apj}\ }\textbf {\bibinfo {volume} {835}},\ \bibinfo {eid} {170} (\bibinfo {year} {2017})}\BibitemShut {NoStop}%
\bibitem [{\citenamefont {{Iwakami}}\ \emph {et~al.}(2009)\citenamefont {{Iwakami}}, \citenamefont {{Kotake}}, \citenamefont {{Ohnishi}}, \citenamefont {{Yamada}},\ and\ \citenamefont {{Sawada}}}]{Iwakami_et_al:2009}%
  \BibitemOpen
  \bibfield  {author} {\bibinfo {author} {\bibfnamefont {W.}~\bibnamefont {{Iwakami}}}, \bibinfo {author} {\bibfnamefont {K.}~\bibnamefont {{Kotake}}}, \bibinfo {author} {\bibfnamefont {N.}~\bibnamefont {{Ohnishi}}}, \bibinfo {author} {\bibfnamefont {S.}~\bibnamefont {{Yamada}}},\ and\ \bibinfo {author} {\bibfnamefont {K.}~\bibnamefont {{Sawada}}},\ }\href {https://doi.org/10.1088/0004-637X/700/1/232} {\bibfield  {journal} {\bibinfo  {journal} {\apj}\ }\textbf {\bibinfo {volume} {700}},\ \bibinfo {pages} {232} (\bibinfo {year} {2009})}\BibitemShut {NoStop}%
\bibitem [{\citenamefont {{Marek}}\ and\ \citenamefont {{Janka}}(2009)}]{Marek_Janka:2009}%
  \BibitemOpen
  \bibfield  {author} {\bibinfo {author} {\bibfnamefont {A.}~\bibnamefont {{Marek}}}\ and\ \bibinfo {author} {\bibfnamefont {H.-T.}\ \bibnamefont {{Janka}}},\ }\href {https://doi.org/10.1088/0004-637X/694/1/664} {\bibfield  {journal} {\bibinfo  {journal} {\apj}\ }\textbf {\bibinfo {volume} {694}},\ \bibinfo {pages} {664} (\bibinfo {year} {2009})}\BibitemShut {NoStop}%
\bibitem [{\citenamefont {{Pajkos}}\ \emph {et~al.}(2021{\natexlab{a}})\citenamefont {{Pajkos}}, \citenamefont {{Warren}}, \citenamefont {{Couch}}, \citenamefont {{O'Connor}},\ and\ \citenamefont {{Pan}}}]{Pajkos_et_al:2021}%
  \BibitemOpen
  \bibfield  {author} {\bibinfo {author} {\bibfnamefont {M.~A.}\ \bibnamefont {{Pajkos}}}, \bibinfo {author} {\bibfnamefont {M.~L.}\ \bibnamefont {{Warren}}}, \bibinfo {author} {\bibfnamefont {S.~M.}\ \bibnamefont {{Couch}}}, \bibinfo {author} {\bibfnamefont {E.~P.}\ \bibnamefont {{O'Connor}}},\ and\ \bibinfo {author} {\bibfnamefont {K.-C.}\ \bibnamefont {{Pan}}},\ }\href {https://doi.org/10.3847/1538-4357/abfb65} {\bibfield  {journal} {\bibinfo  {journal} {\apj}\ }\textbf {\bibinfo {volume} {914}},\ \bibinfo {eid} {80} (\bibinfo {year} {2021}{\natexlab{a}})}\BibitemShut {NoStop}%
\bibitem [{\citenamefont {{Pajkos}}\ \emph {et~al.}(2026)\citenamefont {{Pajkos}}, \citenamefont {{Boyeneni}},\ and\ \citenamefont {{Eggenberger Andersen}}}]{Pajkos_Boyeneni_EggenbergerAndersen:2026}%
  \BibitemOpen
  \bibfield  {author} {\bibinfo {author} {\bibfnamefont {M.~A.}\ \bibnamefont {{Pajkos}}}, \bibinfo {author} {\bibfnamefont {S.}~\bibnamefont {{Boyeneni}}},\ and\ \bibinfo {author} {\bibfnamefont {O.}~\bibnamefont {{Eggenberger Andersen}}},\ }\href {https://doi.org/10.1103/wrnn-y4jg} {\bibfield  {journal} {\bibinfo  {journal} {\prd}\ }\textbf {\bibinfo {volume} {113}},\ \bibinfo {eid} {063051} (\bibinfo {year} {2026})}\BibitemShut {NoStop}%
\bibitem [{\citenamefont {{Fryer}}\ and\ \citenamefont {{Heger}}(2000)}]{Fryer_Heger:2000}%
  \BibitemOpen
  \bibfield  {author} {\bibinfo {author} {\bibfnamefont {C.~L.}\ \bibnamefont {{Fryer}}}\ and\ \bibinfo {author} {\bibfnamefont {A.}~\bibnamefont {{Heger}}},\ }\href {https://doi.org/10.1086/309446} {\bibfield  {journal} {\bibinfo  {journal} {\apj}\ }\textbf {\bibinfo {volume} {541}},\ \bibinfo {pages} {1033} (\bibinfo {year} {2000})}\BibitemShut {NoStop}%
\bibitem [{\citenamefont {{Dimmelmeier}}\ \emph {et~al.}(2002)\citenamefont {{Dimmelmeier}}, \citenamefont {{Font}},\ and\ \citenamefont {{M{\"u}ller}}}]{Dimmelmeier_Font_Mueller:2002}%
  \BibitemOpen
  \bibfield  {author} {\bibinfo {author} {\bibfnamefont {H.}~\bibnamefont {{Dimmelmeier}}}, \bibinfo {author} {\bibfnamefont {J.~A.}\ \bibnamefont {{Font}}},\ and\ \bibinfo {author} {\bibfnamefont {E.}~\bibnamefont {{M{\"u}ller}}},\ }\href {https://doi.org/10.1051/0004-6361:20021053} {\bibfield  {journal} {\bibinfo  {journal} {\aap}\ }\textbf {\bibinfo {volume} {393}},\ \bibinfo {pages} {523} (\bibinfo {year} {2002})}\BibitemShut {NoStop}%
\bibitem [{\citenamefont {{Morlet}}\ \emph {et~al.}(1982)\citenamefont {{Morlet}}, \citenamefont {{Arens}}, \citenamefont {{Forgeau}},\ and\ \citenamefont {{Giard}}}]{Morlet_Arens_Forgeau_Giard:1982}%
  \BibitemOpen
  \bibfield  {author} {\bibinfo {author} {\bibfnamefont {J.}~\bibnamefont {{Morlet}}}, \bibinfo {author} {\bibfnamefont {G.}~\bibnamefont {{Arens}}}, \bibinfo {author} {\bibfnamefont {I.}~\bibnamefont {{Forgeau}}},\ and\ \bibinfo {author} {\bibfnamefont {D.}~\bibnamefont {{Giard}}},\ }\href {https://doi.org/10.1190/1.1441328} {\bibfield  {journal} {\bibinfo  {journal} {Geophysics}\ }\textbf {\bibinfo {volume} {47}},\ \bibinfo {pages} {203} (\bibinfo {year} {1982})}\BibitemShut {NoStop}%
\bibitem [{\citenamefont {{Dimmelmeier}}\ \emph {et~al.}(2007)\citenamefont {{Dimmelmeier}}, \citenamefont {{Ott}}, \citenamefont {{Janka}}, \citenamefont {{Marek}},\ and\ \citenamefont {{M{\"u}ller}}}]{Dimmelmeier_Ott_Janka_Marek_Mueller:2007}%
  \BibitemOpen
  \bibfield  {author} {\bibinfo {author} {\bibfnamefont {H.}~\bibnamefont {{Dimmelmeier}}}, \bibinfo {author} {\bibfnamefont {C.~D.}\ \bibnamefont {{Ott}}}, \bibinfo {author} {\bibfnamefont {H.~T.}\ \bibnamefont {{Janka}}}, \bibinfo {author} {\bibfnamefont {A.}~\bibnamefont {{Marek}}},\ and\ \bibinfo {author} {\bibfnamefont {E.}~\bibnamefont {{M{\"u}ller}}},\ }\href {https://doi.org/10.1103/PhysRevLett.98.251101} {\bibfield  {journal} {\bibinfo  {journal} {\prl}\ }\textbf {\bibinfo {volume} {98}},\ \bibinfo {eid} {251101} (\bibinfo {year} {2007})}\BibitemShut {NoStop}%
\bibitem [{\citenamefont {{Dimmelmeier}}\ \emph {et~al.}(2006)\citenamefont {{Dimmelmeier}}, \citenamefont {{Stergioulas}},\ and\ \citenamefont {{Font}}}]{Dimmelmeier_Stergiouslas_Font:2006}%
  \BibitemOpen
  \bibfield  {author} {\bibinfo {author} {\bibfnamefont {H.}~\bibnamefont {{Dimmelmeier}}}, \bibinfo {author} {\bibfnamefont {N.}~\bibnamefont {{Stergioulas}}},\ and\ \bibinfo {author} {\bibfnamefont {J.~A.}\ \bibnamefont {{Font}}},\ }\href {https://doi.org/10.1111/j.1365-2966.2006.10274.x} {\bibfield  {journal} {\bibinfo  {journal} {\mnras}\ }\textbf {\bibinfo {volume} {368}},\ \bibinfo {pages} {1609} (\bibinfo {year} {2006})}\BibitemShut {NoStop}%
\bibitem [{\citenamefont {{Pajkos}}\ \emph {et~al.}(2021{\natexlab{b}})\citenamefont {{Pajkos}}, \citenamefont {{Warren}}, \citenamefont {{Couch}}, \citenamefont {{O'Connor}},\ and\ \citenamefont {{Pan}}}]{Pajkos_Warren_Couch_OConnor_Pan:2021}%
  \BibitemOpen
  \bibfield  {author} {\bibinfo {author} {\bibfnamefont {M.~A.}\ \bibnamefont {{Pajkos}}}, \bibinfo {author} {\bibfnamefont {M.~L.}\ \bibnamefont {{Warren}}}, \bibinfo {author} {\bibfnamefont {S.~M.}\ \bibnamefont {{Couch}}}, \bibinfo {author} {\bibfnamefont {E.~P.}\ \bibnamefont {{O'Connor}}},\ and\ \bibinfo {author} {\bibfnamefont {K.-C.}\ \bibnamefont {{Pan}}},\ }\href {https://doi.org/10.3847/1538-4357/abfb65} {\bibfield  {journal} {\bibinfo  {journal} {\apj}\ }\textbf {\bibinfo {volume} {914}},\ \bibinfo {eid} {80} (\bibinfo {year} {2021}{\natexlab{b}})}\BibitemShut {NoStop}%
\bibitem [{\citenamefont {{Bugli}}\ \emph {et~al.}(2023{\natexlab{a}})\citenamefont {{Bugli}}, \citenamefont {{Guilet}}, \citenamefont {{Foglizzo}},\ and\ \citenamefont {{Obergaulinger}}}]{Bugli_et_al:2023}%
  \BibitemOpen
  \bibfield  {author} {\bibinfo {author} {\bibfnamefont {M.}~\bibnamefont {{Bugli}}}, \bibinfo {author} {\bibfnamefont {J.}~\bibnamefont {{Guilet}}}, \bibinfo {author} {\bibfnamefont {T.}~\bibnamefont {{Foglizzo}}},\ and\ \bibinfo {author} {\bibfnamefont {M.}~\bibnamefont {{Obergaulinger}}},\ }\href {https://doi.org/10.1093/mnras/stad496} {\bibfield  {journal} {\bibinfo  {journal} {\mnras}\ }\textbf {\bibinfo {volume} {520}},\ \bibinfo {pages} {5622} (\bibinfo {year} {2023}{\natexlab{a}})}\BibitemShut {NoStop}%
\bibitem [{\citenamefont {{Powell}}\ \emph {et~al.}(2023)\citenamefont {{Powell}}, \citenamefont {{M{\"u}ller}}, \citenamefont {{Aguilera-Dena}},\ and\ \citenamefont {{Langer}}}]{Powell_Mueller_AguileraDena_Langer:2023}%
  \BibitemOpen
  \bibfield  {author} {\bibinfo {author} {\bibfnamefont {J.}~\bibnamefont {{Powell}}}, \bibinfo {author} {\bibfnamefont {B.}~\bibnamefont {{M{\"u}ller}}}, \bibinfo {author} {\bibfnamefont {D.~R.}\ \bibnamefont {{Aguilera-Dena}}},\ and\ \bibinfo {author} {\bibfnamefont {N.}~\bibnamefont {{Langer}}},\ }\href {https://doi.org/10.1093/mnras/stad1292} {\bibfield  {journal} {\bibinfo  {journal} {\mnras}\ }\textbf {\bibinfo {volume} {522}},\ \bibinfo {pages} {6070} (\bibinfo {year} {2023})}\BibitemShut {NoStop}%
\bibitem [{\citenamefont {{Couch}}\ and\ \citenamefont {{O'Connor}}(2014)}]{Couch_OConnor:2014}%
  \BibitemOpen
  \bibfield  {author} {\bibinfo {author} {\bibfnamefont {S.~M.}\ \bibnamefont {{Couch}}}\ and\ \bibinfo {author} {\bibfnamefont {E.~P.}\ \bibnamefont {{O'Connor}}},\ }\href {https://doi.org/10.1088/0004-637X/785/2/123} {\bibfield  {journal} {\bibinfo  {journal} {\apj}\ }\textbf {\bibinfo {volume} {785}},\ \bibinfo {eid} {123} (\bibinfo {year} {2014})}\BibitemShut {NoStop}%
\bibitem [{\citenamefont {{Vartanyan}}\ \emph {et~al.}(2019)\citenamefont {{Vartanyan}}, \citenamefont {{Burrows}}, \citenamefont {{Radice}}, \citenamefont {{Skinner}},\ and\ \citenamefont {{Dolence}}}]{Vartanyan_Burrows_Radice_Skinner_Dolence:2019}%
  \BibitemOpen
  \bibfield  {author} {\bibinfo {author} {\bibfnamefont {D.}~\bibnamefont {{Vartanyan}}}, \bibinfo {author} {\bibfnamefont {A.}~\bibnamefont {{Burrows}}}, \bibinfo {author} {\bibfnamefont {D.}~\bibnamefont {{Radice}}}, \bibinfo {author} {\bibfnamefont {M.~A.}\ \bibnamefont {{Skinner}}},\ and\ \bibinfo {author} {\bibfnamefont {J.}~\bibnamefont {{Dolence}}},\ }\href {https://doi.org/10.1093/mnras/sty2585} {\bibfield  {journal} {\bibinfo  {journal} {\mnras}\ }\textbf {\bibinfo {volume} {482}},\ \bibinfo {pages} {351} (\bibinfo {year} {2019})}\BibitemShut {NoStop}%
\bibitem [{\citenamefont {{Kuroda}}\ \emph {et~al.}(2025)\citenamefont {{Kuroda}}, \citenamefont {{Kawaguchi}},\ and\ \citenamefont {{Shibata}}}]{Kuroda_Kawaguchi_Shibata:2025}%
  \BibitemOpen
  \bibfield  {author} {\bibinfo {author} {\bibfnamefont {T.}~\bibnamefont {{Kuroda}}}, \bibinfo {author} {\bibfnamefont {K.}~\bibnamefont {{Kawaguchi}}},\ and\ \bibinfo {author} {\bibfnamefont {M.}~\bibnamefont {{Shibata}}},\ }\href {https://doi.org/10.1093/mnras/staf1065} {\bibfield  {journal} {\bibinfo  {journal} {\mnras}\ }\textbf {\bibinfo {volume} {541}},\ \bibinfo {pages} {1649} (\bibinfo {year} {2025})}\BibitemShut {NoStop}%
\bibitem [{\citenamefont {{Shibagaki}}\ \emph {et~al.}(2026)\citenamefont {{Shibagaki}}, \citenamefont {{Takiwaki}}, \citenamefont {{Kotake}}, \citenamefont {{Kuroda}},\ and\ \citenamefont {{Fischer}}}]{Shibagaki_et_al:2026}%
  \BibitemOpen
  \bibfield  {author} {\bibinfo {author} {\bibfnamefont {S.}~\bibnamefont {{Shibagaki}}}, \bibinfo {author} {\bibfnamefont {T.}~\bibnamefont {{Takiwaki}}}, \bibinfo {author} {\bibfnamefont {K.}~\bibnamefont {{Kotake}}}, \bibinfo {author} {\bibfnamefont {T.}~\bibnamefont {{Kuroda}}},\ and\ \bibinfo {author} {\bibfnamefont {T.}~\bibnamefont {{Fischer}}},\ }\href {https://doi.org/10.48550/arXiv.2603.20473} {\bibfield  {journal} {\bibinfo  {journal} {arXiv e-prints}\ ,\ \bibinfo {eid} {arXiv:2603.20473}} (\bibinfo {year} {2026})},\ \Eprint {https://arxiv.org/abs/2603.20473} {arXiv:2603.20473 [astro-ph.HE]} \BibitemShut {NoStop}%
\bibitem [{\citenamefont {{Bugli}}\ \emph {et~al.}(2023{\natexlab{b}})\citenamefont {{Bugli}}, \citenamefont {{Guilet}}, \citenamefont {{Foglizzo}},\ and\ \citenamefont {{Obergaulinger}}}]{Bugli_Guilet_Foglizzo_Obergaulinger:2023}%
  \BibitemOpen
  \bibfield  {author} {\bibinfo {author} {\bibfnamefont {M.}~\bibnamefont {{Bugli}}}, \bibinfo {author} {\bibfnamefont {J.}~\bibnamefont {{Guilet}}}, \bibinfo {author} {\bibfnamefont {T.}~\bibnamefont {{Foglizzo}}},\ and\ \bibinfo {author} {\bibfnamefont {M.}~\bibnamefont {{Obergaulinger}}},\ }\href {https://doi.org/10.1093/mnras/stad496} {\bibfield  {journal} {\bibinfo  {journal} {\mnras}\ }\textbf {\bibinfo {volume} {520}},\ \bibinfo {pages} {5622} (\bibinfo {year} {2023}{\natexlab{b}})}\BibitemShut {NoStop}%
\bibitem [{\citenamefont {{Takiwaki}}\ \emph {et~al.}(2021)\citenamefont {{Takiwaki}}, \citenamefont {{Kotake}},\ and\ \citenamefont {{Foglizzo}}}]{Takiwaki_Kotake_Foglizzo:2021}%
  \BibitemOpen
  \bibfield  {author} {\bibinfo {author} {\bibfnamefont {T.}~\bibnamefont {{Takiwaki}}}, \bibinfo {author} {\bibfnamefont {K.}~\bibnamefont {{Kotake}}},\ and\ \bibinfo {author} {\bibfnamefont {T.}~\bibnamefont {{Foglizzo}}},\ }\href {https://doi.org/10.1093/mnras/stab2607} {\bibfield  {journal} {\bibinfo  {journal} {\mnras}\ }\textbf {\bibinfo {volume} {508}},\ \bibinfo {pages} {966} (\bibinfo {year} {2021})}\BibitemShut {NoStop}%
\bibitem [{\citenamefont {{Shibagaki}}\ \emph {et~al.}(2020)\citenamefont {{Shibagaki}}, \citenamefont {{Kuroda}}, \citenamefont {{Kotake}},\ and\ \citenamefont {{Takiwaki}}}]{Shibagaki_Kuroda_Kotake_Takiwaki:2020}%
  \BibitemOpen
  \bibfield  {author} {\bibinfo {author} {\bibfnamefont {S.}~\bibnamefont {{Shibagaki}}}, \bibinfo {author} {\bibfnamefont {T.}~\bibnamefont {{Kuroda}}}, \bibinfo {author} {\bibfnamefont {K.}~\bibnamefont {{Kotake}}},\ and\ \bibinfo {author} {\bibfnamefont {T.}~\bibnamefont {{Takiwaki}}},\ }\href {https://doi.org/10.1093/mnrasl/slaa021} {\bibfield  {journal} {\bibinfo  {journal} {\mnras}\ }\textbf {\bibinfo {volume} {493}},\ \bibinfo {pages} {L138} (\bibinfo {year} {2020})}\BibitemShut {NoStop}%
\bibitem [{\citenamefont {{Shapiro}}\ and\ \citenamefont {{Teukolsky}}(1983)}]{Shapiro_Teukolsky:1983}%
  \BibitemOpen
  \bibfield  {author} {\bibinfo {author} {\bibfnamefont {S.~L.}\ \bibnamefont {{Shapiro}}}\ and\ \bibinfo {author} {\bibfnamefont {S.~A.}\ \bibnamefont {{Teukolsky}}},\ }\bibinfo {title} {Rotation and magnetic fields},\ in\ \href {https://doi.org/https://doi.org/10.1002/9783527617661.ch7} {\emph {\bibinfo {booktitle} {Black Holes, White Dwarfs, and Neutron Stars}}}\ (\bibinfo  {publisher} {John Wiley \& Sons, Ltd},\ \bibinfo {year} {1983})\ Chap.~\bibinfo {chapter} {7}, pp.\ \bibinfo {pages} {162--187}\BibitemShut {NoStop}%
\bibitem [{\citenamefont {{Balbus}}\ and\ \citenamefont {{Hawley}}(1991)}]{Balbus_Hawley:1991}%
  \BibitemOpen
  \bibfield  {author} {\bibinfo {author} {\bibfnamefont {S.~A.}\ \bibnamefont {{Balbus}}}\ and\ \bibinfo {author} {\bibfnamefont {J.~F.}\ \bibnamefont {{Hawley}}},\ }\href {https://doi.org/10.1086/170270} {\bibfield  {journal} {\bibinfo  {journal} {\apj}\ }\textbf {\bibinfo {volume} {376}},\ \bibinfo {pages} {214} (\bibinfo {year} {1991})}\BibitemShut {NoStop}%
\bibitem [{\citenamefont {{Akiyama}}\ \emph {et~al.}(2003)\citenamefont {{Akiyama}}, \citenamefont {{Wheeler}}, \citenamefont {{Meier}},\ and\ \citenamefont {{Lichtenstadt}}}]{Akiyama_Wheeler_Meier_Lichtenstadt:2003}%
  \BibitemOpen
  \bibfield  {author} {\bibinfo {author} {\bibfnamefont {S.}~\bibnamefont {{Akiyama}}}, \bibinfo {author} {\bibfnamefont {J.~C.}\ \bibnamefont {{Wheeler}}}, \bibinfo {author} {\bibfnamefont {D.~L.}\ \bibnamefont {{Meier}}},\ and\ \bibinfo {author} {\bibfnamefont {I.}~\bibnamefont {{Lichtenstadt}}},\ }\href {https://doi.org/10.1086/344135} {\bibfield  {journal} {\bibinfo  {journal} {\apj}\ }\textbf {\bibinfo {volume} {584}},\ \bibinfo {pages} {954} (\bibinfo {year} {2003})}\BibitemShut {NoStop}%
\bibitem [{\citenamefont {{Obergaulinger}}\ \emph {et~al.}(2009)\citenamefont {{Obergaulinger}}, \citenamefont {{Cerd{\'a}-Dur{\'a}n}}, \citenamefont {{M{\"u}ller}},\ and\ \citenamefont {{Aloy}}}]{Obergaulinger_et_al:2009}%
  \BibitemOpen
  \bibfield  {author} {\bibinfo {author} {\bibfnamefont {M.}~\bibnamefont {{Obergaulinger}}}, \bibinfo {author} {\bibfnamefont {P.}~\bibnamefont {{Cerd{\'a}-Dur{\'a}n}}}, \bibinfo {author} {\bibfnamefont {E.}~\bibnamefont {{M{\"u}ller}}},\ and\ \bibinfo {author} {\bibfnamefont {M.~A.}\ \bibnamefont {{Aloy}}},\ }\href {https://doi.org/10.1051/0004-6361/200811323} {\bibfield  {journal} {\bibinfo  {journal} {\aap}\ }\textbf {\bibinfo {volume} {498}},\ \bibinfo {pages} {241} (\bibinfo {year} {2009})}\BibitemShut {NoStop}%
\bibitem [{\citenamefont {{M{\"o}sta}}\ \emph {et~al.}(2015)\citenamefont {{M{\"o}sta}}, \citenamefont {{Ott}}, \citenamefont {{Radice}}, \citenamefont {{Roberts}}, \citenamefont {{Schnetter}},\ and\ \citenamefont {{Haas}}}]{Mosta_et_al:2015}%
  \BibitemOpen
  \bibfield  {author} {\bibinfo {author} {\bibfnamefont {P.}~\bibnamefont {{M{\"o}sta}}}, \bibinfo {author} {\bibfnamefont {C.~D.}\ \bibnamefont {{Ott}}}, \bibinfo {author} {\bibfnamefont {D.}~\bibnamefont {{Radice}}}, \bibinfo {author} {\bibfnamefont {L.~F.}\ \bibnamefont {{Roberts}}}, \bibinfo {author} {\bibfnamefont {E.}~\bibnamefont {{Schnetter}}},\ and\ \bibinfo {author} {\bibfnamefont {R.}~\bibnamefont {{Haas}}},\ }\href {https://doi.org/10.1038/nature15755} {\bibfield  {journal} {\bibinfo  {journal} {\nat}\ }\textbf {\bibinfo {volume} {528}},\ \bibinfo {pages} {376} (\bibinfo {year} {2015})}\BibitemShut {NoStop}%
\bibitem [{\citenamefont {{Maeder}}\ \emph {et~al.}(2013)\citenamefont {{Maeder}}, \citenamefont {{Meynet}}, \citenamefont {{Lagarde}},\ and\ \citenamefont {{Charbonnel}}}]{Maeder_Meynet_Lagarde_Charbonnel}%
  \BibitemOpen
  \bibfield  {author} {\bibinfo {author} {\bibfnamefont {A.}~\bibnamefont {{Maeder}}}, \bibinfo {author} {\bibfnamefont {G.}~\bibnamefont {{Meynet}}}, \bibinfo {author} {\bibfnamefont {N.}~\bibnamefont {{Lagarde}}},\ and\ \bibinfo {author} {\bibfnamefont {C.}~\bibnamefont {{Charbonnel}}},\ }\href {https://doi.org/10.1051/0004-6361/201220936} {\bibfield  {journal} {\bibinfo  {journal} {\aap}\ }\textbf {\bibinfo {volume} {553}},\ \bibinfo {eid} {A1} (\bibinfo {year} {2013})}\BibitemShut {NoStop}%
\bibitem [{\citenamefont {{Tassoul}}(1980)}]{Tassoul:1980}%
  \BibitemOpen
  \bibfield  {author} {\bibinfo {author} {\bibfnamefont {M.}~\bibnamefont {{Tassoul}}},\ }\href {https://doi.org/10.1086/190678} {\bibfield  {journal} {\bibinfo  {journal} {\apjs}\ }\textbf {\bibinfo {volume} {43}},\ \bibinfo {pages} {469} (\bibinfo {year} {1980})}\BibitemShut {NoStop}%
\bibitem [{\citenamefont {{Jakobus}}\ \emph {et~al.}(2025)\citenamefont {{Jakobus}}, \citenamefont {{M{\"u}ller}},\ and\ \citenamefont {{Heger}}}]{Jakobus_Mueller_Heger:2025}%
  \BibitemOpen
  \bibfield  {author} {\bibinfo {author} {\bibfnamefont {P.}~\bibnamefont {{Jakobus}}}, \bibinfo {author} {\bibfnamefont {B.}~\bibnamefont {{M{\"u}ller}}},\ and\ \bibinfo {author} {\bibfnamefont {A.}~\bibnamefont {{Heger}}},\ }\href {https://doi.org/10.1093/mnras/staf868} {\bibfield  {journal} {\bibinfo  {journal} {\mnras}\ }\textbf {\bibinfo {volume} {540}},\ \bibinfo {pages} {3008} (\bibinfo {year} {2025})}\BibitemShut {NoStop}%
\bibitem [{\citenamefont {{Zha}}\ \emph {et~al.}(2024)\citenamefont {{Zha}}, \citenamefont {{Eggenberger Andersen}},\ and\ \citenamefont {{O'Connor}}}]{Zha_EggenbergerAndersen_OConnor:2024}%
  \BibitemOpen
  \bibfield  {author} {\bibinfo {author} {\bibfnamefont {S.}~\bibnamefont {{Zha}}}, \bibinfo {author} {\bibfnamefont {O.}~\bibnamefont {{Eggenberger Andersen}}},\ and\ \bibinfo {author} {\bibfnamefont {E.~P.}\ \bibnamefont {{O'Connor}}},\ }\href {https://doi.org/10.1103/PhysRevD.109.083023} {\bibfield  {journal} {\bibinfo  {journal} {\prd}\ }\textbf {\bibinfo {volume} {109}},\ \bibinfo {eid} {083023} (\bibinfo {year} {2024})}\BibitemShut {NoStop}%
\bibitem [{\citenamefont {{Burrows}}\ \emph {et~al.}(2006)\citenamefont {{Burrows}}, \citenamefont {{Livne}}, \citenamefont {{Dessart}}, \citenamefont {{Ott}},\ and\ \citenamefont {{Murphy}}}]{burrows_06}%
  \BibitemOpen
  \bibfield  {author} {\bibinfo {author} {\bibfnamefont {A.}~\bibnamefont {{Burrows}}}, \bibinfo {author} {\bibfnamefont {E.}~\bibnamefont {{Livne}}}, \bibinfo {author} {\bibfnamefont {L.}~\bibnamefont {{Dessart}}}, \bibinfo {author} {\bibfnamefont {C.~D.}\ \bibnamefont {{Ott}}},\ and\ \bibinfo {author} {\bibfnamefont {J.}~\bibnamefont {{Murphy}}},\ }\href {https://doi.org/10.1086/500174} {\bibfield  {journal} {\bibinfo  {journal} {\apj}\ }\textbf {\bibinfo {volume} {640}},\ \bibinfo {pages} {878} (\bibinfo {year} {2006})}\BibitemShut {NoStop}%
\bibitem [{\citenamefont {{Gossan}}\ \emph {et~al.}(2020)\citenamefont {{Gossan}}, \citenamefont {{Fuller}},\ and\ \citenamefont {{Roberts}}}]{gossan_20}%
  \BibitemOpen
  \bibfield  {author} {\bibinfo {author} {\bibfnamefont {S.~E.}\ \bibnamefont {{Gossan}}}, \bibinfo {author} {\bibfnamefont {J.}~\bibnamefont {{Fuller}}},\ and\ \bibinfo {author} {\bibfnamefont {L.~F.}\ \bibnamefont {{Roberts}}},\ }\href {https://doi.org/10.1093/mnras/stz3243} {\bibfield  {journal} {\bibinfo  {journal} {\mnras}\ }\textbf {\bibinfo {volume} {491}},\ \bibinfo {pages} {5376} (\bibinfo {year} {2020})}\BibitemShut {NoStop}%
\bibitem [{\citenamefont {{Mezzacappa}}\ and\ \citenamefont {{Zanolin}}(2024)}]{Mezzacappa_Zanolin:2024}%
  \BibitemOpen
  \bibfield  {author} {\bibinfo {author} {\bibfnamefont {A.}~\bibnamefont {{Mezzacappa}}}\ and\ \bibinfo {author} {\bibfnamefont {M.}~\bibnamefont {{Zanolin}}},\ }\href {https://doi.org/10.48550/arXiv.2401.11635} {\bibfield  {journal} {\bibinfo  {journal} {arXiv e-prints}\ ,\ \bibinfo {eid} {arXiv:2401.11635}} (\bibinfo {year} {2024})}\BibitemShut {NoStop}%
\bibitem [{\citenamefont {{Murphy}}\ \emph {et~al.}(2025)\citenamefont {{Murphy}}, \citenamefont {{Mezzacappa}}, \citenamefont {{Lentz}},\ and\ \citenamefont {{Marronetti}}}]{Murphy_et_al:2025}%
  \BibitemOpen
  \bibfield  {author} {\bibinfo {author} {\bibfnamefont {R.~D.}\ \bibnamefont {{Murphy}}}, \bibinfo {author} {\bibfnamefont {A.}~\bibnamefont {{Mezzacappa}}}, \bibinfo {author} {\bibfnamefont {E.~J.}\ \bibnamefont {{Lentz}}},\ and\ \bibinfo {author} {\bibfnamefont {P.}~\bibnamefont {{Marronetti}}},\ }\href {https://doi.org/10.48550/arXiv.2503.06406} {\bibfield  {journal} {\bibinfo  {journal} {arXiv e-prints}\ ,\ \bibinfo {eid} {arXiv:2503.06406}} (\bibinfo {year} {2025})}\BibitemShut {NoStop}%
\bibitem [{\citenamefont {{Westernacher-Schneider}}(2020)}]{WesternacherSchneider:2020}%
  \BibitemOpen
  \bibfield  {author} {\bibinfo {author} {\bibfnamefont {J.~R.}\ \bibnamefont {{Westernacher-Schneider}}},\ }\href {https://doi.org/10.1103/PhysRevD.101.083021} {\bibfield  {journal} {\bibinfo  {journal} {\prd}\ }\textbf {\bibinfo {volume} {101}},\ \bibinfo {eid} {083021} (\bibinfo {year} {2020})}\BibitemShut {NoStop}%
\bibitem [{\citenamefont {{Torres-Forn{\'e}}}\ \emph {et~al.}(2019)\citenamefont {{Torres-Forn{\'e}}}, \citenamefont {{Cerd{\'a}-Dur{\'a}n}}, \citenamefont {{Obergaulinger}}, \citenamefont {{M{\"u}ller}},\ and\ \citenamefont {{Font}}}]{Torres-Forne_et_al:2019}%
  \BibitemOpen
  \bibfield  {author} {\bibinfo {author} {\bibfnamefont {A.}~\bibnamefont {{Torres-Forn{\'e}}}}, \bibinfo {author} {\bibfnamefont {P.}~\bibnamefont {{Cerd{\'a}-Dur{\'a}n}}}, \bibinfo {author} {\bibfnamefont {M.}~\bibnamefont {{Obergaulinger}}}, \bibinfo {author} {\bibfnamefont {B.}~\bibnamefont {{M{\"u}ller}}},\ and\ \bibinfo {author} {\bibfnamefont {J.~A.}\ \bibnamefont {{Font}}},\ }\href {https://doi.org/10.1103/PhysRevLett.123.051102} {\bibfield  {journal} {\bibinfo  {journal} {\prl}\ }\textbf {\bibinfo {volume} {123}},\ \bibinfo {eid} {051102} (\bibinfo {year} {2019})}\BibitemShut {NoStop}%
\bibitem [{\citenamefont {{Stergioulas}}(2003)}]{Stergioulas_2003}%
  \BibitemOpen
  \bibfield  {author} {\bibinfo {author} {\bibfnamefont {N.}~\bibnamefont {{Stergioulas}}},\ }\href {https://doi.org/10.12942/lrr-2003-3} {\bibfield  {journal} {\bibinfo  {journal} {Living Reviews in Relativity}\ }\textbf {\bibinfo {volume} {6}},\ \bibinfo {eid} {3} (\bibinfo {year} {2003})}\BibitemShut {NoStop}%
\bibitem [{\citenamefont {{Ackley}}\ \emph {et~al.}(2020)\citenamefont {{Ackley}}, \citenamefont {{Adya}}, \citenamefont {{Agrawal}}, \citenamefont {{Altin}}, \citenamefont {{Ashton}}, \citenamefont {{Bailes}}, \citenamefont {{Baltinas}}, \citenamefont {{Barbuio}}, \citenamefont {{Beniwal}}, \citenamefont {{Blair}}, \citenamefont {{Blair}}, \citenamefont {{Bolingbroke}}, \citenamefont {{Bossilkov}}, \citenamefont {{Shachar Boublil}}, \citenamefont {{Brown}}, \citenamefont {{Burridge}}, \citenamefont {{Calderon Bustillo}}, \citenamefont {{Cameron}}, \citenamefont {{Tuong Cao}}, \citenamefont {{Carlin}}, \citenamefont {{Chang}}, \citenamefont {{Charlton}}, \citenamefont {{Chatterjee}}, \citenamefont {{Chattopadhyay}}, \citenamefont {{Chen}}, \citenamefont {{Chi}}, \citenamefont {{Chow}}, \citenamefont {{Chu}}, \citenamefont {{Ciobanu}}, \citenamefont {{Clarke}}, \citenamefont {{Clearwater}}, \citenamefont {{Cooke}}, \citenamefont {{Coward}}, \citenamefont {{Crisp}}, \citenamefont {{Dattatri}}, \citenamefont
  {{Deller}}, \citenamefont {{Dobie}}, \citenamefont {{Dunn}}, \citenamefont {{Easter}}, \citenamefont {{Eichholz}}, \citenamefont {{Evans}}, \citenamefont {{Flynn}}, \citenamefont {{Foran}}, \citenamefont {{Forsyth}}, \citenamefont {{Gai}}, \citenamefont {{Galaudage}}, \citenamefont {{Galloway}}, \citenamefont {{Gendre}}, \citenamefont {{Goncharov}}, \citenamefont {{Goode}}, \citenamefont {{Gozzard}}, \citenamefont {{Grace}}, \citenamefont {{Graham}}, \citenamefont {{Heger}}, \citenamefont {{Hernandez Vivanco}}, \citenamefont {{Hirai}}, \citenamefont {{Holland}}, \citenamefont {{Holmes}}, \citenamefont {{Howard}}, \citenamefont {{Howell}}, \citenamefont {{Howitt}}, \citenamefont {{H{\"u}bner}}, \citenamefont {{Hurley}}, \citenamefont {{Ingram}}, \citenamefont {{Jaberian Hamedan}}, \citenamefont {{Jenner}}, \citenamefont {{Ju}}, \citenamefont {{Kapasi}}, \citenamefont {{Kaur}}, \citenamefont {{Kijbunchoo}}, \citenamefont {{Kovalam}}, \citenamefont {{Kumar Choudhary}}, \citenamefont {{Lasky}}, \citenamefont
  {{Lau}}, \citenamefont {{Leung}}, \citenamefont {{Liu}}, \citenamefont {{Loh}}, \citenamefont {{Mailvagan}}, \citenamefont {{Mandel}}, \citenamefont {{McCann}}, \citenamefont {{McClelland}}, \citenamefont {{McKenzie}}, \citenamefont {{McManus}}, \citenamefont {{McRae}}, \citenamefont {{Melatos}}, \citenamefont {{Meyers}}, \citenamefont {{Middleton}}, \citenamefont {{Miles}}, \citenamefont {{Millhouse}}, \citenamefont {{Lun Mong}}, \citenamefont {{Mueller}}, \citenamefont {{Munch}}, \citenamefont {{Musiov}}, \citenamefont {{Muusse}}, \citenamefont {{Nathan}}, \citenamefont {{Naveh}}, \citenamefont {{Neijssel}}, \citenamefont {{Neil}}, \citenamefont {{Ng}}, \citenamefont {{Oloworaran}}, \citenamefont {{Ottaway}}, \citenamefont {{Page}}, \citenamefont {{Pan}}, \citenamefont {{Pathak}}, \citenamefont {{Payne}}, \citenamefont {{Powell}}, \citenamefont {{Pritchard}}, \citenamefont {{Puckridge}}, \citenamefont {{Raidani}}, \citenamefont {{Rallabhandi}}, \citenamefont {{Reardon}}, \citenamefont {{Riley}},
  \citenamefont {{Roberts}}, \citenamefont {{Romero-Shaw}}, \citenamefont {{Roocke}}, \citenamefont {{Rowell}}, \citenamefont {{Sahu}}, \citenamefont {{Sarin}}, \citenamefont {{Sarre}}, \citenamefont {{Sattari}}, \citenamefont {{Schiworski}}, \citenamefont {{Scott}}, \citenamefont {{Sengar}}, \citenamefont {{Shaddock}}, \citenamefont {{Shannon}}, \citenamefont {{SHI}}, \citenamefont {{Sibley}}, \citenamefont {{Slagmolen}}, \citenamefont {{Slaven-Blair}}, \citenamefont {{Smith}}, \citenamefont {{Spollard}}, \citenamefont {{Steed}}, \citenamefont {{Strang}}, \citenamefont {{Sun}}, \citenamefont {{Sunderland}}, \citenamefont {{Suvorova}}, \citenamefont {{Talbot}}, \citenamefont {{Thrane}}, \citenamefont {{T{\"o}yr{\"a}}}, \citenamefont {{Trahanas}}, \citenamefont {{Vajpeyi}}, \citenamefont {{van Heijningen}}, \citenamefont {{Vargas}}, \citenamefont {{Veitch}}, \citenamefont {{Vigna-Gomez}}, \citenamefont {{Wade}}, \citenamefont {{Walker}}, \citenamefont {{Wang}}, \citenamefont {{Ward}}, \citenamefont {{Ward}},
  \citenamefont {{Webb}}, \citenamefont {{Wen}}, \citenamefont {{Wette}}, \citenamefont {{Wilcox}}, \citenamefont {{Winterflood}}, \citenamefont {{Wolf}}, \citenamefont {{Wu}}, \citenamefont {{Jet Yap}}, \citenamefont {{You}}, \citenamefont {{Yu}}, \citenamefont {{Zhang}}, \citenamefont {{Zhang}}, \citenamefont {{Zhao}},\ and\ \citenamefont {{Zhu}}}]{ackley_20}%
  \BibitemOpen
  \bibfield  {author} {\bibinfo {author} {\bibfnamefont {K.}~\bibnamefont {{Ackley}}}, \bibinfo {author} {\bibfnamefont {V.~B.}\ \bibnamefont {{Adya}}}, \bibinfo {author} {\bibfnamefont {P.}~\bibnamefont {{Agrawal}}}, \bibinfo {author} {\bibfnamefont {P.}~\bibnamefont {{Altin}}}, \bibinfo {author} {\bibfnamefont {G.}~\bibnamefont {{Ashton}}}, \bibinfo {author} {\bibfnamefont {M.}~\bibnamefont {{Bailes}}}, \bibinfo {author} {\bibfnamefont {E.}~\bibnamefont {{Baltinas}}}, \bibinfo {author} {\bibfnamefont {A.}~\bibnamefont {{Barbuio}}}, \bibinfo {author} {\bibfnamefont {D.}~\bibnamefont {{Beniwal}}}, \bibinfo {author} {\bibfnamefont {C.}~\bibnamefont {{Blair}}}, \bibinfo {author} {\bibfnamefont {D.}~\bibnamefont {{Blair}}}, \bibinfo {author} {\bibfnamefont {G.~N.}\ \bibnamefont {{Bolingbroke}}}, \bibinfo {author} {\bibfnamefont {V.}~\bibnamefont {{Bossilkov}}}, \bibinfo {author} {\bibfnamefont {S.}~\bibnamefont {{Shachar Boublil}}}, \emph {et~al.},\ }\href {https://doi.org/10.1017/pasa.2020.39} {\bibfield
  {journal} {\bibinfo  {journal} {\pasa}\ }\textbf {\bibinfo {volume} {37}},\ \bibinfo {eid} {e047} (\bibinfo {year} {2020})}\BibitemShut {NoStop}%
\end{thebibliography}%

\end{document}